\documentclass[showpacs,preprintnumbers,amsmath,amssymb,prb]{revtex4}

\usepackage{graphicx}
\usepackage{natbib}
\usepackage{amssymb}
\usepackage{amsmath}  
\usepackage{amsbsy}
\usepackage{subfig}
\renewcommand{\Re}{{Re}}
\newcommand{\Ha}{{Ha}}
\newcommand{\del}[1]{\partial_{#1}}

\begin{document}

\title[Laminar and transitional duct flow near a magnetic dipole]{Laminar and transitional liquid metal duct flow near a magnetic point dipole}

\author{S. Tympel, T. Boeck and J. Schumacher}%
\affiliation{Institut f\"ur Thermo- und Fluiddynamik, Technische Universit\"at Ilmenau, Postfach 100565, D-98684 Ilmenau,  Germany}  

\date{\today}

\begin{abstract}
The flow transformation and the generation of vortex structures by a strong  magnetic dipole field in a liquid metal duct flow is studied by means of three-dimensional
direct numerical simulations. The dipole is considered as the paradigm for a magnetic obstacle which will deviate the streamlines due
to Lorentz forces acting on the fluid elements. The duct is of square cross-section. The dipole is located above the top wall and is centered in  spanwise direction.
Our model uses the quasi-static approximation which is applicable in the limit of small 
magnetic Reynolds numbers. 
The analysis covers the stationary flow regime at small hydrodynamic Reynolds numbers $Re$ as well as the transitional 
time-dependent regime at higher values which may generate a turbulent flow in the wake of the magnetic obstacle. We present a systematic 
study of these two basic flow regimes and their dependence on $Re$ and on the Hartmann number $Ha$, a measure of the strength of the magnetic dipole field. 
Furthermore, three orientations of the dipole are compared:  streamwise, spanwise and  wall-normal oriented dipole axes. The most efficient generation of turbulence 
at a fixed distance above the duct follows for the spanwise orientation,  which is caused by a certain configuration of Hartmann layers and reversed flow at the top plate.  The enstrophy in the turbulent wake grows linearly with $Ha$ which is connected with a dominance of the wall-normal 
derivative of the streamwise velocity.
\end{abstract}


\maketitle 

\section{Introduction}\label{sec:intro} 
Fluid motion which is interacting with electromagnetic fields plays
 a central role for the dynamics in the interior and atmospheres of stars 
\citep{Biskamp1993, Ruediger2004} or in nuclear fusion \citep{Niu1989}.  Although less spectacular, such interactions
also occur in  flows of molten metals 
in technological applications ranging from the generation of monocrystals for silicon wafers to the production of steel and 
complex alloys for industrial manufacturing \citep{Davidson1999}.  Understanding the basic mechanisms of how an electromagnetic field 
generates or suppresses fluid motion is a prerequisite for
the efficient electromagnetic control of such technological processes. Applying 
this knowledge may for example improve the purity of the materials. 

Apart from flow control, electromagnetic fields  can also be employed for flow measurement in conducting liquids.  Inductive flow meters determine the fluid
velocity from the voltage induced across the flow as it traverses a  magnetic field \citep{Shercliff62}. This
 method is well-established and very
accurate. However, it is not contactless but requires electrodes in the liquid. It is therefore not suited for hot and aggressive molten metals.
Several efforts of the last decade have, therefore,  been devoted to the development of contact-less methods for such non-transparent liquids based on electromagnetic induction.
Examples of such methods are the contactless  
inductive flow tomography \citep{Stefani2004}, the rotary flow meter \citep{Priede2011}  and  Lorentz Force Velocimetry (LFV) \citep{Thess06,Thess07}.

In LFV, the  currents which are induced  in the flow by an external magnet system  result in a braking force on the flow. Conversely, the flow exerts a 
reaction force on the magnet on account of Newton's third law. This Lorentz  force on the magnet depends on the velocity magnitude and distribution, 
and can be used for velocity measurement. The state of the art in LFV is  largely limited to global measurements, i.\,e.,  mean velocities or volume fluxes of 
the liquid metal flow. The utility of LFV for local measurement has been demonstrated only very recently for the wake behind a solid obstacle \citep{Heinicke2013}. 
It requires a  localized field that pervades only a small portion of the flow domain, where  the flow can be significantly modified by the  Lorentz force.
Further development of  LFV  for detailed,  spatially resolved measurements requires  a better  understanding of such aspects.  They have to be studied by direct numerical simulations (DNS) since only DNS can provide  the fully resolved space-time evolution of three-dimensional liquid metal flows and Lorentz force components that affect the motion. By means of DNS,  the {\em local impact} of the magnetic field and the related parameter dependencies can be studied systematically in laminar, transient or turbulent cases. Because of the high computational cost of DNS, this  approach will remain limited to a moderate range of dimensionless parameters, i.\,e., Reynolds and Hartmann numbers, $Re$ and $Ha$.

In recent years, computations of liquid-metal magnetohydrodynamic (MHD) flows  have become more frequent, and a number of DNS studies have been reported.  Typically, such works were concerned with
 effects arising from the anisotropic nature of the Joule damping and from the specific MHD boundary layers at walls \citep{Boeck2007,Knaepen2008,Vantieghem2009,Chaudhary2010,Krasnov2010,Krasnov2012, Zhao2012}. Instabilities of MHD boundary layers such as the Hartmann layer, and the transition to turbulence have been analyzed  numerically in ducts and channels for homogeneous magnetic fields \citep{Gerard-Varet2002,Airiau2004,Krasnov2004,Kobayashi2008,Shatrov2010,Krasnov2013PRL}. From the LFV perspective, such investigations are of limited interest because   
the total Lorentz force vanishes in the case of a homogeneous field.

The aim of this article is to investigate the structure formation and the locally resolved Lorentz force fields in a liquid metal duct flow in the presence of a 
magnetic point dipole field. The flow  and magnetic field configurations are similar to  recent laboratory experiments on LFV using liquid InGaSn alloy at room temperature \citep{Heinicke2012}. We  conduct three-dimensional DNS based on the quasi-static approximation of the induction equation. 
This approximation applies for liquid metal flows with a high conductivity which are considered here and can be found typically  in laboratory experiments and 
industrial applications \citep{Davidson}. The induced magnetic field is then weak and does not modify the stronger primary 
magnetic field significantly. Hence, the induced currents depend linearly on the velocity field and can be computed from Ohm's law for a moving 
conductor with the electric field represented by a potential gradient.

In terms of the flow geometry, we will focus on a straight duct flow with square cross section as
one of the simplest shear flow configurations. Our choice of a magnetic dipole field is also guided by mathematical simplicity and by its rapid decay. 
A drawback is that it  represents the far field of a typical permanent magnet, i.\,e., it will not provide a 
very good approximation when such a magnet is placed close to the duct~\citep{Heinicke2012}.
Because of the rapid decay, the magnetohydrodynamic interaction of the dipole field with the flow is  strongly localized and
three-dimensional. We  will  demonstrate that it can cause a locally pronounced  deflection of the velocity field for laminar flows, and that 
the wake of the dipole is characterized by vortical structures at higher velocities.
For particular orientations of the magnetic dipole moment,
transition to turbulence in the wake is  observed at moderate Reynolds numbers. In order to keep the parameter space manageable we only consider three basic orientations of the magnetic dipole moment, namely along the streamwise, spanwise and wall-normal directions.

The present study can be regarded as a continuation of  the theoretical and numerical investigations reported in  \cite{Heinicke2012}. In this previous work,  dynamic effects of the Lorentz force on the flow were only significant in the creeping flow regime. As before, we shall not only consider the flow modification but also determine the electromagnetic force and torque on the dipole, which are  of great interest from the perspective of  LFV.

 Without reference to LFV, the flow modification by a localized magnetic field has also been investigated by other authors.  \cite{Cuevas}
have termed it magnetic obstacle  in analogy with the flow around a solid obstacle. 
Similar configurations 
have been 
investigated both experimentally and numerically \citep{Andreev2006,VotyakovPRL,Andreev2009,VotyakovJFM,Kenjeres2012}. The magnetic fields in 
these works typically had a lower  degree of spatial non-uniformity, and cannot be characterized in such as simple way as a point dipole. A point dipole has 
been used for analytic investigations in Cuevas et al. (2006)~\cite{Cuevas2006}, but for a quasi two-dimensional problem. In addition,  force and torque on the magnet system were 
typically also not studied, and  three-dimensional simulations were performed with relatively coarse grids (e.\,g. Votyakov et al. (2008)~\cite{VotyakovJFM}).
We therefore also attempt to significantly improve on these earlier works in terms of numerical resolution.
 
\begin{figure}
\centering
\includegraphics[width=0.7\linewidth]{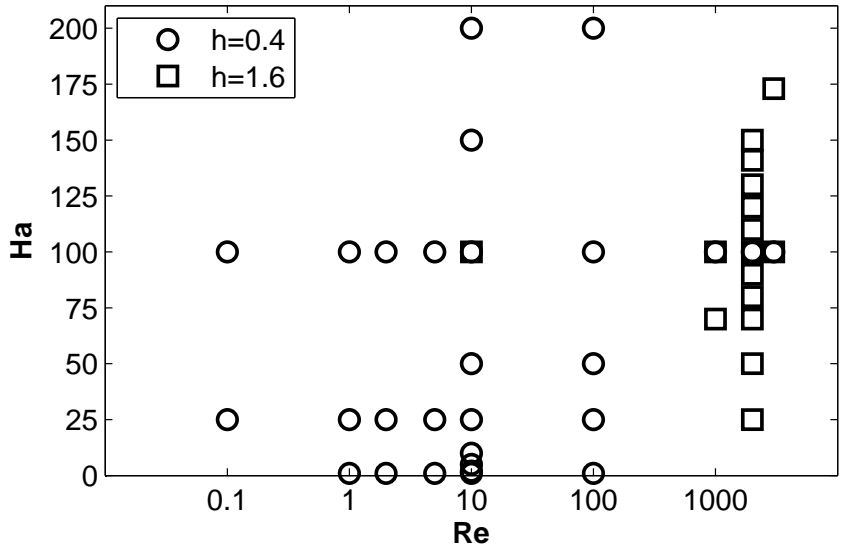}
\caption{Parameter sets of Hartmann and Reynolds numbers, $Ha$ and $Re$ investigated. Each point denotes a separate direct 
numerical simulation. The distances $h$ of the dipole from the top surface of the duct is $h=0.4$ (circles) or $h=1.6$ (squares) measured 
in units of the half width of the duct.}\label{fig:HaRe}
\end{figure}

Naturally, the geometric parameters such as position and orientation of the magnetic dipole, have a strong influence on the acting forces. 
However, a detailed investigation of this point would go beyond the scope of this article. Indeed, we will focus on the influence of the hydrodynamic 
Reynolds number $Re$ and the Hartmann number $Ha$. In brief, the 
Reynolds number describes the ratio of the inertial to the 
viscous forces in the flow (mathematical definitions of both numbers will be given in Eqns. (\ref{eq:ReDef}) and (\ref{eq:HaDef})). The Hartmann number is a measure for 
the strength of the magnetic field and its influence on the flow.  Both parameters are investigated in a wide range, as displayed in 
Fig.~\ref{fig:HaRe}.  Since it is understood  that the position of the dipole is important for the spatial development and 
transformation of the flow, we consider  two particular distances between dipole and duct. In particular, a very short distance is 
used for the lower Reynolds numbers, as this allows a direct comparison with recent laboratory experiments by Heinicke et al. (20012)~\cite{Heinicke2012}. 
At higher Reynolds numbers the magnetic dipole is positioned further away from the top wall of the duct. This changes the magnetic field configuration 
inside the flow and excites vortices that will  eventually cause the flow to become turbulent.

The article is organized as follows. First, we introduce the geometry and setting of the problem under consideration 
in Sec.~\ref{sec:defin} and explain the numerical method in Sec.~\ref{sec:nummethods}. Second, we present the results 
for low Reynolds number to describe the basic deformations of the flow field in Sec.~\ref{sec:steady_state_solutions} 
in the stationary regime. This regime exists for the lower Reynolds numbers and persists when both, $Ha$ and $Re$, 
are increased to a certain point. In Sec.~\ref{sec:wake_instabilities}, we discuss the transition to turbulence when the 
Reynolds number is further increased. Conclusions and a brief outlook are given in Sec.~\ref{sec:conclusion}.

\section{Definition of the problem}

\subsection{Equations of motion and setup}\label{sec:defin}
 \begin{figure}
  \hfill %
  \subfloat[]{\includegraphics[width=0.47\linewidth]{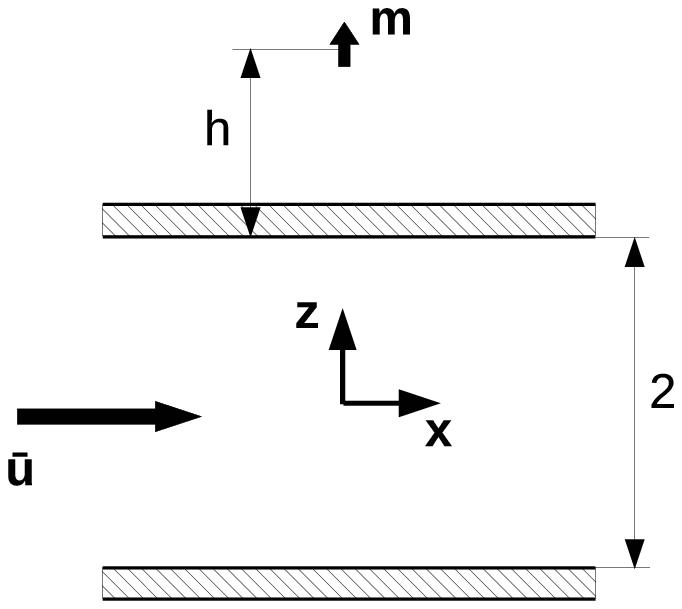}}
  \hfill %
  \subfloat[]{\hfill
\includegraphics[width=0.47\linewidth]{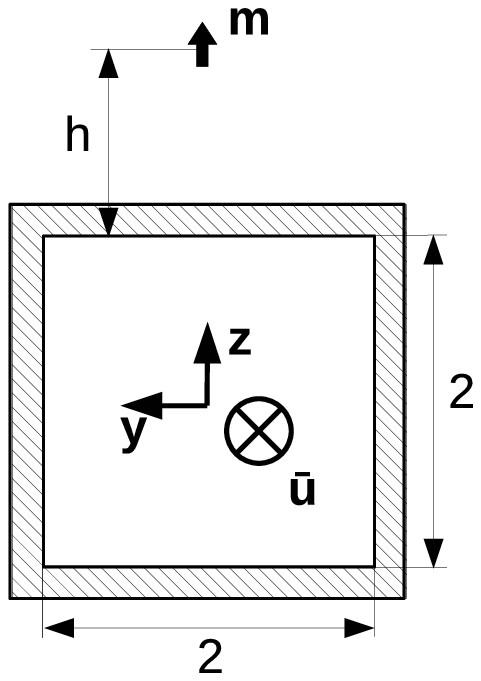}}
  \hfill %
 \caption{Setup of the problem. (a) Side view. (b) Rear view. -- The mean velocity $\bar u$ of the duct flow points in positive 
 $x$-direction. The origin of the coordinate system is placed at the center of the duct. The distance between magnetic point 
 dipole and fluid surface is denoted as $h$. The characteristic length scale of this setup is chosen to be the half-width of the duct. 
 All lengths are expressed in this unit.}
\label{fig:setup}
\end{figure}
We consider the flow of an electrically conducting fluid
in a square duct  of side length $2L$, i.\,e., the characteristic length $L$ will be defined as  the half width of the duct.   
We use Cartesian coordinates, where the streamwise direction is denoted as $x$, the spanwise direction as $y$ and 
the vertical, wall-normal direction as $z$.  The duct is exposed to 
the  inhomogeneous magnetic field of  a point dipole  located at a vertical distance $H$ above the top surface of the duct. 
In dimensional units, the dipole field at position $\vec x$ 
is given by  \citep{Jackson}
\begin{equation}
\label{eq:dipolefield_dimensional}
\vec B= \frac{\mu_0 M} {4\pi}   \left( \frac{3 \vec m \cdot \vec r }{|\vec r|^5} \vec r - \frac{\vec m}{|\vec r|^3}\right),
\end{equation}
 where $M$ denotes the magnetic dipole moment and the unit vector $\vec m$ its orientation. The other quantities
are   the vacuum permeability $\mu_0$  and the  distance $\vec r = \vec x - \vec r_0$ between $\vec x$ and the
dipole position at $\vec r_0$. For nondimensionalization of the problem we use 
the maximum value
\begin{equation}
\label{eq:dipolefield_bmax}
B_{max}= \frac{\mu_0 M } {4\pi H^3} \sqrt{m_x^2 + m_y^2 + 4m_z^2}\,
\end{equation}
 of the dipole field within the duct as unit of magnetic induction. 

 Based on $B_{max}$ and $L$ as unit of length, the 
nondimensional magnetic flux density of the dipole is 
\begin{equation}
\label{eq:dipolefield}
 \vec B^{\, \prime } ( \vec x^{\, \prime }) = \frac{h^3}{\sqrt{m_x^2 + m_y^2 + 4m_z^2}} \left( \frac{3 \vec m \cdot \vec r^{\, \prime }}{|\vec r^{\, \prime }|^5} \vec r^{\, \prime } - \frac{\vec m}{|\vec r^{\, \prime }|^3}\right),
\end{equation}
where $h=H/L$ is the nondimensional distance between dipole and duct and  nondimensional quantities are indicated by a prime. In the following, we shall use nondimensional quantities exclusively and will therefore omit the prime in subsequent equations. 

A sketch of the setup in dimensionless units is shown in Fig.~\ref{fig:setup}. The origin is chosen to be at the centerline of the 
duct. Thus the duct extends in $y$ and $z$-direction from -1 to 1. The position of the dipole is given by $\vec r_0 = (0,0,h+1)$, 
i.\,e., in the  vertical center plane. 
A key parameter of our study is the orientation of the magnetic moment of the point dipole, which is referred to as the dipole orientation. 
We will call the orientation streamwise if $\vec m =  \vec e_x$, spanwise  if $\vec m =  \vec e_y$ and vertical if $\vec m =  \vec e_z$. 
In this article, we only consider these three main orientations, but preliminary studies have shown that oblique orientations lead to 
more complex structures in the flow which will be discussed elsewhere.

As mentioned above, the dipole field approximates the far field of a small permanent magnet. 
Close to the permanent magnet  the dipole is not a very accuate model for the actual field.
Nevertheless, the differences are not substantial. We have measured
the  field distribution around a cube magnet of 1\,cm side length and compared it with a point dipole.
The dipole moment was chosen such that the magnetic induction agrees between the cube magnet
and the dipole at  1\,cm distance from its center. The magnetic induction was compared along the
dipole axis for distances greater than 1\,cm. At identical positions, the relative error in magnetic induction 
between dipole and cube magnet never exceeded 30\%. Details can be found in Tympel (2013)~\cite{Tympel_Thesis}.

Before  listing the equations that will be solved numerically, we want to recall the basic physical principle which causes the deflection of 
the fluid motion. Inside the square duct, there is a laminar pressure-driven flow of an electrically conducting liquid.
The magnetic dipole close to the duct induces electric currents, which are confined to the flow 
due to the insulating walls. These currents  give rise to  two effects. 
First, they induce a secondary magnetic field. This secondary magnetic 
field will be much weaker than the primary  field of the point dipole.  In many applications in metallurgy, one may assume that the 
secondary magnetic field is negligible. This idea coincides with the assumption that the {\em magnetic} Reynolds number $Rm$ 
\citep{Davidson, Moreau} is  very small. Therefore, we can neglect the secondary field and apply the quasistatic approximation. Physically, 
this means that the flow is unable to deform the field lines of the magnetic dipole. There is no feedback from the 
flow onto the magnetic field. 
 
Second, the currents induced by the primary magnetic field of the dipole generate a Lorentz force within the liquid. 
It gives rise to a strong deformation of the velocity profiles (cf.  Sec.~\ref{sec:steady_state_solutions}) and even triggers 
a transition to turbulence (cf. Sec.~\ref{sec:wake_instabilities}). Due to Newton's third law, a counter force  acts on the magnetic dipole 
which is of same magnitude, but of opposite sign as the total force $\vec F$ on the liquid. Recent works by Heinicke et al. (2012)~\cite{Heinicke2012} were 
focused on the determination of this force and the derivation of mean flow properties from its magnitude. Here, we will focus on the 
particular impact of the Lorentz force on the local velocity field and the resulting flow transformation.

In addition to the geometry parameters,  the problem depends on  two dimensionless parameters $Re$ and $Ha$. The Reynolds number $Re$ 
is defined as
\begin{equation}\label{eq:ReDef}
Re \equiv \frac{\bar{u}L}{\nu}\,,
\end{equation}
where $\bar{u}$ is a mean streamwise velocity
and $\nu$ 
the kinematic viscosity. The strength of the magnetic field is  quantified by the Hartmann number 
\begin{equation}\label{eq:HaDef}
Ha \equiv B_{max} L \sqrt{\frac{\sigma}{\rho\nu}}\,,
\end{equation}
where 
$\rho$ is the mass density and $\sigma$ the electric conductivity. 
In contrast to  the case of a uniform magnetic field, the definition of $Ha$ in the present case 
 is ambiguous because of the non-uniform magnetic flux density.  We 
choose the maximum of the magnetic flux density $B_{max}$ inside the duct. It  occurs at the upper boundary of the duct right  below the dipole at point 
$x=y=0$ and $z=1$. 

The governing equations of the problem will be given in nondimensional form below. We use    $\bar{u}$  
as unit of velocity and $L$ as unit of length. The pressure,  electric current and potential are given in units of $\rho \bar u^2$, $\sigma \bar u B_{max}$ and 
$\bar u B_{max}L$, respectively. To model the interaction between the Lorentz force $\vec f$ and the conducting liquid, we solve the 
incompressible Navier-Stokes equations 
\begin{equation}\label{eq:navier-stokes-with-force}
  \partial_t \vec u + (\vec u\cdot\nabla) \vec u  =
  -\nabla p+ \frac{1}{Re}  \nabla^2 \vec u +  \vec f, 
\end{equation}
\begin{equation}
\vec f=  \frac{Ha^2}{Re} \vec j \times \vec B,
\end{equation}
\begin{equation}\label{eq:incompressibility}
\nabla\cdot\vec u  =0.
\end{equation}
The induced currents are given by Ohm's law
\begin{equation}\label{eq:Ohm}
\vec j= -\nabla\phi + \vec u\times \vec B,
\end{equation}
where the electric field is represented by the negative gradient of the  electric potential $\phi$. It is determined by the condition 
$\nabla \cdot \vec j =0$, which corresponds to the Poisson equation
\begin{equation}\label{eq:poisson_potential}
\Delta \phi = \nabla \cdot (\vec u \times \vec B).
\end{equation}
Equations (\ref{eq:Ohm}) and (\ref{eq:poisson_potential}) constitute the quasi-static approximation \citep{Davidson}
with an imposed field $\vec B$. 
 For the velocity, we use no-slip boundary conditions, i.\,e., 
\begin{equation}\label{eq:bc1}
  \vec u  = 0\,
\end{equation}
at all walls. Furthermore, all walls are insulating with zero wall-normal currents, and  we therefore apply
 \begin{equation}\label{eq:bc2}
  \frac{\partial \phi}{\partial \vec n}  =0 \,,
\end{equation}
where $\vec n$ denotes the corresponding normal direction perpendicular to the walls. 
In the streamwise direction, we apply periodic boundary condition in our simulations. 
For higher Reynolds numbers, we model in- and outflow conditions with the help of the so-called fringe 
force as described below in Sec.~\ref{sec:nummethods}.



\subsection{Numerical method using a fringe force}\label{sec:nummethods}
\begin{figure}
\centering
\hfill
\includegraphics[width=1.\linewidth, angle=0]{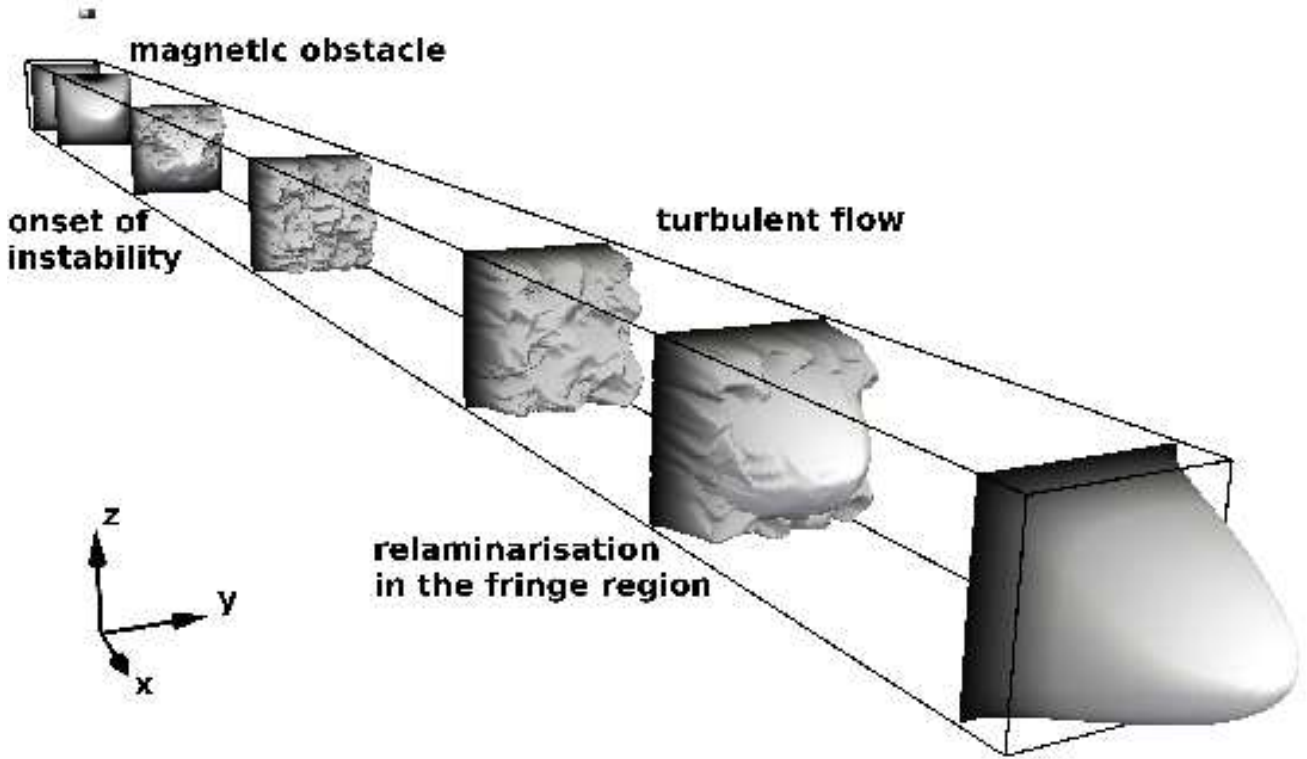}
\hfill %
\caption{Sketch of principle of the fringe force method. The incoming laminar flow is deflected by the dipole 
which acts as a magnetic obstacle. The flow is strongly deformed and can become turbulent in the wake of 
the dipole. In the region close to the end of the duct, the fringe force relaminarises the flow such that the flow 
profile is laminar again at the end of the computational domain. This method allows the usage of periodic 
boundary conditions in the streamwise direction.}
\label{fig:fringe_principle}
\end{figure}
Our DNS are performed using the  code from \cite{Krasnov}. The equations of motion are discretized on a structured 
mesh by an explicit second-order finite-difference  scheme with a collocated grid arrangement. The incompressibility 
constraint is satisfied by a standard projection method. The formulations proposed by \cite{Morinishi} and \cite{Ni} 
are incorporated into the code making the present numerical scheme  conservative for mass, momentum and electric 
charge. We use a  hyperbolic-tangent based stretching to refine the grid in the boundary layers, i.\,e., the grid is  
non-uniform in the spanwise $y$- and wall-normal  $z$-directions.
 In particular, the stretching function is given by \citep{Krasnov} 
\begin{equation}\label{eq:tanh_stretching}
 y = \frac{\tanh(\alpha_y \eta)}{ \tanh(\alpha_y )} 
\end{equation}
where $\eta$ is the coordinate on a uniform grid, $y$ denotes the transformed coordinate, while $\alpha_y$ is  the stretching coefficient. 
Equivalent  stretching  is applied in the $z$-direction.
 In the streamwise $x$-direction, we use periodic 
boundary conditions and a uniform grid spacing, and can thus apply fast Fourier transformations (FFT). This allows 
us to solve the Poisson equations for the pressure and the electric potential with the 2D-Poisson solver FishPack
(\cite{FishPack}). Verifications of the code versus a spectral code and more details on the algorithm can be found in 
\cite{Krasnov}. 

The  periodic streamwise boundary conditions simplify the solution of the Poisson equation.
FishPack is comparatively fast and uses a direct method. The resulting drawback is to ensure that a 
laminar unperturbed duct flow profile is maintained upstream of the magnetic dipole. Thus, the computational 
domain has to be increased with increasing Reynolds number when the flow is below the threshold to 
transition to turbulence. In case of low Reynolds numbers, $Re\lesssim 10$, a periodicity length of $5\pi$ was sufficient 
for the laminar duct profile to be completely reestablished at the end of the domain. For $Re=100$, a  length of $20\pi$ 
was required. This procedure increases therefore the computational cost rapidly when $Re$ is enhanced. 
 
For higher Reynolds numbers, e.\,g. $Re=2000$, the flow may become turbulent and this strategy breaks down completely. 
The conflict with the periodic boundary conditions is circumvented by the application of an additional fringe force which 
acts in the final downstream section of the duct and relaminarises the transient or turbulent flow \citep{Nordstroem1999}. 
It is sometimes also referred to as the sponge method and is frequently used in studies of turbulent 
boundary layers \citep{Simens2009, Albrecht2006}. 

The fringe method is based on an artificial body force $\vec F_{Fringe}=\lambda (\vec u_{laminar} -\vec u)$. 
This fringe force is added to the right hand side of the Navier-Stokes equations~(\ref{eq:navier-stokes-with-force}). It 
affects the flow only when the prefactor $\lambda(x)$ is non-zero.  The shape of the prefactor function is in our cases 
given by a step function with  smoothed edges \citep{Albrecht2006}. More precisely, it is given by
\[
 \lambda ( x)= \lambda_{max} \left[ S\left(\frac{x-x_{start}}{\Delta_{rise}}\right) - S\left(\frac{x-x_{end}}{\Delta_{fall}} +1\right)  \right] 
\]
with 
\[
 S(\xi) = \begin{cases}
         0, & \xi \leq 0\\ 
         0.5 - 0.5\cos (\xi\pi) , & 0<\xi<1\\
         1, & \xi\geq 1.
        \end{cases}
\]
We studied the influence of the parameters, i.\,e., the maximal amplitude $ \lambda_{max}$, 
steepness of the curve in the beginning $ \Delta_{rise}$ and in the end $\Delta_{fall} $ as well as  the total length 
$|x_{end}-x_{start} | $ of the fringe zone. These tests showed that $\lambda(x)$ has to be smooth 
and periodic. The particular shape is not important. Typical values in the calculation are $\lambda_{max}=1$,
$ \Delta_{rise}=\pi$, $\Delta_{fall}= \pi/2$ and $|x_{end}-x_{start} |=3\pi $. 
Preliminary results  with  this method were already presented in Tympel et al. (2012)~\cite{Tympel2012}.
An example for the resulting velocity profiles is shown in Figure~\ref{fig:fringe_principle}.   Here, the transformation of 
the flow by the magnetic dipole for $h=1.6$, $Re=3000$ and $Ha=100$ is illustrated. This computation was performed with 
$4096 \times 192\times 192$ grid points in $x,y$ and $z$ directions, stretched with a coefficent of $\alpha_y=\alpha_z=2$. The streamwise extent of the computational domain is $30\pi$.
 Due to high computational costs, it is not possible to perform a detailed parameter study with such fine grids. 

We performed a grid sensitivity study for three typical parameter sets which are summarized in Table \ref{tab:grid}. 
In order to resolve the boundary layers and the steep gradients of the magnetic field close to the point dipole properly, we apply a strong   
grid clustering in $y$- and $z$-direction. To elucidate the impact, the corresponding minimal and maximal grid spacing are given.  
We use three characteristic values to determine whether the resolution is sufficient or not.  
The first is  the total streamwise  Lorentz force $F_x$ (cf. Eq. (\ref{eq:total_Lorentz_force})) as it is highly dependent on the velocity and the magnetic field. 
Second, we compare two integrals  over the cross section  at $x=0$: 
the change of the streamwise momentum flux $I_2$ and the wall shear stresses $I_4$ (cf. Eqns. (\ref{I2}) and (\ref{I4})).

\begin{table}
 \begin{center}
\begin{tabular}{llllllllll}
&& $n_x\times n_y\times n_z$  & $L_x$ & $\alpha_{y/z}$ & min$\Delta_{y/z}$& max$\Delta_{y/z}$ & $|F_x|$ & $|I_2(0)|$  & $|I_4(0)|$ \\\hline 
 $Re=10$,        ~~& $\star$  &$1024\times96\times96$ & $5\pi$ & $1.5$ &0.0064& 0.0345& 1.1734   &0.2390&3.5758\\ 
               $h=0.4$  & &$4096\times256\times256$ & $5\pi$ & $1.5$ &0.0024&0.0129 & 1.1497  &0.2354&3.5623\\\hline 
$Re=1000$,   ~~         & &$2048\times48\times48$ & $30\pi$ & $2.0$ &0.0078 &0.0862 & 0.3050   & 0.1545 & 0.1080\\  
 $h=1.6$                & &$3072\times64\times64$ & $30\pi$ & $2.0$ &0.0055 & 0.0647& 0.3080   & 0.1544 & 0.1089\\ 
                        & &$4096\times96\times96$ & $30\pi$ & $2.0$ &  0.0034& 0.0432 & 0.3101 & 0.1545 & 0.1096\\ 
                 &$\star$ &$2048\times96\times96$ & $15\pi$ & $2.0$  & 0.0034 & 0.0432 & 0.3102& 0.1545 &0.1096 \\ 
                        & &$6144\times128\times128$ & $30\pi$ & $2.0$ &0.0024 &0.0324 & 0.3111 & 0.1547 & 0.1099\\ 
                        &  &$8192\times192\times192$ & $30\pi$ & $2.0$ &0.0016 &0.0216 & 0.3120 & 0.1550 &0.1101 \\ \hline
$Re=2000$,              &  &$1024\times48\times48$ & $30\pi$ & $2.0$ & 0.0078 &0.0862 & 0.1612  & 0.0386 & 0.0252\\ 
 $h=1.6$                &  &$1728\times80\times80$ & $30\pi$ & $2.0$ &0.0038 & 0.0518& 0.1635   & 0.0365  & 0.0255\\ 
                        &  &$2048\times96\times96$ & $30\pi$ & $2.0$ & 0.0034 & 0.0432 & 0.1639 & 0.0367 & 0.0256\\ 
                 & $\star$ &$2048\times96\times96$ & $15\pi$ & $2.0$ & 0.0034 & 0.0432 & 0.1639 & 0.0367 & 0.0256\\ 
                        &  &$4096\times96\times96$ & $30\pi$ & $2.0$ & 0.0034 & 0.0432 &  0.1639& 0.0367  & 0.0256 \\ 
                        &  &$4096\times128\times128$ & $30\pi$ & $2.0$ & 0.0024 &0.0324 & 0.1645& 0.0367 & 0.0256\\ 
                        &  &$8192\times256\times256$ & $30\pi$ & $1.5$ & 0.0024 & 0.0129& 0.1650&0.0366 & 0.0257 \\ 
                        &  &$8192\times256\times256$ & $30\pi$ & $1.75$ &0.0017 & 0.0145 &0.1650& 0.0366 & 0.0257
 \end{tabular}
\caption{Results of the resolution study for three different parameter sets with spanwise oriented dipole at $Ha=100$. The symbol $\star$ indicates the resolution that is applied in the parameter studies. Properties of the grids are the number of grid points $n_x\times n_y\times n_z$, streamwise length of computational domain $L_x$, the used the stretching coefficient $\alpha_y=\alpha_z$ in $y$ and $z$-direction with resulting minimal and maximal grid spacing min$\Delta_{y/z}$ and max$\Delta_{y/z}$, respectively. 
Grid convergence is studied with the streamwise Lorentz force $F_x$ (cf. Eq. (\ref{eq:total_Lorentz_force})), the change of the streamwise momentum flux $I_2$ and the wall shear stresses $I_4$ in a cross section at $x=0$ (cf. Eqns. (\ref{I2}) and (\ref{I4})). Values for $Re=2000$ are time-averaged as done in Sec. \ref{sec:wake_instabilities}. }\label{tab:grid}
 \end{center}
\end{table}

The study confirms that the resolution used in our investigations is sufficient to capture the flow dynamics. 
For instance in case of  $Re=10$, $Ha=100$, $h=0.4$ and a spanwise oriented dipole -- which represents the basic setting in Sec. \ref{sec:steady_state_solutions} --
the  difference between the resulting Lorentz forces of the applied grid with $1024\times96^2$ and the finest grid with $4096\times256^2$  points is approximately 2\,\%.  
For the transient flow at higher Reynolds numbers and the  parameter studies in Sec.~\ref{sec:wake_instabilities}, we attempt to achieve  resolutions similar to 
Gavrilakis (1992)~\cite{Gavrilakis1992} or Huser and Biringen (1993)~\cite{Huser1993}. This results in  a resolution of  $2048 \times 96^2$ grid points for a  streamwise domain  length of $15\pi$. In addition,  the total force of this grid  
is only  0.6\% lower than for the finest grid. 
Furthermore, the resolution  study shows that the same physical effects (e.\,g. the vortex shedding that is described in Sec.  \ref{sec:wake_instabilities})
 are still obtained with a coarser resolution of $1024 \times 48^2$. 
All values for $Re =2000$ are time-averaged over 3168 snapshots. 
To exclude errors due to the time averaging,  the grid sensitivity study was repeated for $Re=1000$. 
Again, the total force differs only by 0.7 \% for the used grid. 
In addition, the velocity field itself was compared along a line right below the dipole ($x=0$ and $z=0.98$): 
The maximal error is less than $0.1\%$ for all velocity components and less than $0.2\%$ for all components of the Lorentz force density, respectively.


\section{Stationary flow at lower Reynolds numbers}\label{sec:steady_state_solutions}

\subsection{Mechanisms of flow profile deformation}\label{sec:steady_mechanism}
In the following, we discuss the behavior of the flow in case of low Reynolds numbers.  As a first step, we take a 
closer look at the particular deformation of the velocity profile for a Reynolds number $Re= 10$ and a Hartmann 
number $Ha= 100$. It is observed that differences in the deflection depend on the orientation of the magnetic 
moment of the dipole. 
\begin{figure}
\centering
\includegraphics[width=0.99\textwidth]{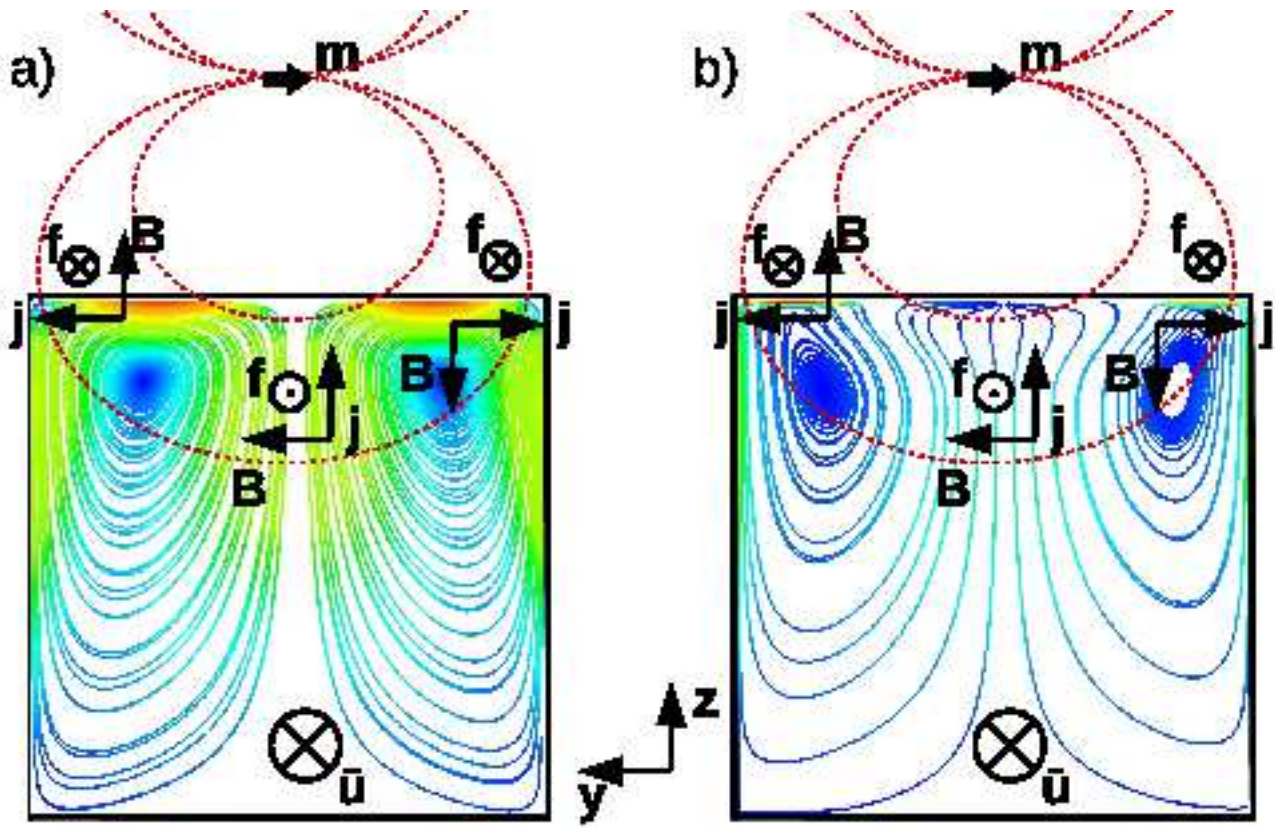}
\caption{(colour online) Sketch of principle for the acting forces in case of spanwise oriented magnetic dipole for (a) a laminar base flow and (b) the deflected flow. 
Streamlines (solid) represent the electric currents in the cross section with $x=0$ for Reynolds number $Re=2000$,  
Hartmann number $Ha=100$ and dipole  distance of $h=1.6$. Red dashed lines indicate the magnetic field lines. 
Arrows denote the local main direction of the magnetic field and the currents. In the Hartmann layers (top corners of 
the duct), the resulting Lorentz force points then in the same direction as the mean velocity. In the centerline, the Lorentz 
force brakes the fluid and -- if strong enough -- drives the vortex formation.}
\label{fig:sketch_of_principle}
\end{figure}

\begin{figure}
\includegraphics[height=\textwidth, angle=0]{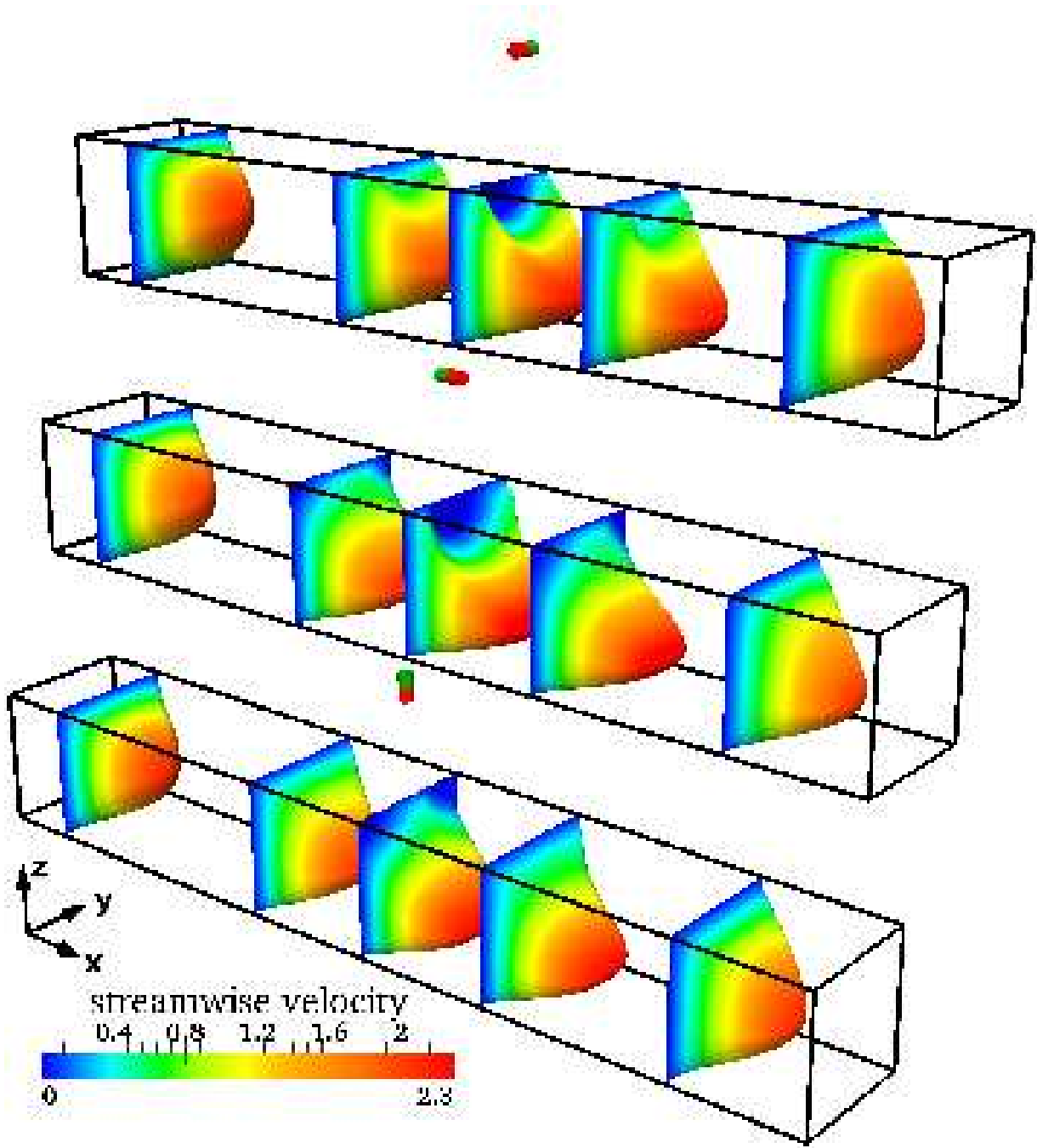}
\caption{ The point dipole as a magnetic obstacle: Contours of streamwise velocity at various positions along the duct illustrate the effect of a point dipole on the 
duct flow dynamics. From left to right: $x=-3,\,-1,\,0,\,1,\,2.5$. Total duct length is $5\pi $ and $Re=10$, $Ha=100$, $h=1.6$. Three different orientations of 
the dipole are presented: top -- spanwise, middle -- streamwise and bottom -- vertical magnetic moment}\label{fig:distortion}
\end{figure}
\begin{figure}
 \centering
  \hfill %
  \subfloat[ ]{\includegraphics[width =  \textwidth]{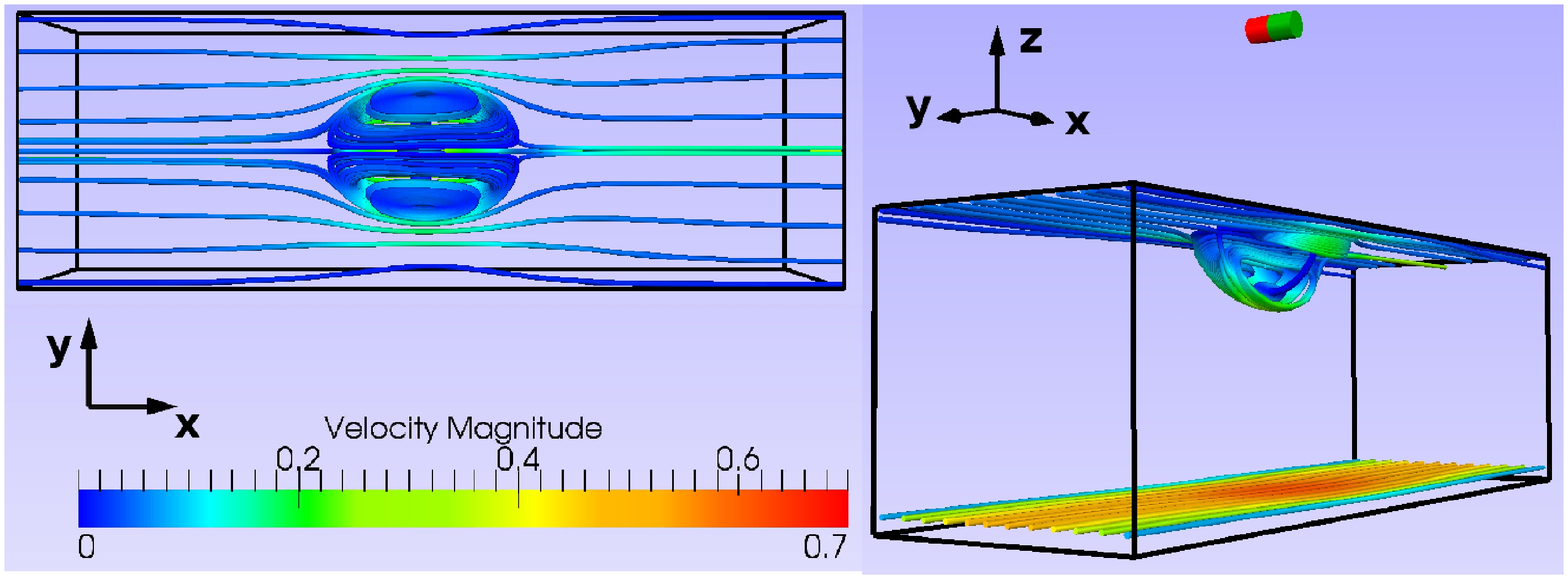}}
  \hfill %
\\
  \hfill %
  \subfloat[ \label{fig:low_Re_stream_top_view}]{\includegraphics[width =  \textwidth]{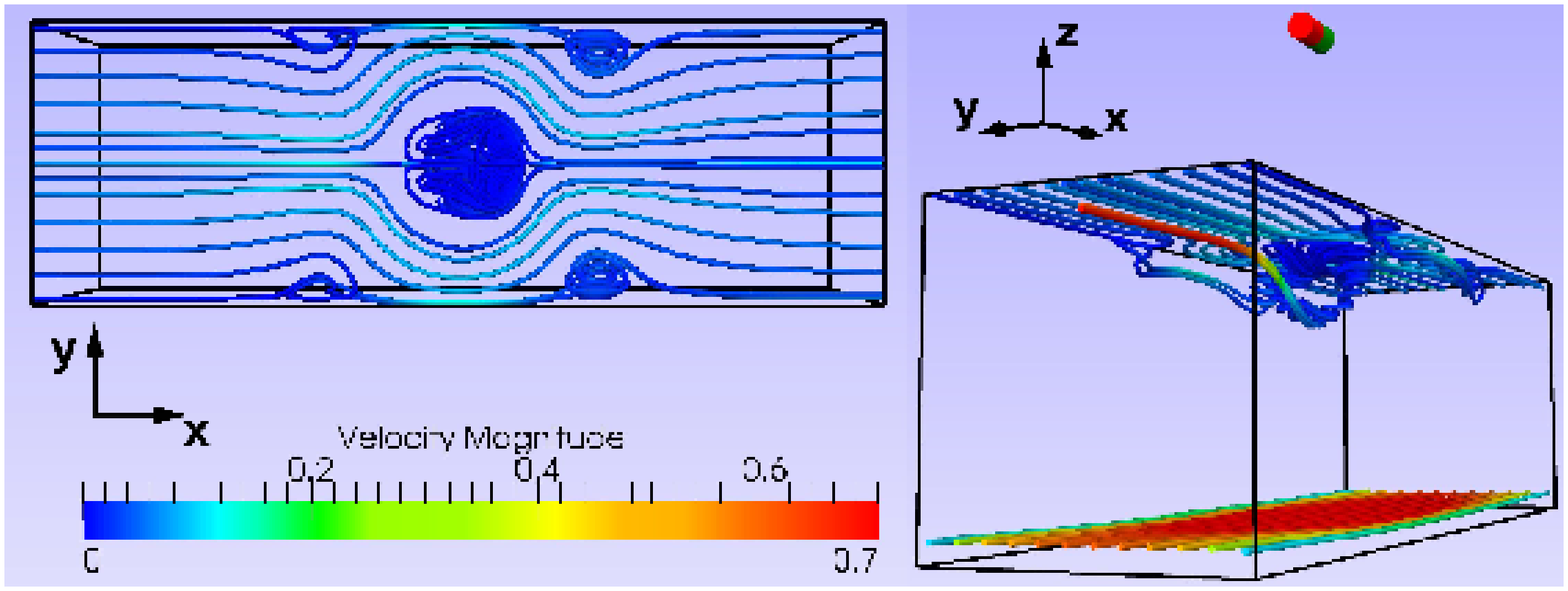}}
  \hfill %
\\
  \hfill %
   \subfloat[]{\includegraphics[width =  \textwidth]{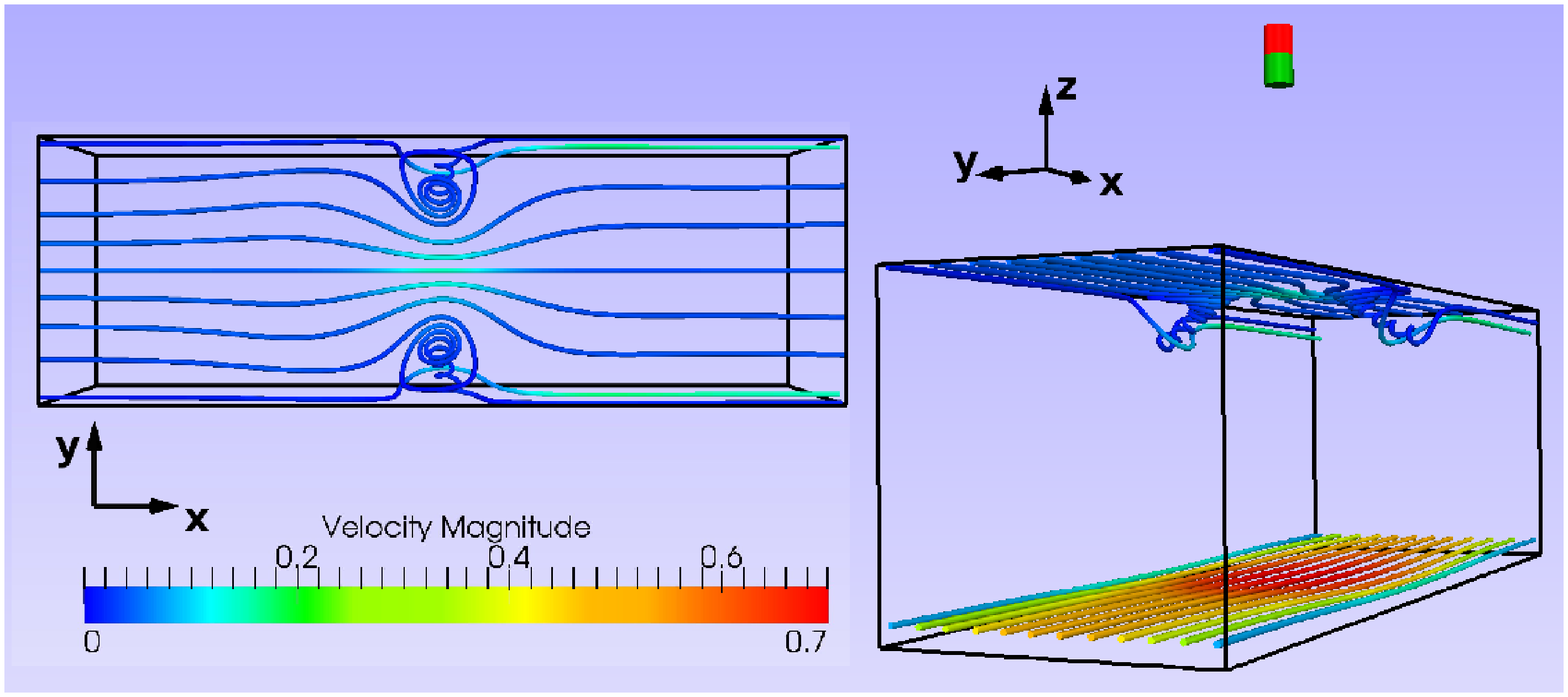}}
  \hfill %
\caption{(colour online) Velocity field streamlines for different orientations of the dipole at $Re=10$, $Ha=100$, $h=1.6$. Dipole orientations are 
a) spanwise, b) streamwise, c) vertical. The local Hartmann layer is visible due to higher velocities at the top surface. Vortices and flow structures 
are always three-dimensional. All three configurations are shown from two different perspectives in each case: left is a top view and right a side view.
The box indicates the duct from $x=-3$ to $x=3$. The total length of the duct was $5\pi$.}
\label{fig:low_Re_streamlines_verschiedene_orientierungen}
\end{figure}


For low Reynolds numbers,   pressure gradient and Lorentz forces balance in the core of the flow provided that the magnetic field
is sufficiently strong.  This solution  is usually not compatible with the conditions at walls. When the magnetic field has a sufficiently strong wall-normal component,
the tangential force balance at a wall gives rise to an electromagnetic boundary layer called  Hartmann  layer. The velocity difference across this layer
is related to an electric current parallel to the wall and perpendicular to the outer velocity \citep{Shercliff1965}. 
Hartmann layers can be formed whenever  their  thickness is small compared with the lateral variation of the flow and magnetic field. Their appearance is therefore not restricted to uniform fields. 

 In the problem at hand, the Hartmann layers are  regions where the velocity at the outer edge is controlled by the current  in the layer. These currents in the Hartmann layers are generated by the  electromotive force $\vec u\times \vec B$ in the core. Since the flow is dominated by the streamwise velocity component, we can infer the  current pattern  from the distribution of the magnetic dipole field. The Hartmann layers then typically indicate  a local  acceleration of the flow, which affects the velocity distribution in its vicinity.
In the following, we shall focus on the characteristic current distribution for a velocity field dominated by the streamwise component. The associated Lorentz forces and Hartmann layers for this current distribution indicate how the flow field is transformed.

Let us consider  the case of a magnetic dipole  in spanwise direction, 
i.\,e., $\vec m\parallel \vec e_y$. The magnetic field is perpendicular to the flow direction and parallel to the top wall 
when the dipole is not too close to the duct. 
The magnetic field  is of high magnitude and strongly inhomogeneous in the cross section directly below the dipole, i.\,e., 
at $x=0$ (Fig.~\ref{fig:sketch_of_principle}).  The induced current density  is the result of the electromotive force
$\vec u\times \vec B$ and the induced  electric field represented by the gradient of the electric potential. In the bulk, the currents
qualitatively conform to the electromotive force $\vec u\times \vec B$. Differences arise near the walls where the currents have to close
because of $\nabla \cdot \vec j=0$.  Figure~\ref{fig:sketch_of_principle}~(a) shows that there are two stagnation  points in the streamline pattern of the electric  currents based on the laminar velocity distribution.  They are close to the surface.  Figure ~\ref{fig:sketch_of_principle}~(b)  shows that the
 essential properties of the current distribution are preserved for the modified flow field. The streamlines have been obtained from a numerical simulation and are projected into the cross section. What may look like a sink or a source in Fig.~\ref{fig:sketch_of_principle}~(b) can be explained as a result of the projection on the two-dimensional cross section.

With the obtained current density, one may now estimate the resulting Lorentz force distribution  $\vec f \sim \vec j \times \vec B$ in 
the duct which will result in a braking force in the bulk of the flow. It should be emphasized that this braking  force is also present 
in the region directly below the dipole. Here, the magnetic field has the largest magnitude. If this force is strong enough, which 
is the case when the Hartmann number is sufficiently large, a local flow reversal is observed. The resulting vortices in Figs.~\ref{fig:distortion} 
(top) and~\ref{fig:low_Re_streamlines_verschiedene_orientierungen} (a) mark now the areas with strong spanwise magnetic field.
In the top corners, the situation is  different. As the current density field lines close here, the resulting Lorentz force accelerates  the flow. 
Thus,  local Hartmann layers are created. These layers are of interest when the influence of the Hartmann number is investigated 
in Sec. \ref{sec:turb_ha}. Contrary to the classical Hartmann flow configuration  \citep{HartmannI, HartmannII}, the local Hartmann layer is 
present at the top wall and practically absent at the bottom wall.
As a consequence, 
the flow distortion decreases rapidely when moving from the top to the bottom wall. 
This is illustrated  in Fig.~\ref{fig:low_Re_streamlines_verschiedene_orientierungen}
by a second set of  streamlines seeded at a distance of 0.05 length units above the bottom wall. Nonetheless,
local Hartmann layers also appear for the other dipole orientations where the normal component of the magnetic field is 
sufficiently strong at a wall.
In case 
of a streamwise dipole orientation, strong Hartmann layers are found at the centerline ahead and behind the dipole position. 
Much less pronounced  layers are also found at the sides.  

For the same reasons, one observes  a well pronounced local Hartmann layer positioned directly below a vertical dipole, i.\,e., 
$\vec m\parallel\vec e_z$. Here, the main component of the magnetic field is in vertical direction and thus perpendicular to the flow 
direction and  to the top wall. These layers can be observed in Figs.~\ref{fig:distortion} (bottom)  and 
\ref{fig:low_Re_streamlines_verschiedene_orientierungen} (c). In addition, two vortices are created in the top corners.  
One may understand the case of the vertical dipole as counterpart of the spanwise case as it forms Hartmann layers where the 
other shows areas of reversed flow and vice versa. Moreover, the vertical dipole induces spanwise oriented  currents in the bulk as
it is the case for a homogeneous vertical field. Consequently, the total Lorentz force appears again as a braking force.

In the case of streamwise magnetic moment, i.\,e., $\vec m\parallel \vec e_x$, one observes five well pronounced vortices at the 
top as shown in Fig.~\ref{fig:low_Re_streamlines_verschiedene_orientierungen}~(b). The four small vortices in 
Fig.~\ref{fig:low_Re_streamlines_verschiedene_orientierungen}~(b) are located up- and downstream  of the dipole at positions where local Hartmann layers appear
along the centerline. In this region, the currents in the bulk are in the spanwise direction. They  close near the upper and lower walls.
This is the reason for a  braking Lorentz force near the lateral walls by the spanwise component of the magnetic field. At the
centerline, the flow is accelerated in  the Hartmann layer. This Lorentz force distribution causes  regions of reversed
and accelerated flow.
The strongly pronounced vortex directly below the dipole is not directly generated by the 
magnetic field. The main component of the magnetic field in this area is a streamwise component and thus parallel to the mean flow 
direction and the top wall. A streamwise magnetic field cannot produce an electric current and therefore the action of the Lorentz force 
vanishes right below the streamwise oriented dipole. Nevertheless, vortices like the one just discussed have been observed for the streamwise 
oriented dipole at several Reynolds numbers and strong Hartmann numbers.  The origin of this vortex may be found in the local 
Hartmann layers that surround the vortex. The Lorentz forces affect the liquid at the top as follows. The flow is first accelerated at the 
centerline, then pushed aside to the two Hartmann layers close to the edges and finally again accelerated at the centerline. This forcing 
leaves the liquid right below the dipole no other possibility, but to swirl and to form a vortex. One has to mention here that this explanation 
is more or less descriptive  and that the flow is always fully three-dimensional as indicated in  
Figs.~\ref{fig:low_Re_streamlines_verschiedene_orientierungen}~(b).

An alternative explanation of the flow modification might be given by means of the characteristic surfaces in creeping magnetohydrodynamic flows, which have been introduced by \cite{Kulikovskii1968}.  These surfaces can be defined when pressure gradient and Lorentz force balance in the core of the flow. They are spanned by the parts of the magnetic field lines within the flow that have a constant value of the line integral $\int ds/\| \vec B\|$. Under certain additional assumptions, the flow will be parallel to these surfaces \citep{Alboussiere1996, Alboussiere2001}.  One may therefore expect the flow to avoid the regions enclosed by field lines with a certain magnitude of  the magnetic field. For the stream- and spanwise orientations this would be a roughly hemispherical region directly beneath the dipole. For the vertical dipole orientation, there would be two such domains to the left and to the  right of the dipole.  This is also in qualitative agreement with the specific arguments based on the current distribution that we have discussed before.

\begin{figure}
  \hfill %
  \subfloat[]{\includegraphics[width = 0.32 \textwidth]{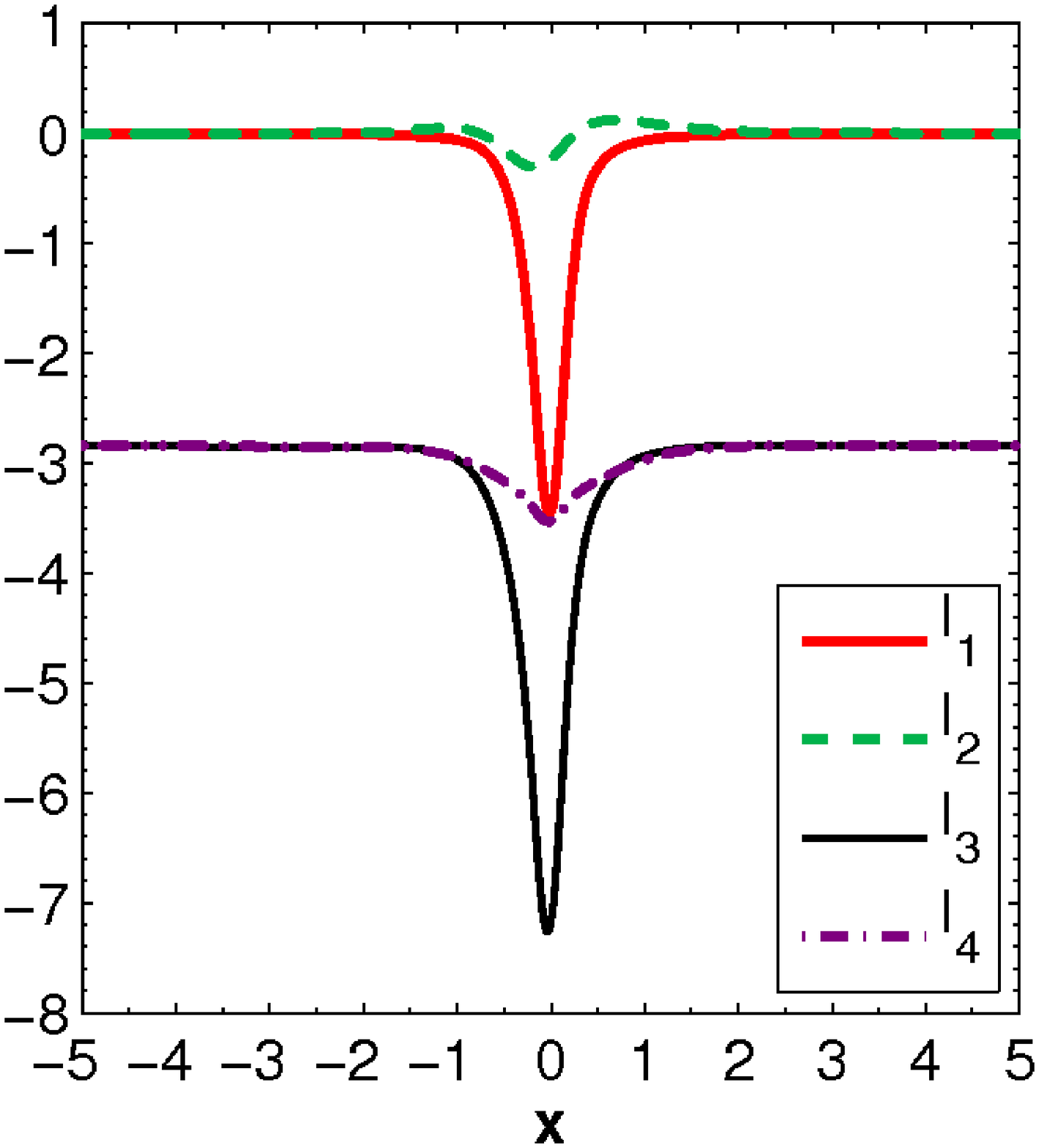}}
  \hfill %
  \subfloat[]{\includegraphics[width = 0.32 \textwidth]{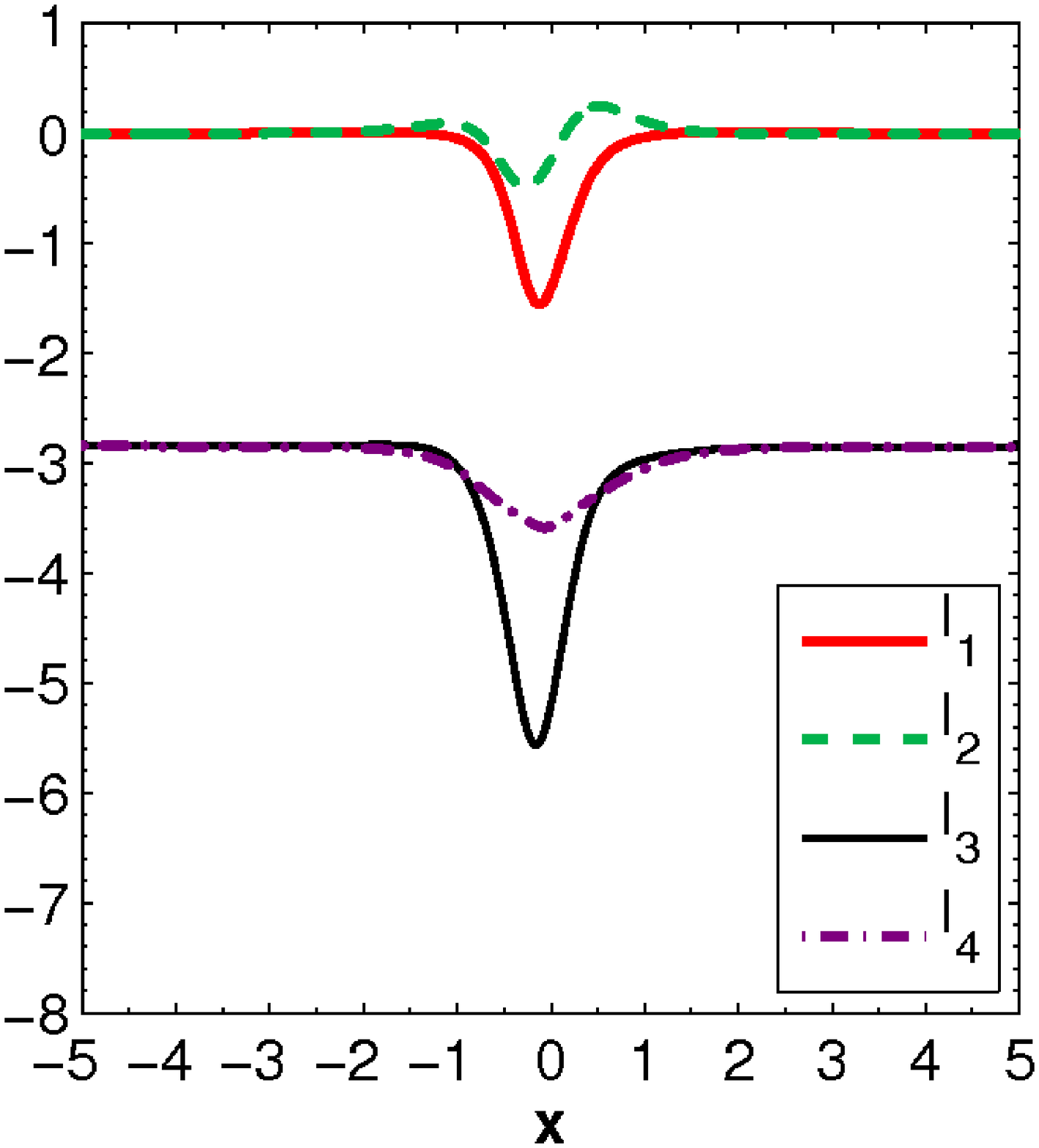}}
  \hfill %
  \subfloat[]{\includegraphics[width = 0.32 \textwidth]{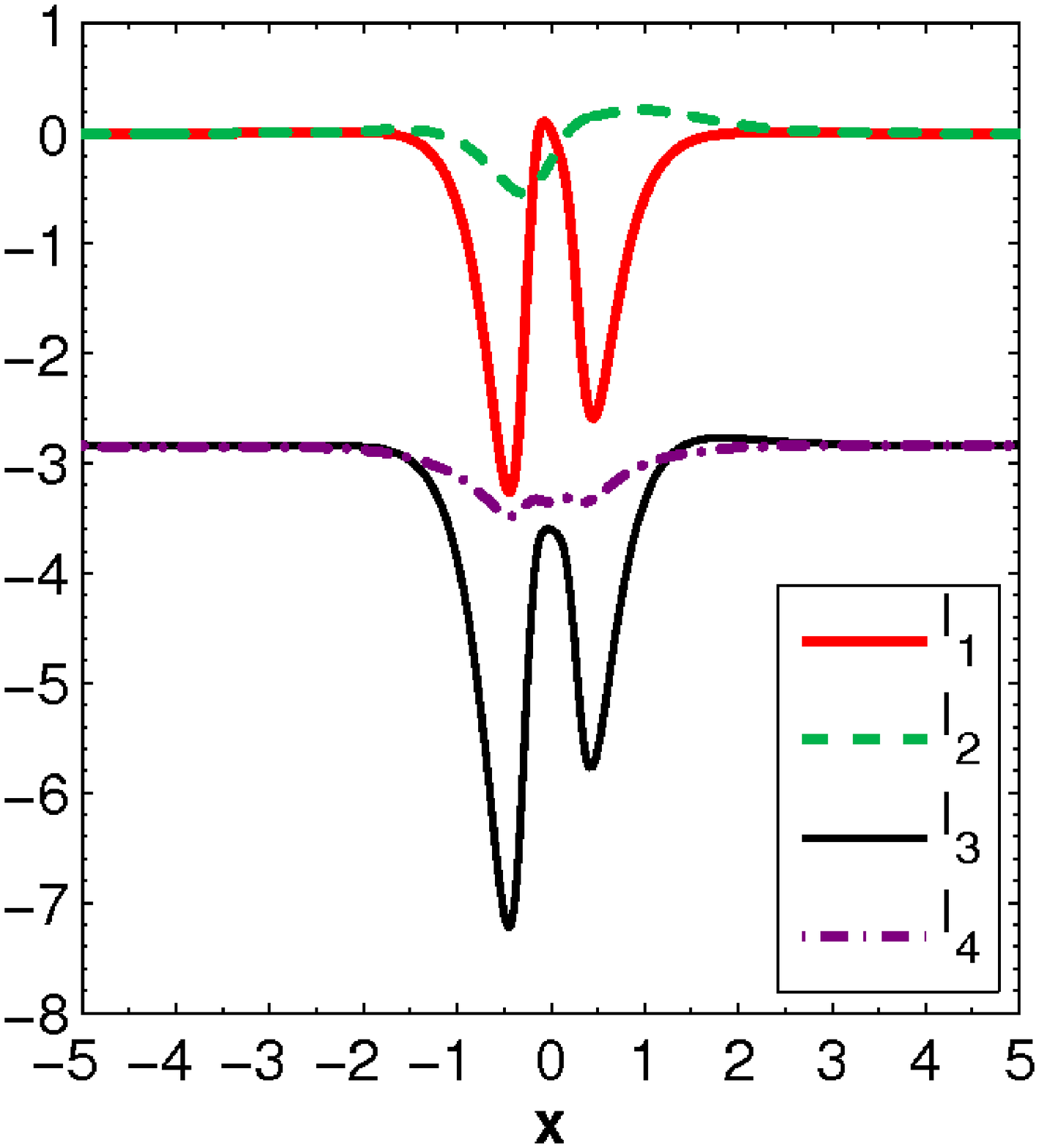}}
  \hfill %
 \caption{(colour online) Terms of the integrated momentum balance as given by Eqns. (\ref{I1}) to (\ref{I4})  at $Re=10$, 
$Ha=100$, $h=0.4$. Streamwise profiles for different dipole orientations are shown: (a) vertical, (b) spanwise, (c) streamwise.}
\label{fig:momentum_balance_low_Re}
\end{figure}

So far, we have reported the velocity profiles and the streamlines of the flow. These plots give us information on the flow structures, but 
not on the Lorentz force components in the flow. In order to proceed, we  quantify the influence of the dipole on the flow 
with an integral criterion for the balance of Lorentz force, pressure gradient and wall stresses. Such a balance equation is obtained 
directly from the Navier-Stokes equation~(\ref{eq:navier-stokes-with-force}). We integrate the $x$-component of~(\ref{eq:navier-stokes-with-force}) 
with respect to $y$ and $z$ and apply the assumptions that the velocity field is smooth and stationary and that the 
volume flux is constant and normalized to 1. The balance equation is
\begin{align}\label{eq:integrals_bilanz}
I_1+I_2-I_3+I_4=0.
\end{align}
It contains the Lorentz force term 
\begin{equation}
I_1 = \iint_{[-1,1]^2} f_x\, d y\, d z, 
\label{I1}
\end{equation}
 the change of the streamwise momentum flux 
\begin{equation}
I_2=-\iint_{[-1,1]^2} \del x (u_x^2)\, d y\, d z\,,
\label{I2}
\end{equation}
the friction coefficient 
\begin{equation}
I_3=\iint_{[-1,1]^2} \del x p \, d y\, d z \,
\label{I3}
\end{equation}
and the contribution of the wall shear stresses 
\begin{equation}
I_4=\frac 1 {Re} \left( \int_{-1}^1\left[\del y u_x\right]_{y=\pm 1}\, d z +\int_{-1}^1 \left[\del z u_x\right]_{z=\pm 1}\, d y\right)\,.
\label{I4}
\end{equation}
For a homogeneous magnetic field, the integral of the Lorentz force per cross section vanishes. In case of a laminar flow, 
wall stresses are balanced by the pressure gradient. With an inhomogeneous magnetic field present, the question of which 
hydrodynamic forces balance the appearing Lorentz forces arises. Figure~\ref{fig:momentum_balance_low_Re} shows all 
terms of the balance equation (\ref{eq:integrals_bilanz}) for the three dipole orientations. 

We can see that the contribution to the Lorentz force is mainly based on the presence of the local Hartmann layers. 
Therefore, the forces are much stronger for the vertical case than for the spanwise one. The strongest contribution of the 
Lorentz force comes from the streamwise oriented dipole. Here, two Hartmann layers are clearly visible, ahead and behind the 
dipole positioned at $x=0$. It also becomes clear that it is mainly the pressure gradient which balances the Lorentz force. 
The nonlinear term $I_2$  is small due to the low Reynolds number of $Re=10$. In Sec.~\ref{sec:wake_instabilities}, we will come
back to this point and will see that the nonlinear term has a stronger influence as expected for higher Reynolds numbers.
Before increasing the Reynolds number, we study the influence of the Hartmann number at fixed $Re=10$. 

\subsection{Hartmann number  dependence at fixed Reynolds number}
The following scalar integral measure 
\begin{equation}\label{eq:distortion_meassure}
\mathcal A(x)= \dfrac{\iint |\vec u - \vec u_{lam}| \,d y\,d z}{\iint |\vec u_{lam}| \,d y\,d z}.
\end{equation}
will be used for quantification of the distortion of the laminar profile $\vec u_{lam}$
caused by the magnetic obstacle.
\begin{figure}
\centering
\includegraphics[width = 0.65 \textwidth]{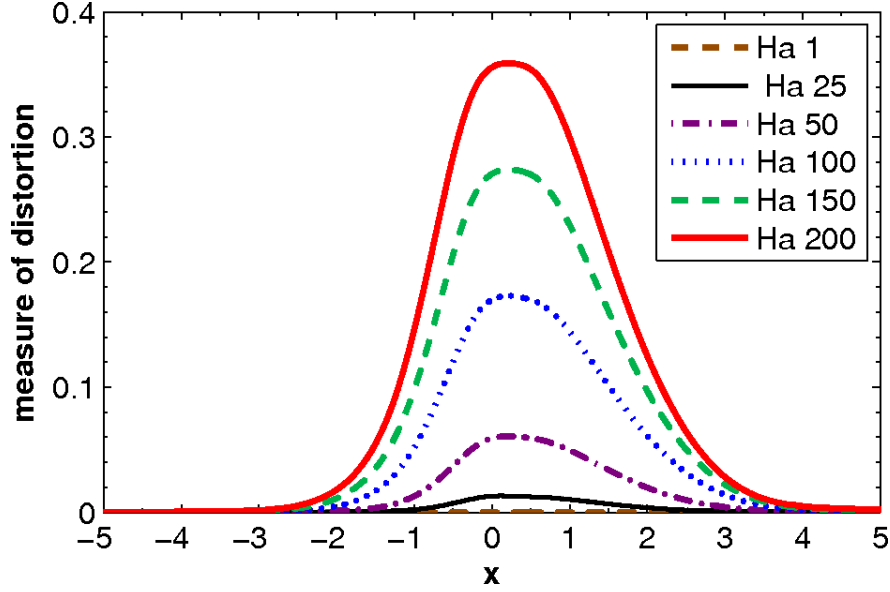}
\caption{(colour online) Distortion of the laminar duct flow as a function of the Hartmann number for a low Reynolds number duct flow at $Re=10$. 
Deflection of the flow in dependence of Hartmann number as quantified by $\mathcal{A}$.}
\label{fig:distortion_per_Ha}
\end{figure}
An example for the distortion is shown in Fig.~\ref{fig:distortion_per_Ha} for a Reynolds number of 10 and a wall-normal oriented dipole in a distance 
$h=0.4$. The graphs compare the distortion for several Hartmann numbers which are indicated in the legend.  Two effects are observed. 
First, there is almost no distortion of the flow for $Ha <25$. In this range,  the resulting forces behave to a good approximation
as in the kinematic case for an unperturbed flow. They are also comparable with the experiments by Heinicke et al. (2012)~\cite{Heinicke2012} in this parameter range.  
Second, the maximal amplitude of $\mathcal A$ increases approximately linearly with the Hartmann number for $Ha > 25$. Nevertheless, 
one can assume a saturation of this deformation for very high Hartmann numbers due to the finite duct geometry.  
Furthermore, the starting point of the deflection is shifted upstream with increasing Hartmann number. 
In particular, it changes from $x\approx -1$ for $Ha=25$ to $ x\approx -3$ for $Ha=200$. 
In all cases the distortion subsides at approximately four 
characteristic lengths downstream of the dipole position. 
Therefore, the curves become slightly 
more symmetric for higher Hartmann numbers.
\begin{figure}
\centering
  \hfill %
  \subfloat[]{\includegraphics[width = 0.49 \textwidth]{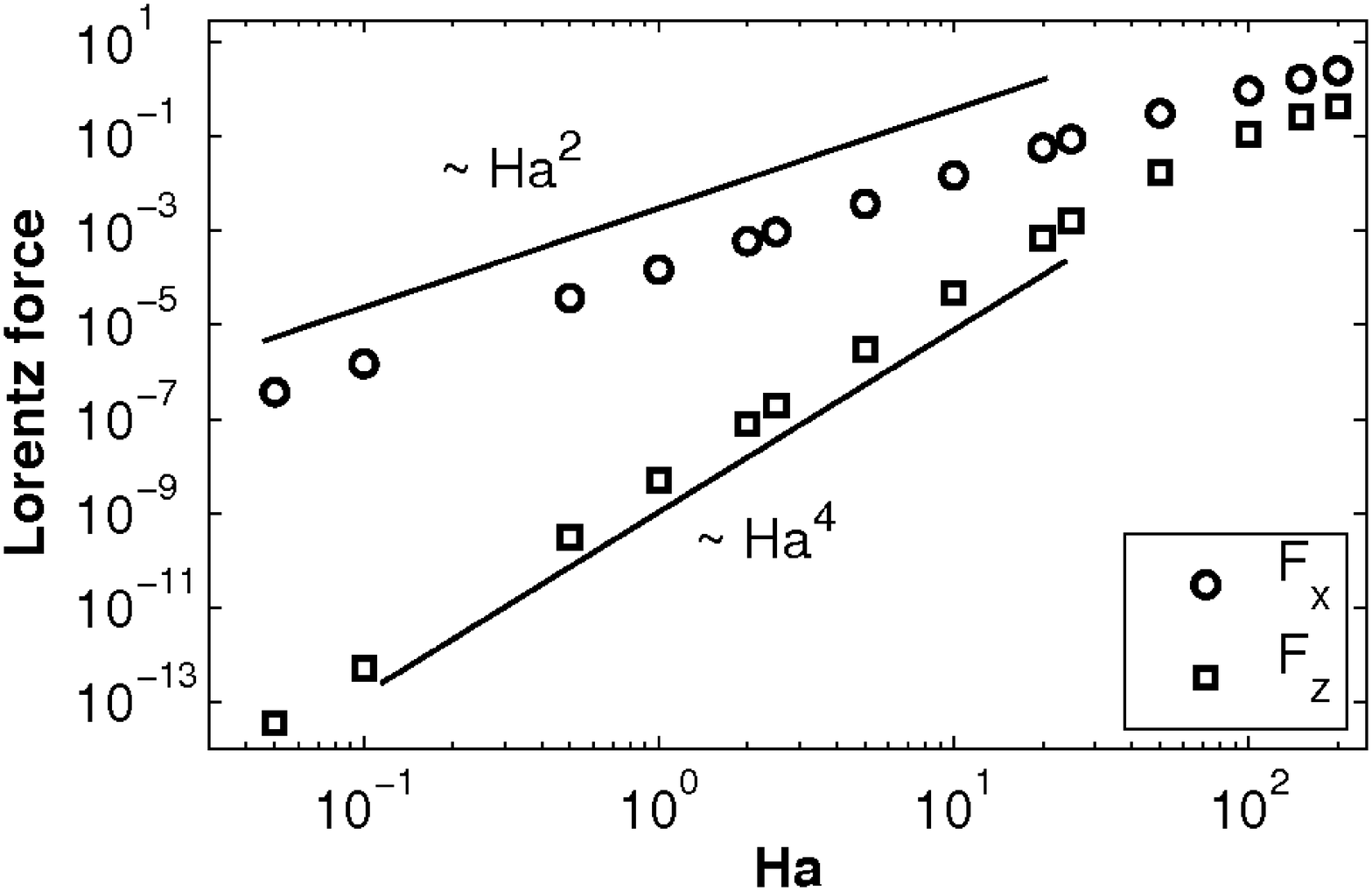}}
  \hfill %
  \subfloat[]{\includegraphics[width = 0.49 \textwidth]{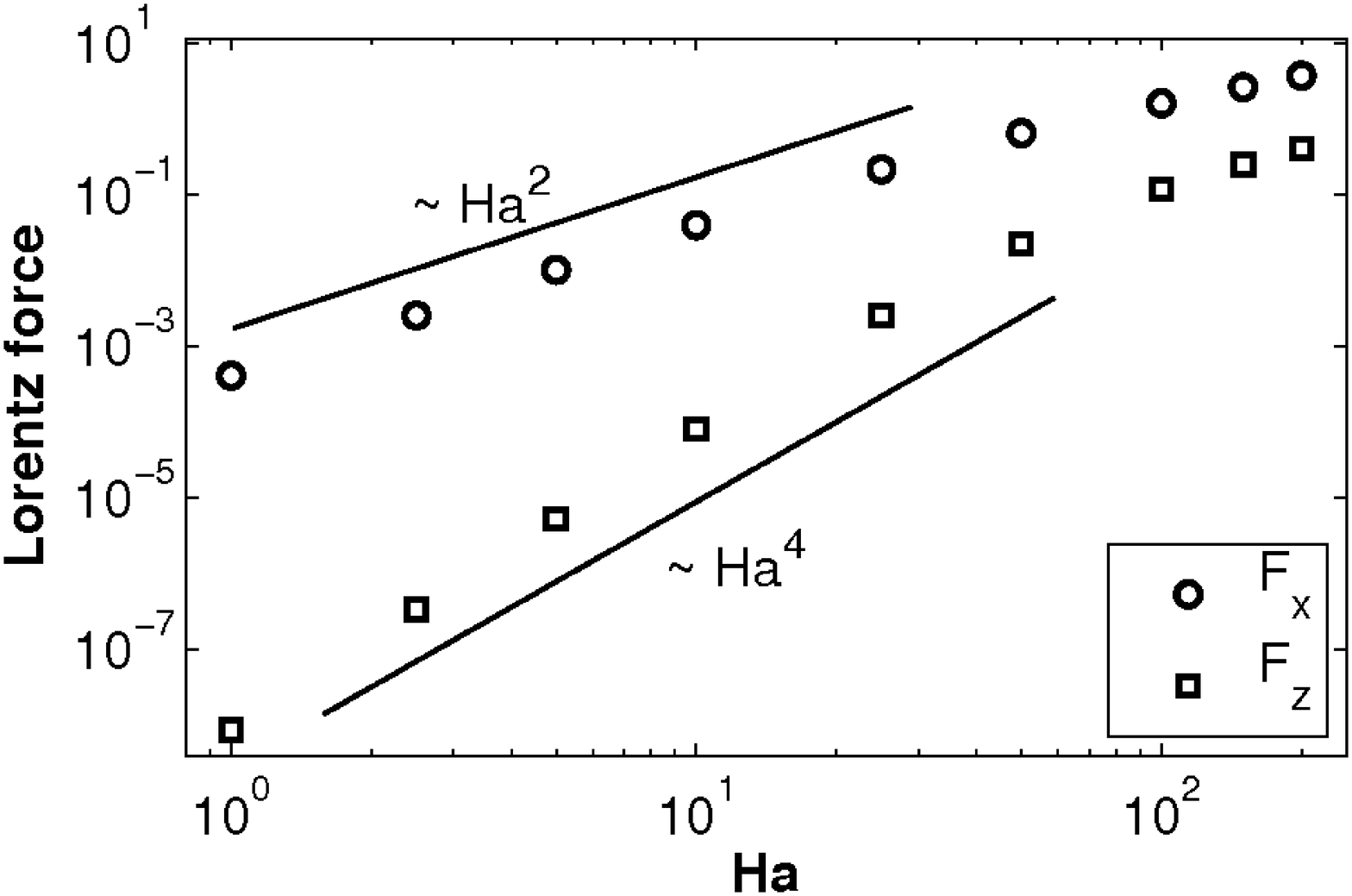}}
  \hfill %
  \subfloat[]{
\includegraphics[width = 0.49 \textwidth]{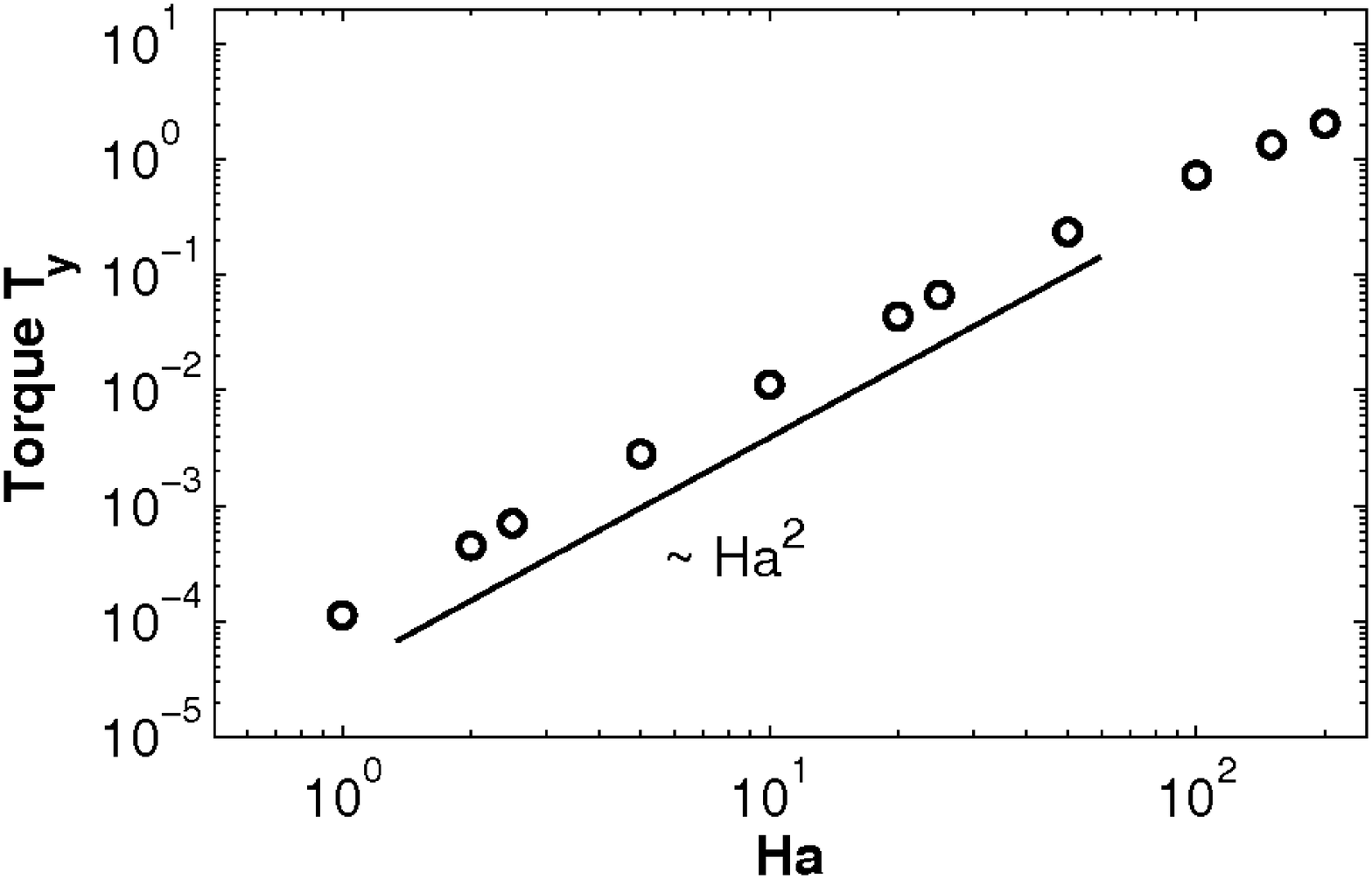}}
  \hfill %
  \subfloat[]{\includegraphics[width = 0.49 \textwidth]{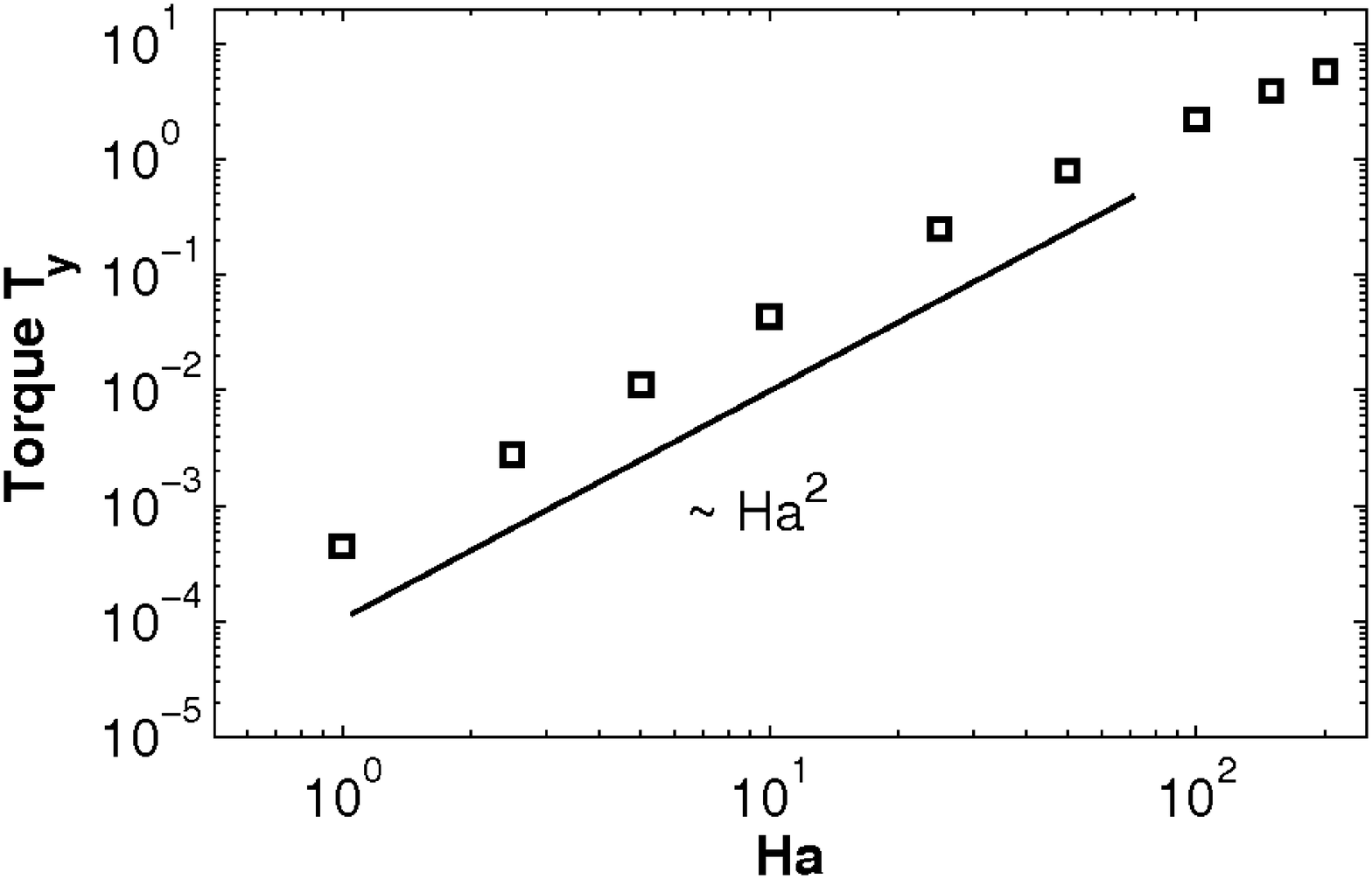}}
  \hfill %
  \caption{Drag and lift forces and resulting torque as a function of the Hartmann number for $Re = 10$ and $h=0.4$.  
(a,c) wall-normal orientation of the dipole. (b,d) streamwise orientation of the dipole.}
\label{fig:low_Re_FxFzTy_per_Ha}
\end{figure}

Let us now focus on the integral forces and torques. 
Therefore, we define the total Lorentz force by
\begin{equation}\label{eq:total_Lorentz_force}
\vec F= \int_V \vec f \,d V=\int_V  \frac{Ha^2}{Re} (-\nabla\phi + \vec u\times \vec B)\times \vec B\,d V\,, 
\end{equation}
and the total torque as
\begin{equation}\label{eq:total_torque}
\vec T=\int_V \vec r\times \vec f \,d V
\end{equation} 
with $ \vec r = \vec x - (h+1)\vec e_z\,$.
 Calculations are done for the three main dipole orientations. The dependence of 
the integral forces on Reynolds and Hartmann number was found to be practically the same for all orientations. More 
interestingly,  the absolute values of the forces were found to differ at the same values of Hartmann and Reynolds numbers. 
The streamwise oriented dipole gave always stronger forces than the wall-normal vertical one. It has to be recalled that the 
Hartmann number is based on $B_{max}$, the maximal value of the magnetic field inside the duct, and not on the magnetic 
moment of the dipole. Thus, in an experiment using a real magnet, it is always found  that a rotation of the magnetic dipole 
from vertical to streamwise orientation will cause a decrease of the forces simply because such a rotation will decrease the 
Hartmann number by a factor of two in correspondence with Eq. (\ref{eq:dipolefield_bmax}). In all cases, the spanwise oriented dipole will give the 
weakest force. The integral torque component behaves in the same way as the forces. The torque is zero for the spanwise 
dipole orientation due to the symmetry of the problem.

In Fig.~\ref{fig:low_Re_FxFzTy_per_Ha}, we compare the total drag force,  i.\,e., the streamwise force component,  the total lift force  i.\,e., the vertical force component
  (upper panels), and the total torque 
(lower panels) for the wall-normal and streamwise dipole. We observe that the drag force is higher than the lift force. The data in 
Fig.~\ref{fig:low_Re_FxFzTy_per_Ha} reveal three power laws which seem to be valid up to $Ha \approx 25$. For higher Hartmann 
number values the growth with $Ha$ becomes weaker.  As expected, one finds that the drag force $F_x\sim Ha^2$ and that the 
torque component $T_y\sim Ha^2$. These power laws are caused by the fact that $Ha^2/Re$ is a prefactor to the Lorentz force 
term in the momentum balance. The third power law for the lift force is found as  $F_z\sim Ha^4$. This behavior of the lift force is 
not immediately obvious. We explain it by the following argument. 

Let  $\vec u$ be a solution to the Navier Stokes equation~(\ref{eq:navier-stokes-with-force}) for a given Hartmann and Reynolds number. 
Similar to a perturbation expansion in weakly nonlinear flows, we divide the velocity field in three parts,  
\begin{equation}
\vec u = \vec u_0+\vec u_1+ \vec u_2\,.
\label{expansion}
\end{equation}
Here, $\vec u_0$ is the base flow, $\vec u_1$ describes the distortion of the base flow 
and $\vec u_2$ stands for a higher-order nonlinear term.  The base flow $\vec u_0$ is the flow in the limit of 
small interaction parameter $Ha^2/Re$ and  low Reynolds number. It is supposed to be a steady flow which is 
solely driven by the pressure gradient. Thus, $\vec u_0$ is the solution of 
\begin{equation}
0 =  -\nabla p+ \frac{1}{Re}  \nabla^2 \vec u_0.
\end{equation}
This solution is the laminar flow $\vec u_{lam}$ which  can be derived analytically~\citep{Pozrikidis97}. 
As it was shown in \cite{Heinicke2012}, the Lorentz forces for a given laminar profile and for the full Navier-Stokes 
equation~(\ref{eq:navier-stokes-with-force}) are almost the same in case of Hartmann and Reynolds 
numbers below certain thresholds. The present case of $Ha=25$ and $Re = 10$ satisfies these thresholds and the 
approximation with a  laminar profile $\vec u_{lam}$ gives a good agreement for the drag component of the Lorentz 
force. However, the lift force is zero for a laminar velocity profile. 

The second term $  \vec u_1$ describes the deflection due to the dipole.  It is again an approximation which holds 
for small Reynolds numbers. Again, this flow is steady.  Viscous forces are balanced  by the Lorentz force 
\begin{equation}
\label{eq:u_2}
\frac{1}{Re}  \nabla^2 \vec u_1 \approx- \frac{Ha^2}{Re} (-\nabla\phi_0 + \vec u_0\times \vec B)\times \vec B \,.
\end{equation}
It is clear from this equation that the amplitude $u_1 \sim \Ha^2$. We note that an additional pressure correction is necessary 
to maintain incompressibility. The contribution $\vec u_1$ does not give a contribution to the lift force. This 
can be shown with the following argumentation based on Stokes flow. Assume that there is a non-zero lift force $F_z$. 
On the one hand, the linearity of the equations implies that a sign reversal of pressure and $\vec u_0+ \vec u_1$, respectively, 
results in an opposite force. In particular, the lift force would reverse, i.\,e., one would obtain $-F_z$. On the other 
hand, the problem has mirror symmetry with respect to the $(y,z)$ plane containing the dipole, i.\,e., a reversal of flow 
direction should change the sign of the drag force $F_x$ but not the sign of the lift force. For this reason $F_z=-F_z$, i.\,e., 
the lift force has to vanish. 

The third part of the decomposition, $\vec u_2$,  is the nonlinear part which mainly reflects effects of inertia. This term is 
an approximation, but presents the realistic flow pattern for moderate Reynolds numbers.  Rewriting Eq. (\ref{expansion}) to 
$\vec u_2 = \vec u- \vec u_0- \vec u_1$ in the Navier-Stokes equation~(\ref{eq:navier-stokes-with-force})  
for the steady flow case and neglecting all terms of order $O(Ha^4)$ leads to 
\begin{equation}
\nabla^2 \vec u_2 = Re \left((\vec u_1\cdot\nabla) \vec u_0+ (\vec u_0\cdot\nabla) \vec u_1 \right)\,.
\label{expansion2}
\end{equation}
From this equation, one can estimate that $u_2/\ell^2 \approx Re\,u_1 u_0/\ell \approx Re\,Ha^2 \,u_0^2 \ell$. Thus, 
$u_2\approx Re\,Ha^2 \,u_0^2 \ell^3$ which gives rise to the lift force $F_z \approx (Ha^2/Re) u_2 $ and thus the 
expected Hartmann number dependence of $Ha^4$ for the lift force. We denoted in this estimate by $\ell$ the 
nondimensional characteristic gradient variation scale given in units of the duct half width. It was also numerically 
verified that the lift force vanishes when the nonlinear term $(\vec u\cdot\nabla) \vec u$ is artificially set to zero. In this 
case, the flow was symmetric in streamwise direction. Thus, the numerical test confirms that the nonlinear term is 
responsible for non-symmetric pattern in the flow and for the lift force. We also note that a more formal argument 
could be given using a double expansion in the two small parameters $Ha$ and $Re$. 

\subsection{Reynolds number dependence at fixed Hartmann number}
\begin{figure}
\centering
  \hfill %
  \subfloat[]{\includegraphics[width = 0.32 \textwidth]{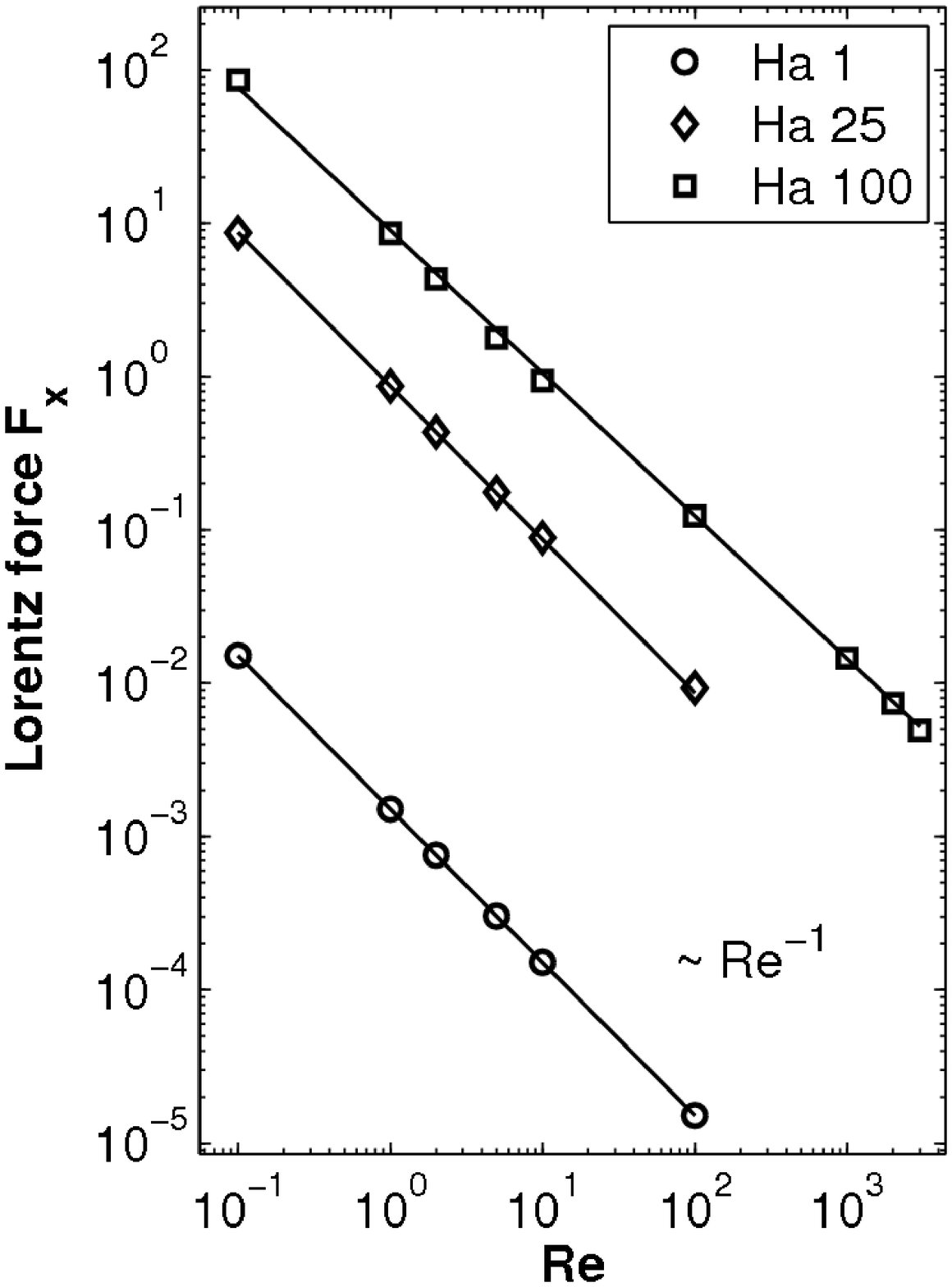}}
  \hfill %
  \subfloat[]{\includegraphics[width = 0.32 \textwidth]{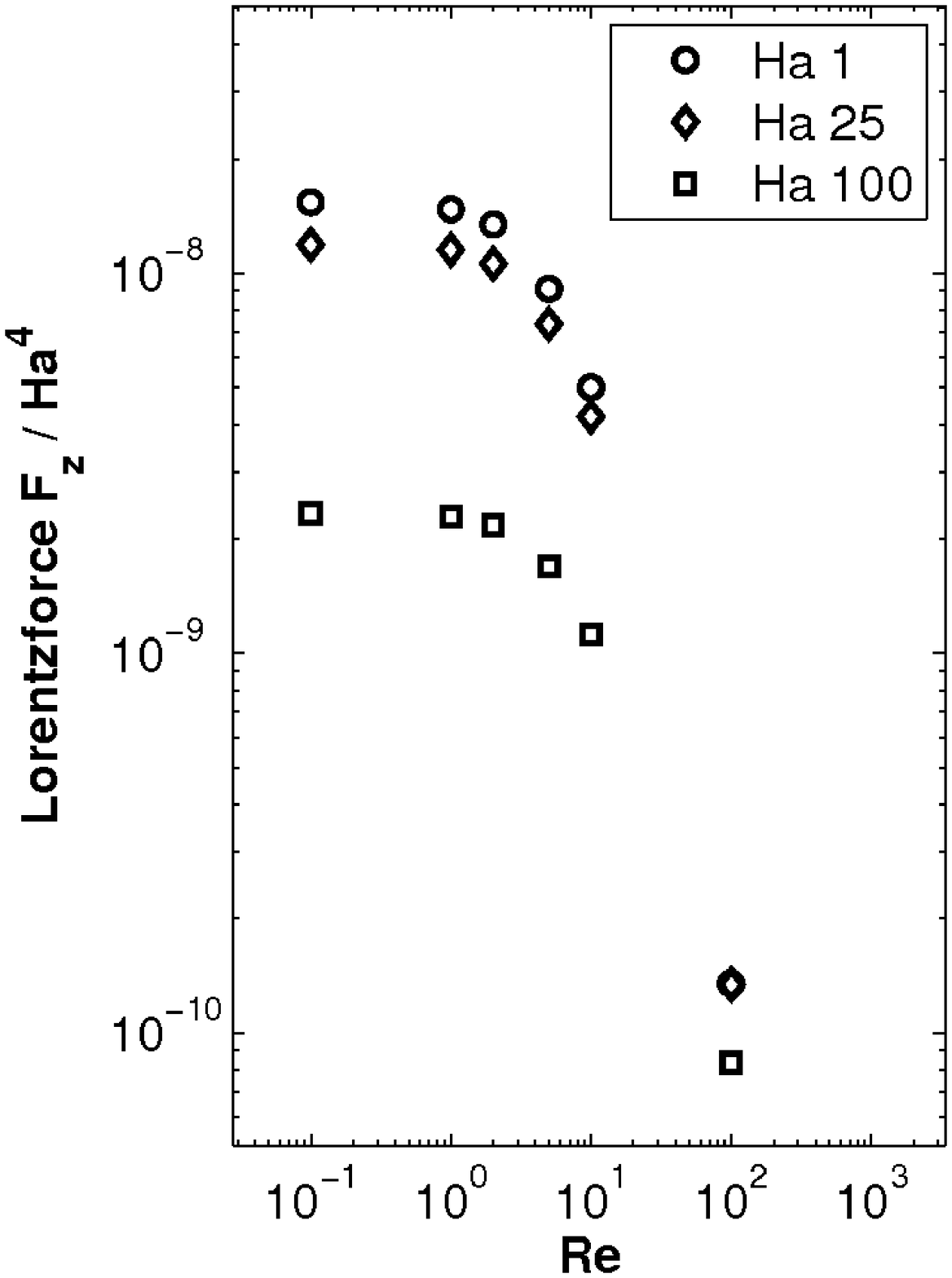}}
  \hfill %
  \subfloat[]{\includegraphics[width = 0.32 \textwidth]{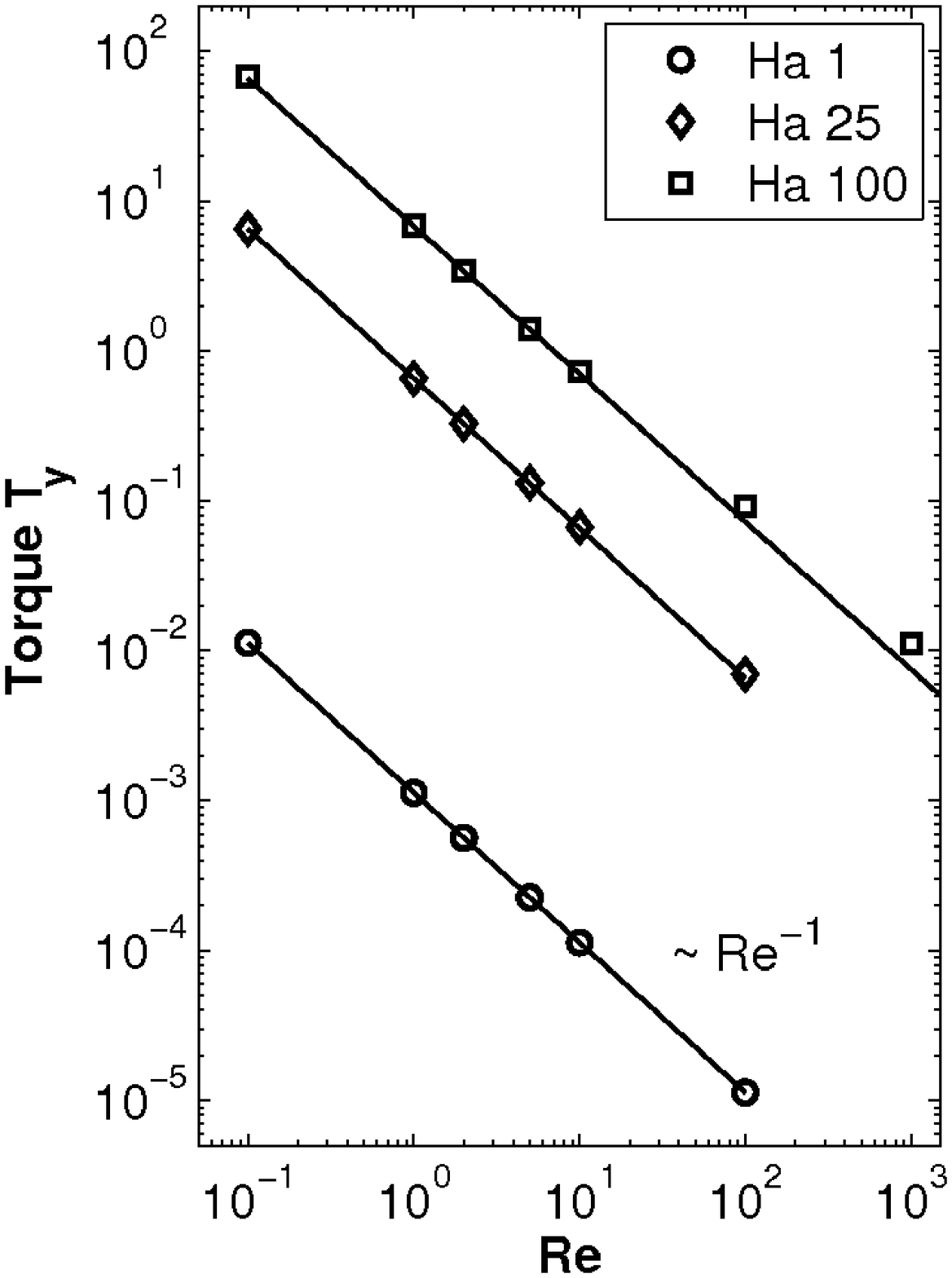}}
  \hfill %
\caption{Reynolds number dependence of the force components and the torque. (a)  drag force $F_x$;  (b) lift force $F_z$ compensated by $Ha^4$;
(c) torque $T_y$. Data are for a wall-normal magnetic point dipole at a distance of $h=0.4$. The dependence $1/Re$ is 
indicated by a solid line for the drag component of the Lorentz force and the torque. Data are obtained for several Hartmann 
numbers as indicated in the legend.}
\label{fig:low_Re_FxFzTy_per_Re}
\end{figure}

Besides the Hartmann number, the second control parameter, $Re$, will affect the structures of the flow.  Following 
directly from the definition of the Lorentz force, we verify the dependencies $F_x\sim 1/Re$ and $T_y \sim 1/Re$, respectively, 
which are shown in Fig.~\ref{fig:low_Re_FxFzTy_per_Re} for several Hartmann numbers. Furthermore, we 
see that the lift force $F_z$ is constant at a given Hartmann number for $Re < 1$ as already discussed 
above.  In case of higher Reynolds number, the lift force data decay steeper than $1/Re$. 

\begin{figure}
\centering
  \hfill %
  \subfloat[]{\includegraphics[width = 0.49 \textwidth]{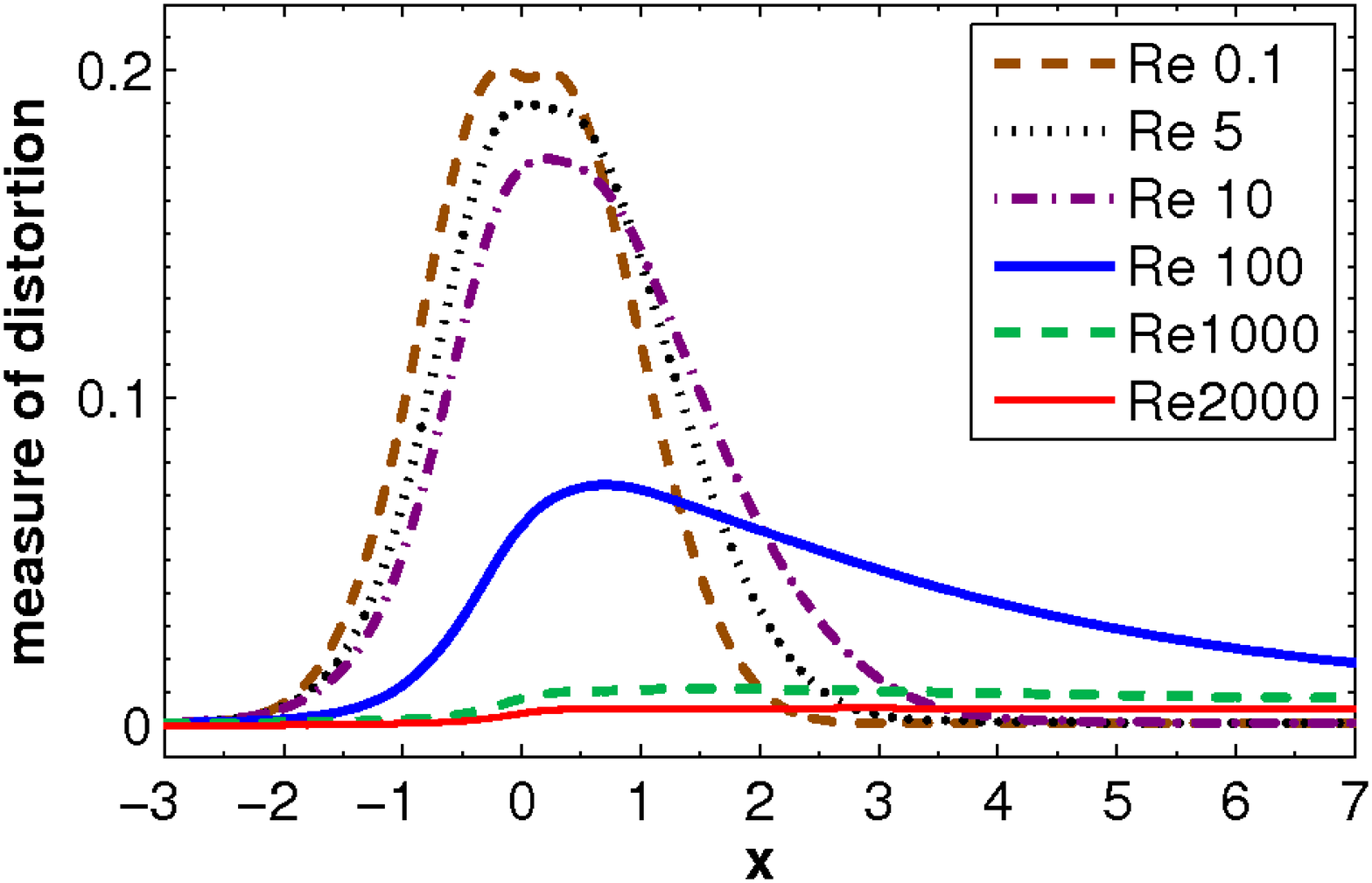}}
  \hfill %
  \subfloat[]{\includegraphics[width = 0.49 \textwidth]{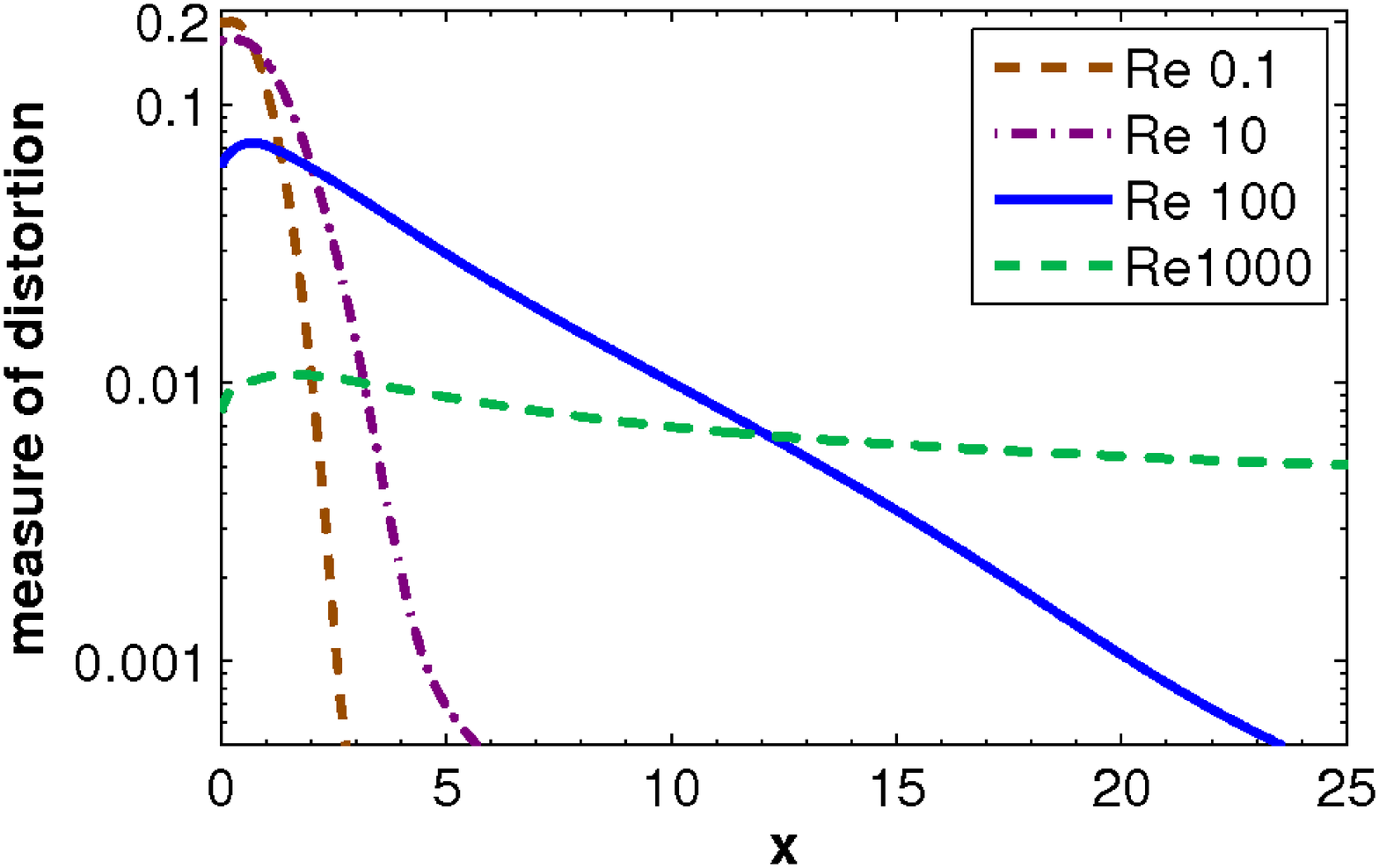}}
  \hfill %
  \subfloat[]{\includegraphics[width = 0.49 \textwidth]{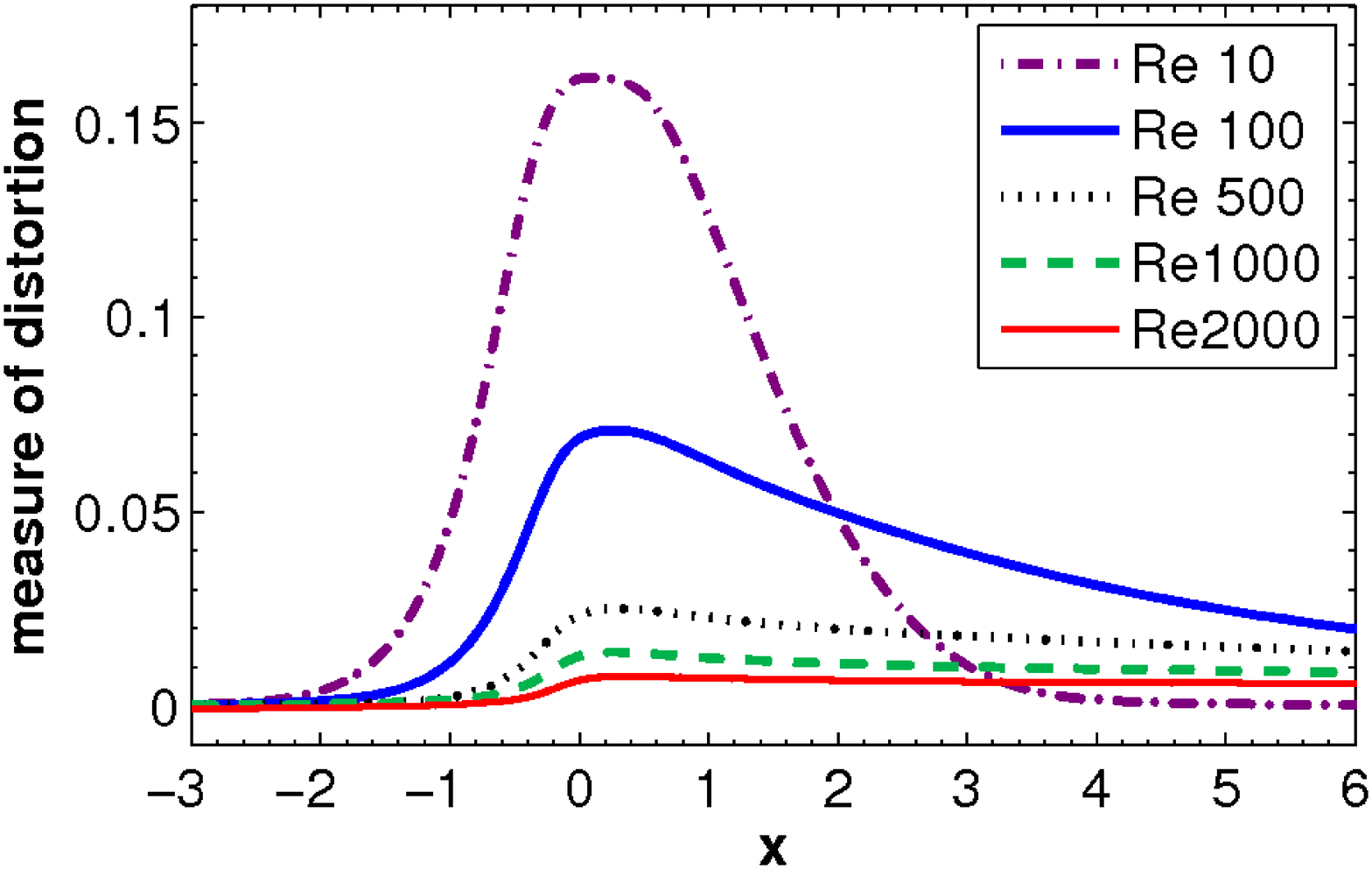}}
  \hfill %
  \subfloat[]{\includegraphics[width = 0.49 \textwidth]{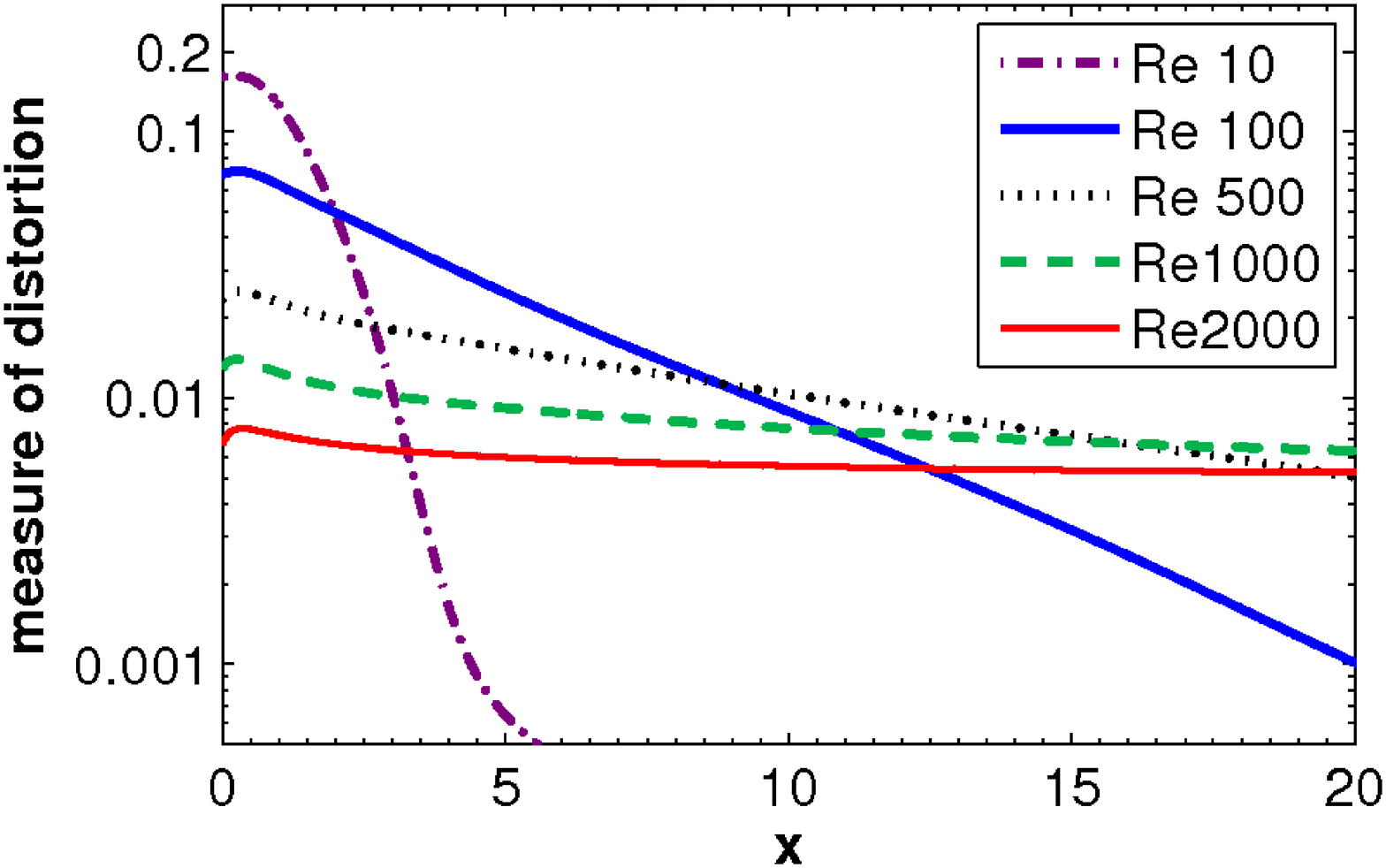}}
  \hfill %
\caption{(colour online)  Flow distortion as quantified by Eq.~(\ref{eq:distortion_meassure}) for several Reynolds numbers. In all 
cases, $Ha = 100$ and $h=0.4$. The dipole is located at $x=0$ with (a) wall-normal magnetic moment and (c) spanwise magnetic 
moment, respectively. Panels (b) and (d) replot the same data on logarithmic-linear axes.}
\label{fig:distortion_Ha100}
\end{figure}

With increasing Reynolds number the contributions of the nonlinear advection term will increase and manifest in an increasing 
distortion of the flow. We observe that the distortion in the wake, as measured by Eq. (\ref{eq:distortion_meassure}), decays  
almost exponentially, i.\,e., as a function $\exp(-\alpha x)$ with respect to the streamwise direction  obeying a spatial decay rate $\alpha$. This is 
demonstrated  in Fig.~\ref{fig:distortion_Ha100} for a Hartmann number of $Ha=100$ in a logarithmic-linear plot. The first observation is 
that the maximum of the distortion is decreasing with increasing Reynolds number -- an effect of the decaying interaction parameter $Ha^2/Re$. 
The decay of the maximal value may also be considered as an effect of the nonlinear term. A numerical test case shows that the 
value does not decrease if the nonlinear term is artificially switched off.  When the data are presented in a logarithmic scale, as in the 
lower right panel of Fig.~\ref{fig:distortion_Ha100}, the exponential decay is clearly visible. The fit of these data with an exponential 
function results in a spatial decay rate which is approximately inversely proportional to the Reynolds number.

The Reynolds number dependence of the decay rate can be rationalized from the following consideration for steady
flow. The Lorentz force term $(Ha^2/Re)(\vec j\times\vec B)$ is a localized force term and will not affect the decay in the 
wake sufficiently far downstream from the dipole position, i.\,e., we can drop it in the wake. The  Navier-Stokes equations 
simplify to 
\begin{equation}
(\vec u\cdot \nabla)  \vec u  = \frac{1}{\Re} \Delta \vec u - \nabla p.
\end{equation}
We now represent the distortion as $\vec v= \vec u - \vec u_{lam}$. If we use this representation, then the dominant term 
on the left hand side becomes $u_{lam}\partial_x \vec v$. On subtracting the equation for the laminar flow itself we obtain
\begin{equation}
u_{lam}\partial_x \vec v  = \frac{1}{\Re} \Delta \vec v - \nabla p^\prime,
\end{equation}
where the pressure contribution $p^\prime$ serves to maintain incompressibility. If we further approximate
the laminar velocity distribution by its mean value $\bar u$ we eventually have 
\begin{equation}
\bar u\partial_x \vec v  = \frac{1}{\Re} \Delta \vec v - \nabla p^\prime.
\end{equation}
Following  usual  boundary-layer approximation ideas based on $Re \gg 1$ one can further neglect the second 
derivative with respect to $x$ in the Laplacian. The problem then effectively reduces to a diffusion problem 
\begin{equation}
\del t \vec v  = \frac{1}{Re}\left(\frac{\partial^2 }{\partial y^2} +\frac{\partial^2 }{\partial z^2}\right) \vec v
\label{diffusive}
\end{equation}
where we have introduced $t= x/\bar u$. Using separation of variables, it is clear that
the decay of $|\vec v|$ with $t$ is  ultimately determined by the largest eigenvalue $-\alpha_1$ of
the two-dimensional Laplacian, i.\,e., 
\begin{equation}
|\vec v| \sim \exp (- \alpha_1  t/Re)=\exp(- \alpha_1 x/Re \bar u).
\end{equation}
Therefore, the measure of distortion has to decay approximately exponentially.
The last steps of our argumentation are 
more heuristic because we neglect the remaining pressure term. A justification of this step might be not straightforward, 
but we suppose that it is justified because it amounts to a projection on the space of solenoidal functions, which
should not interfere with the gist of the argument. We further note that the spatial decay is independent of the orientation 
of the magnetic dipole. The numerical simulations show that this is approximately the case for $Ha = 100$ and $Re$ reaching from 
10 up to 2000. 
We also remark that 
the decay is clearly visible only at large distances from the dipole position. 
For $Re\gtrsim 500$, it is difficult to identify 
 a clear   exponential decay of  $\mathcal A(x)$ (cf. Eq. \ref{eq:distortion_meassure}) within the computational domain because 
 of the slower decay of higher modes.

To summarize the results for the lower Reynolds numbers: we explained why different transformations of the flow are observed for different dipole 
orientations leading to the formation of local Hartmann layers and areas of reversed flow due to strong Lorentz forces. The strength of the forces 
and their effect on the deflection of the flow in dependence on the Hartmann number was also analyzed. The total  drag force is found to be proportional 
to $Ha^2$ and the total lift force is proportional to $Ha^4$ in the present Reynolds number regime. When the Reynolds number is increased, we 
observe that the length of the wake is increased. It was also shown why the spatial downstream decay of the deflected flow in the wake is proportional to $1/Re$.  
In the next section, we will see how the vortex formation process changes when the dipole triggers a transition to turbulence as the Reynolds
number is increased.


\section{Time-dependent flow at higher Reynolds numbers}\label{sec:wake_instabilities}

In this section, we study time-dependent flow structures which start to appear at Reynolds numbers of about 2000 
and higher and for Hartmann numbers above 80. We will observe vortex shedding for cases when the dipole is 
positioned sufficiently far in a distance of $h=1.6$ from the top surface of the liquid.  For smaller distances than 
$h=1.0$ the flow is always stationary in the range of Reynolds numbers which could be covered here $(Re\le 3000)$. 
Therefore, we will restrict our study in this section to the case with $h= 1.6$ which generates qualitatively new features
compared to the last section. 
\begin{figure}
  \hfill %
  \subfloat[]{\includegraphics[width = 0.32 \textwidth]{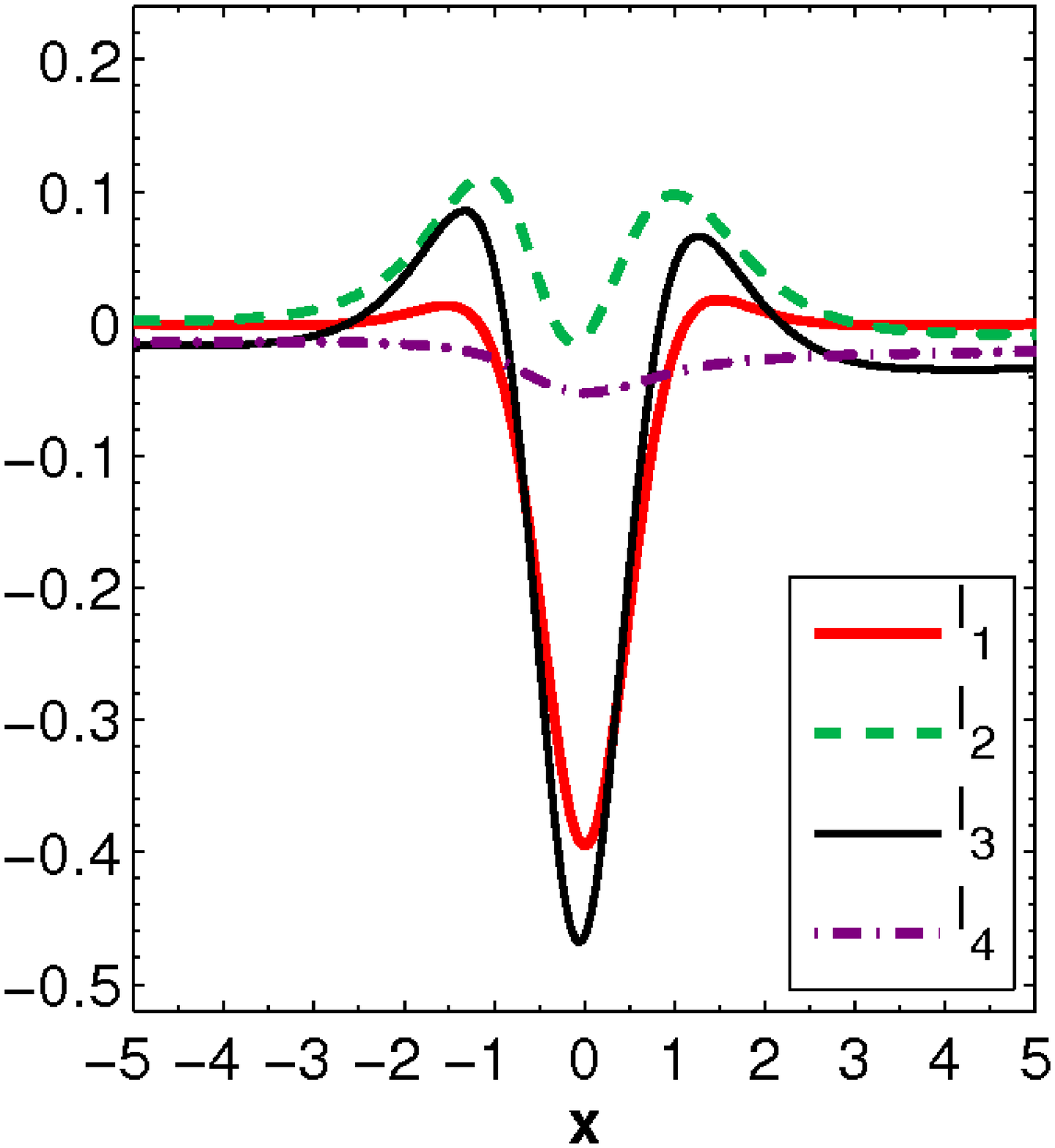}}
  \hfill %
  \subfloat[]{\includegraphics[width = 0.32 \textwidth]{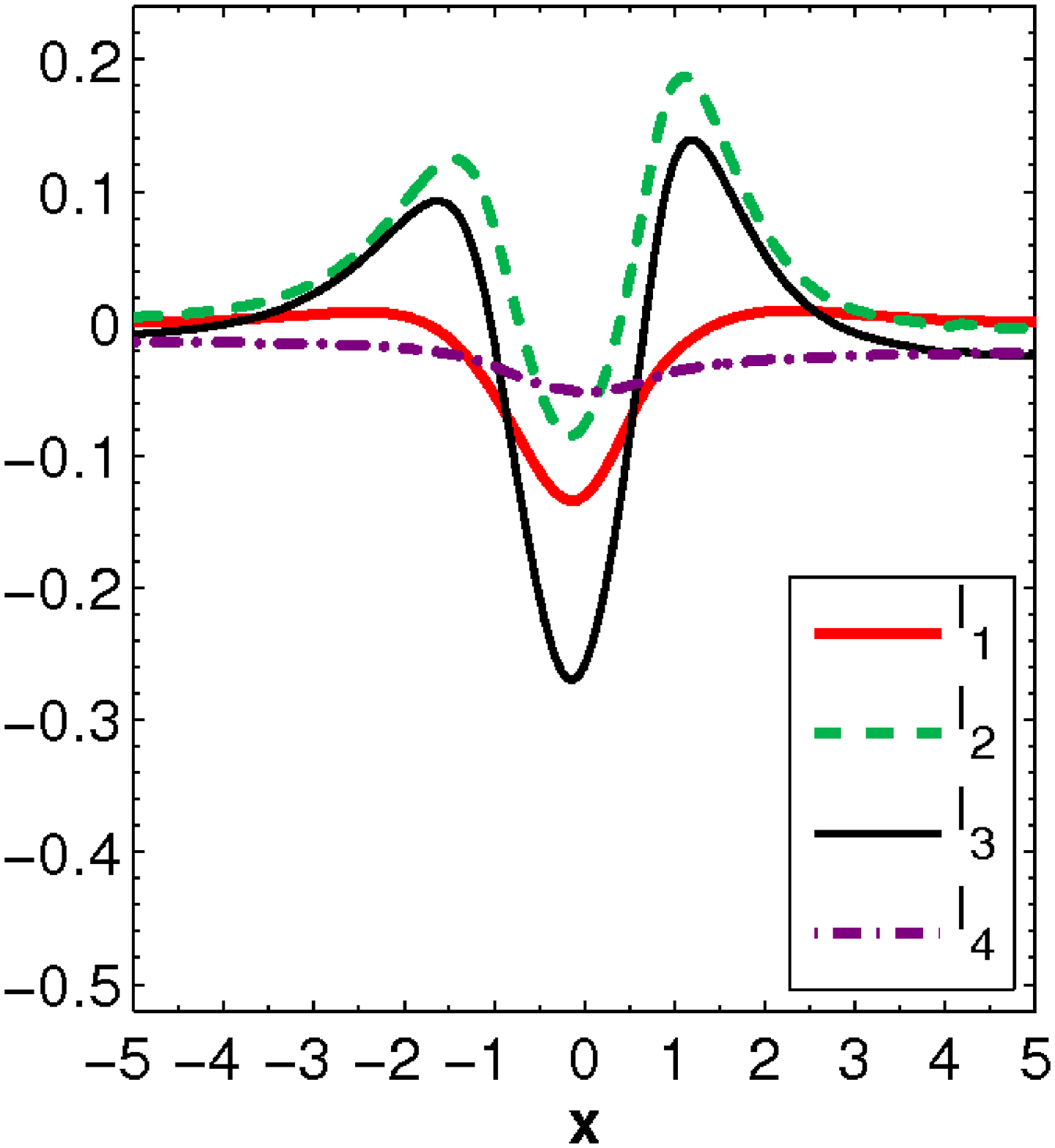}}
  \hfill %
  \subfloat[]{\includegraphics[width = 0.32 \textwidth]{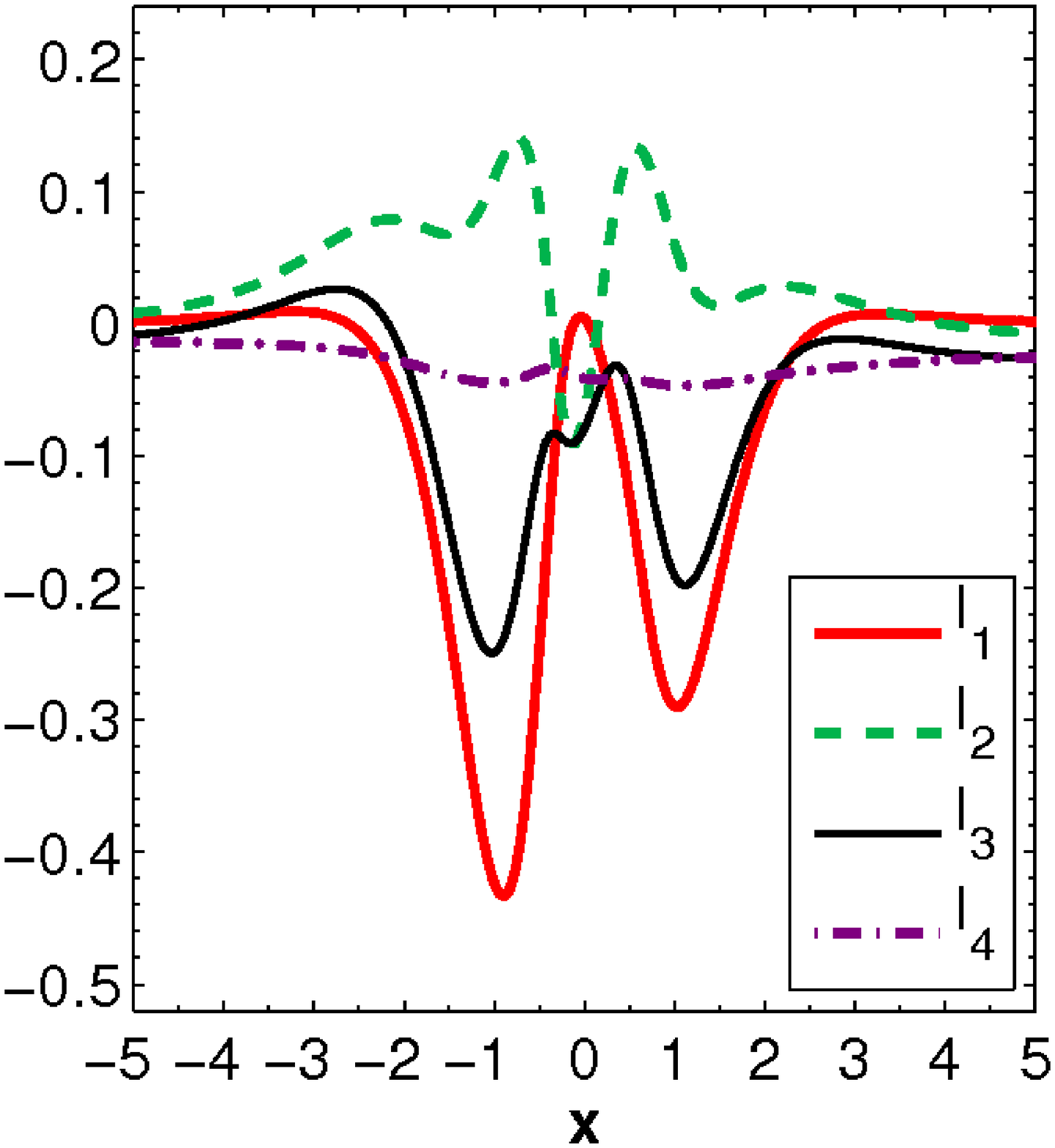}}
  \hfill %
\caption{(colour online) Streamwise variation of the different terms of the  momentum 
balance (\ref{eq:integrals_bilanz}) averaged over the cross section at $Re=2000$, $Ha=100$, $h=1.6$ for  three different 
orientations of the dipole: (a) wall-normal vertical, (b) spanwise, (c) streamwise. Equations (\ref{I1}) to (\ref{I4}) have been 
used again. Since the spanwise case in the mid panel is time dependent,  data are time-averaged over one oscillation time of 3.168 time units.}
\label{fig:momentum_balance_high_Re}
\end{figure}
\subsection{Deformation of the flow}
The analysis of the different contributions to the momentum balance (\ref{eq:integrals_bilanz}) is shown in 
Fig.~\ref{fig:momentum_balance_high_Re} for $Re=2000$ and $Ha=100$ for the three dipole orientations: 
streamwise, spanwise and wall-normal vertical. The deformations of the flow field show effects which are 
similar to the low Reynolds number case from Sec.~\ref{sec:steady_state_solutions}. Since the Reynolds 
number is higher, the nonlinear term has a stronger influence on the deflection of the flow as visible by its 
larger magnitude in the figures. The terms of the momentum balance  (\ref{eq:integrals_bilanz}) reveal that the Lorentz force obeys a 
qualitatively similar behavior as in the low Reynolds number cases (cf. also Fig.~\ref{fig:momentum_balance_low_Re}). It is 
again the pressure gradient that balances the additional contribution  that is now produced by the  change of streamwise momentum flux $I_2$.

\begin{figure}
\begin{center}
\includegraphics[width = \textwidth]{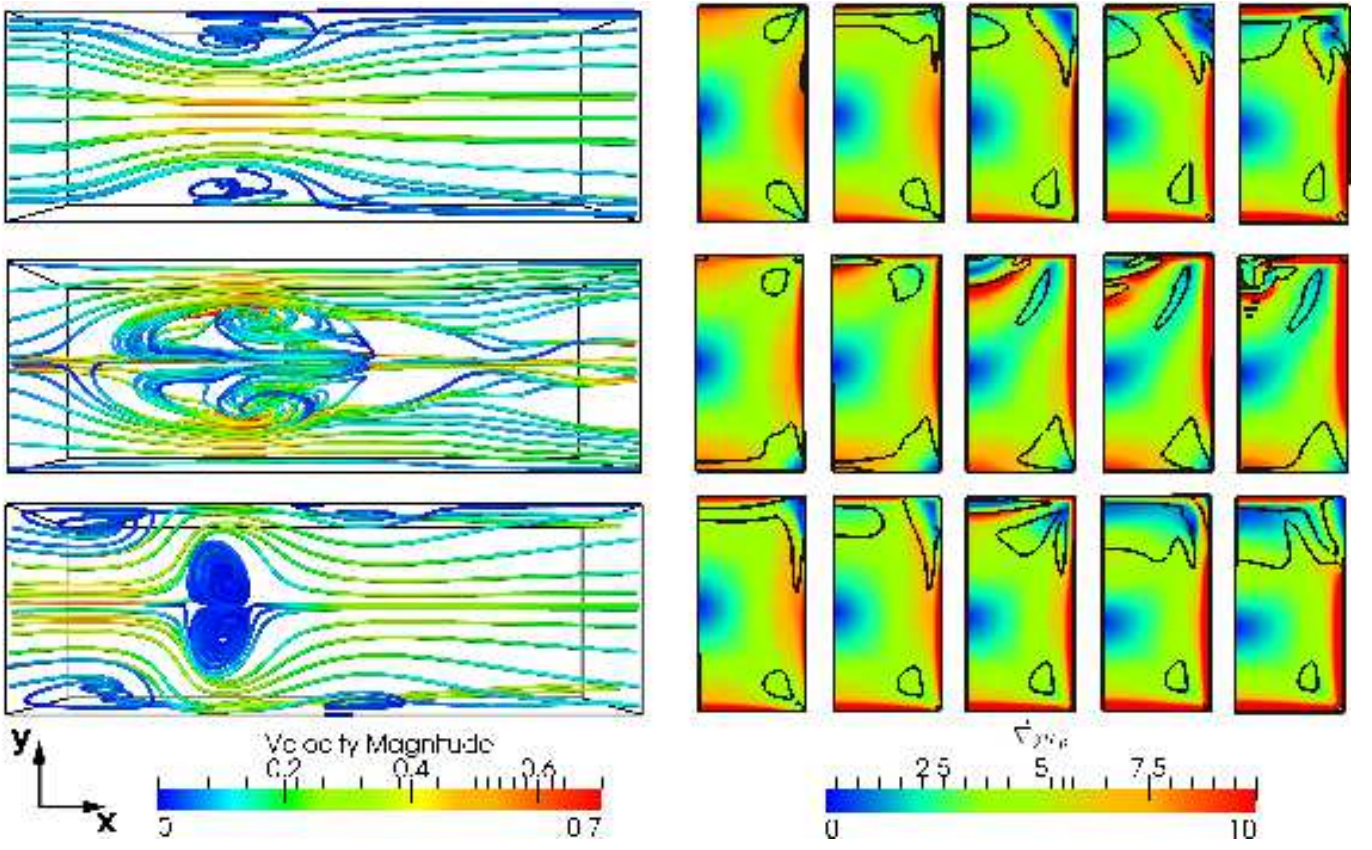}
\caption{ Streamlines (left column) and inflection lines (right column) at $Re=2000$, $Ha=100$ and $h=1.6$ for different 
orientations. Top row: wall-normal vertical dipole. Mid row:  spanwise oriented dipole. Bottom row:  streamwise oriented dipole. 
Streamlines are shown along top  wall from $x=-2$ till $x=5$. Color indicates the velocity magnitude. The panels to the right show the 
corresponding cross sections at streamwise positions $x = -2,-1,0,1$ and $2$. Due to symmetry only half plane is displayed. 
The background colors display the magnitude of the cross-stream gradient $\vec\nabla_{2} u_x$. Black solid lines are the inflection 
lines as defined in Eq. (\ref{inflection}).}
\label{fig:streamlines_und_dudn_verschiedene_orientierungen}
\end{center}
\end{figure}

The velocity field streamlines for the three cases are shown in Fig.~\ref{fig:streamlines_und_dudn_verschiedene_orientierungen}.
We observe again the formation of local Hartmann layers and vortices as described for the low Reynolds number runs in 
Sec.~\ref{sec:steady_state_solutions}. In case of the spanwise oriented dipole, a strong vortex in the center of the duct is formed 
which becomes now time-dependent, i.\,e., a vortex shedding is initiated. For streamwise and wall-normal oriented dipoles the flow 
in the wake remains stationary due to the local Hartmann layers that stabilize the flow in the range of Reynolds numbers accessible here.

Stability investigations for the onset of time-dependent flow  become less straightforward as the base flow depends on the Hartmann number. Thus, a simplified criterion 
for two-dimensional flow is applied here.  A possible explanation for the instability and transition to a time-dependent 
flow is shown in Fig.~\ref{fig:streamlines_und_dudn_verschiedene_orientierungen}. In the right column of the figure, we show 
five cross sections that contain the magnitude of the  cross-stream gradient of the streamwise velocity which is determined by 
$\vec\nabla_{2} u_x$ where $\vec\nabla_2$ denotes the gradient with respect to $y$ and $z$ directions. Furthermore, the inflection 
point criterion \citep{Uhlmann2006} follows with the definition
\begin{equation}
\vec n=\frac{\vec\nabla_{2}u_x}{|\vec\nabla_{2}u_x|}
\label{inflection0}
\end{equation}
to
\begin{equation}
\vec\nabla^2_{\vec n}u_x=0\,.
\label{inflection}
\end{equation}
Criterion (\ref{inflection})  generalizes the inflection points in one-dimensional velocity profiles. It has been used frequently in order to determine 
whether a flow can become unstable or not. Contour lines which satisfy (\ref{inflection}) are added to each of the five cross section plots in 
Fig.~\ref{fig:streamlines_und_dudn_verschiedene_orientierungen}. We observe that in case of a spanwise oriented dipole the magnitude 
of the gradient close to the inflection line is higher compared to the other cases. Thus the probability of  instability increases. As a 
consequence, we will restrict our observations and parameter studies in this section to the spanwise oriented dipole, the most 
interesting case for transition to turbulence. In the subsequent paragraph,  we will be  concerned with the time-dependent behavior 
of the generated  vortex structures, i.\,e., the vortex shedding, and the resulting structures in the wake.

\begin{figure}
\includegraphics[width=1.0\textwidth]{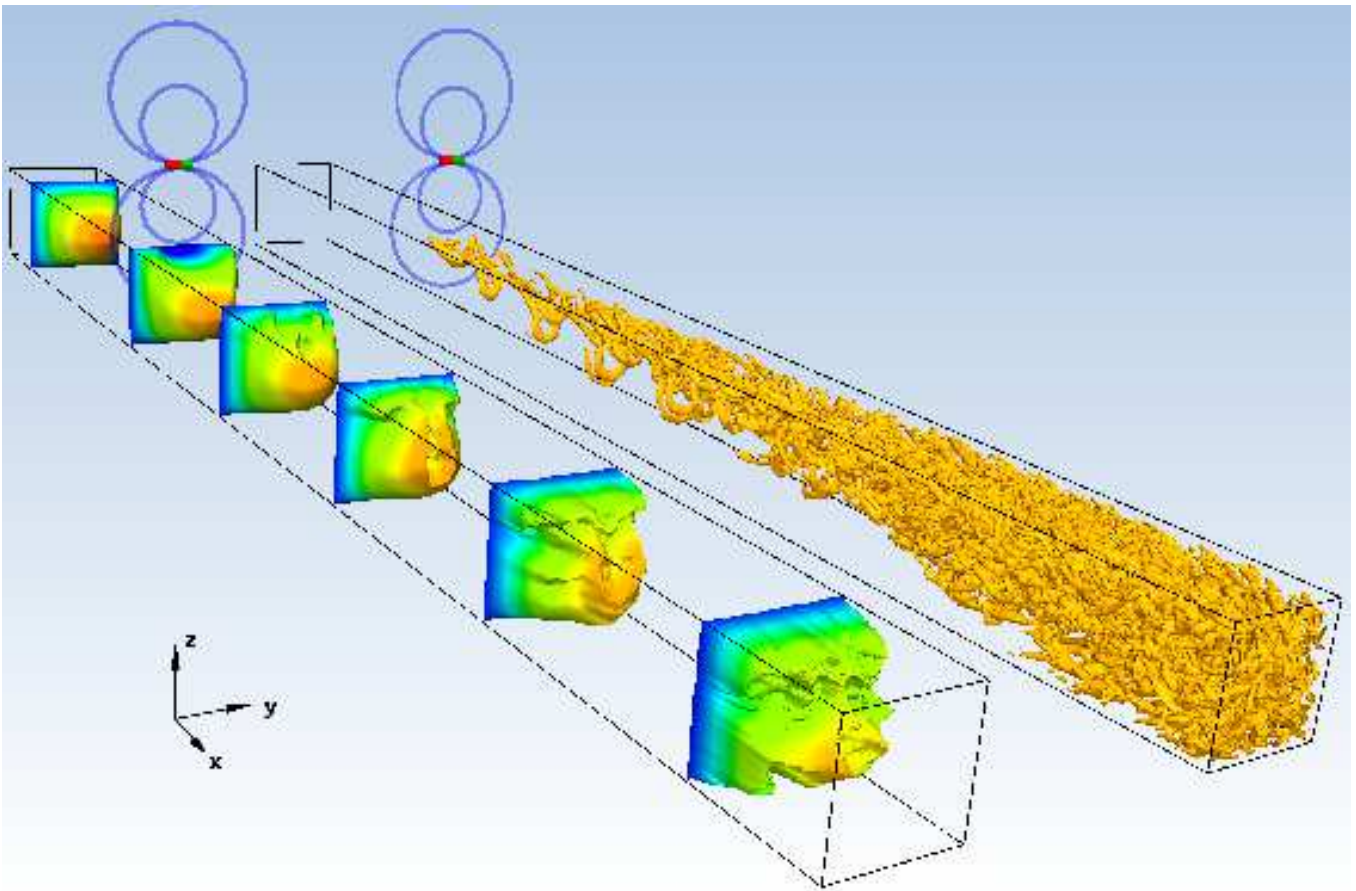}
\caption{Vortex shedding for a spanwise oriented dipole. Both ducts show the same snapshot for a 
Reynolds number $Re=2000$,  a Hartmann number $Ha=100$ and a distance $h=1.6$.
The left duct displays instantaneous velocity profiles taken at 6 different streamwise locations between $x=-10$ and 40. The right
figure shows the isocontours of  $\lambda_2 =-1$.}
\label{fig:vel_und_lambda}
\end{figure}

The vortex shedding generates a turbulent wake which is displayed in Fig.~\ref{fig:vel_und_lambda}. In parallel
to several cross sections of the streamwise velocity  along the duct, isocontours of $\lambda_2=-1$ are shown where
$\lambda_2$ is the second largest eigenvalue ($\lambda_1\ge \lambda_2\ge \lambda_3$) of the symmetric matrix 
\begin{equation}
\label{Lambda} \Lambda_{ij}=S_{ik}S_{kj}+\Omega_{ik}\Omega_{kj} \,
\end{equation} 
which is composed of the rate of strain and vorticity tensors, respectively, 
\begin{equation}
S_{ij}=\frac{1}{2}\left(\frac{\partial u_i}{\partial x_j}+\frac{\partial u_j}{\partial x_i}\right)\;\;\;\;\mbox{and}\;\;\;\;
\Omega_{ij}=\frac{1}{2}\left(\frac{\partial u_i}{\partial x_j}-\frac{\partial u_j}{\partial x_i}\right) \,.
\end{equation} 
Negative values, $\lambda_2<0$, denote vortex cores \citep{Jeong1995} which can be clearly 
identified as so-called hairpin structures in the figure. They are found close to the top wall in the beginning of the wake 
before the whole duct is filled with turbulent flow patterns further downstream.

Alternatively, the vortex shedding can be observed by the time signal of the Lorentz force itself. Both, the drag force 
component $ F_x$ and the lift force component $ F_z$, show a periodic sinusoidal time dependence. The temporal 
modulation of the signal is weak compared to the absolute magnitude (cf. Fig.~\ref{fig:force_per_t}  below). 
For example, it is  $F_x = -1.64\cdot 10^{-1}\pm 4.67\cdot 10^{-4}$ and $F_z = 1.74\cdot 10^{-2}\pm 6.75 \cdot 10^{-4}$ 
for $Re=2000$ and  $Ha=100$. Both forces oscillate with a frequency of 
$3.15\cdot 10^{-1}$ inverse time units.

\subsection{Reynolds number  dependence at fixed Hartmann number}

\begin{figure}
\centering
 \includegraphics[width=1.0\textwidth]{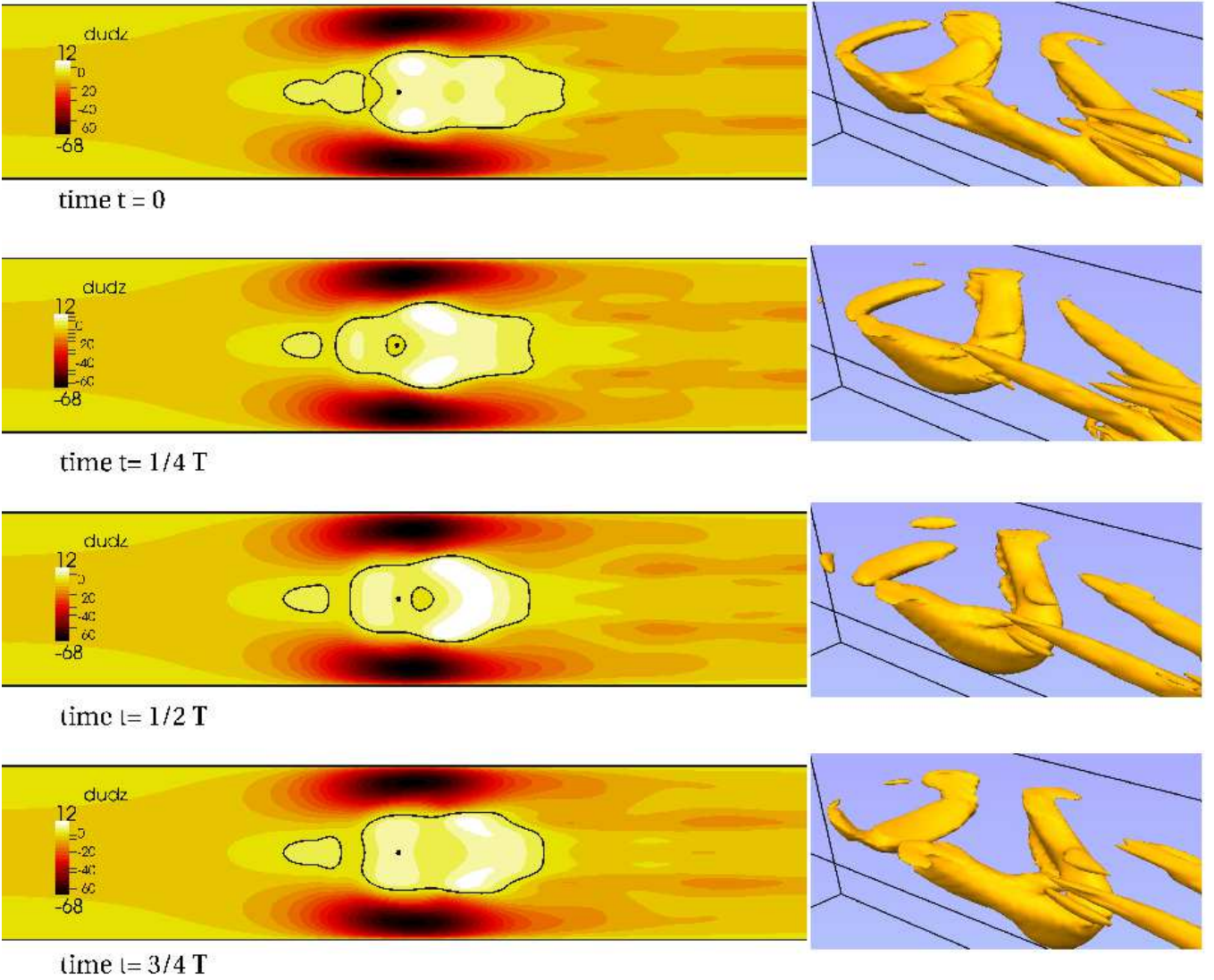}
\caption{ (colour online)  
Visualization of the time-dependent structures for $Re=2000$, $\Ha=100$, $h=1.6$ and spanwise  
magnetic moment. Four snapshots are shown every 0.8 convective time units which cover one 
oscillation time of 3.168 time units. Left: Top surface with velocity gradient $\del z u$. Position of 
the dipole is indicated with a black dot.  The black solid line indicates the line of separation $\del z u = 0$.
Right: isocontours of  $\lambda_2 =-0.5$ at same time.  Direction of view in the picture is the same as 
in Fig.~\ref{fig:streamlines_Re}~b.}
\label{fig:seperationline_Re2000_time}
\end{figure}

The time-dependent structures are shown in Fig.~\ref{fig:seperationline_Re2000_time} for $Re = 2000$ in four 
snapshots in  time intervals of approximately a quarter of the oscillation time. 
The  figure displays the top view of the duct with the position of the dipole marked as a small black dot. 
The line of separation marks the region of locally reversed flow, i.\,e.,  points with $\del z u = 0$. 
Along this line the flow detaches from the surface. This criterion was also used by Mistrangelo (2011)~\cite{Mistrangelo2011} 
to determine the size of the vortex that occurs in a duct with sudden expansion in dependence on the applied 
homogeneous magnetic field. There, the magnetic field  damps the vortex,  in contrast to our investigation 
with the magnetic field being the  source of the vortex formation.
 
In addition, Fig.~\ref{fig:seperationline_Re2000_time} displays   the structures in the wake. Here, 
 the presented top surface is colored with $\del z u $. This gradient reaches its maximal value in the Hartmann layers. 
It should be noted here that the Hartmann layers and the vortex are not completely independent of each other. 
However, the mechanism that describes the interaction of both -- and possibly drives the vortex shedding -- has still 
to be determined. 

To characterize the three-dimensional structure of the vortices, the $\lambda_2$-criterion is used. 
The pictures on the right-hand side of Fig.~\ref{fig:seperationline_Re2000_time} show the isocontours of 
$\lambda_2=-0.5$. Each period a hairpin vortex structure is produced. It starts at time $t=0$ with a small vortex 
that rotates clockwise about  $y$, i.\,e., $(\nabla \times \vec u )_y>0$. This vortex is then advected 
further and surrounded by reversed flow until it passes the  position of the dipole at $x=0$ at time $t=1/4\, T$, 
$T$ being the oscillation time. The vortex causes an increase of the width of the area of reversed flow as the 
back flow has to circumvent the vortex. In parallel, close to the corners of the duct, the Hartmann layer accelerates the 
flow. The accelerated fluid is blocked by the increase of the width of the reversed flow. This creates two extensions 
to the vortex at the sides, that are placed at $x=0$ at time $t=1/2\, T$.  The  vortex is then pushed into the bulk away 
from the top surface (time $t=3/4 \,T$). In this way, the backflow can move freely and recover the high magnitude 
at the beginning of the area of reversed flow. A next vortex is formed at time $t= T$ and the process repeats. 
 A simple interpretation of this dynamics might therefore  be that  vortices are produced periodically by  roll-up of a 
shear layer (Kelvin-Helmholtz instability) at the edge of 
the roughly  hemispherical zone shielded by the dipole field from the main flow.

\begin{figure}
\centering
 \includegraphics[width=1.0\textwidth]{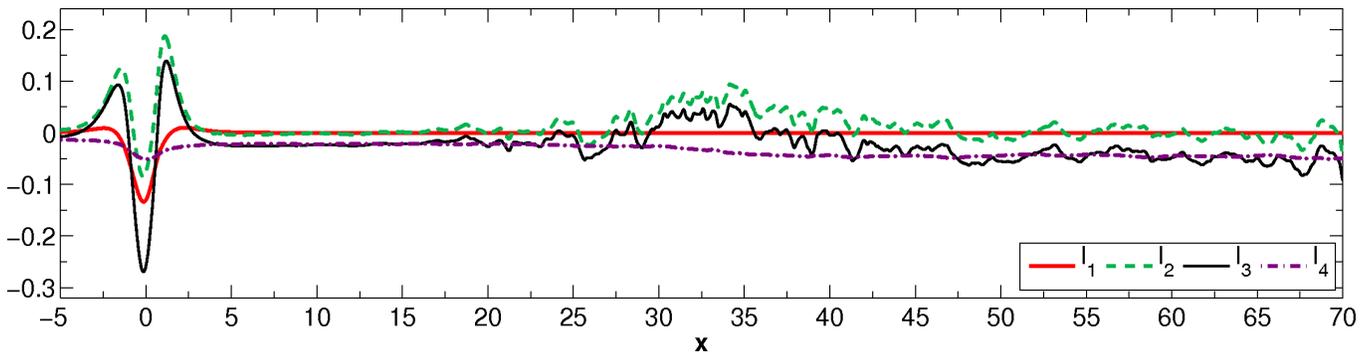}
\caption{(colour online)  Streamwise variation of the different terms of the cross section averaged momentum 
balance (\ref{eq:integrals_bilanz})  at $Re=2000$, $Ha=100$, $h=1.6$ for  spanwise orientation of the 
dipole. The data are time-averaged over one period as in Fig.~\ref{fig:momentum_balance_high_Re}~b. The four
terms $I_1$ to $I_4$ are given again by Eqns. (\ref{I1}) to (\ref{I4})}.
\label{fig:wake_bild}
\end{figure}
\begin{figure}

 \includegraphics[width = \textwidth]{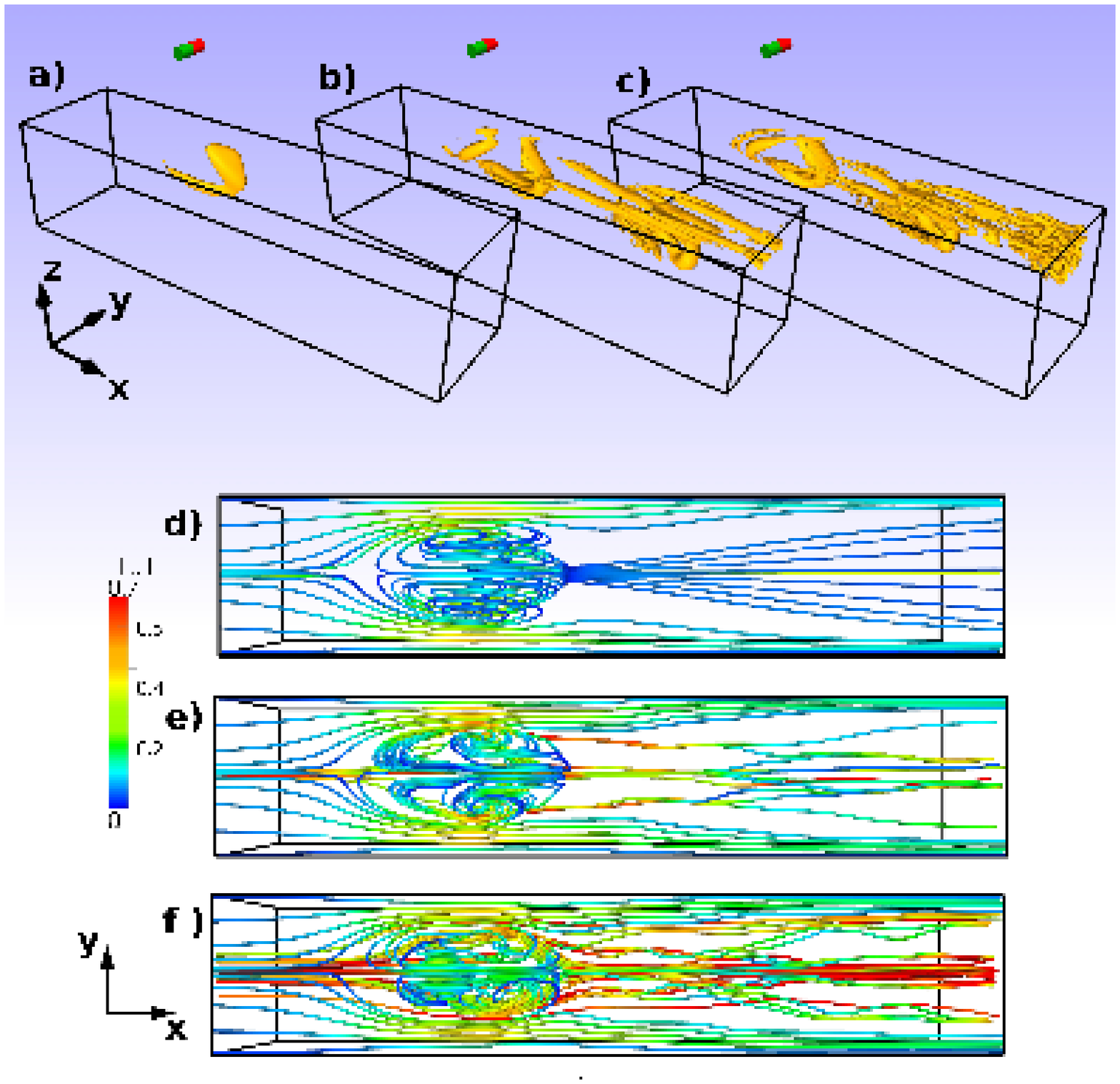}
\caption{(colour online) 
Influence of a spanwise oriented dipole on the flow for  $Ha=100$ and $h=1.6$ at different Reynolds numbers:
(a,d) is for $Re=1000$, (b,e) for $Re =2000$ and (c,f) for $Re=3000$. 
  Snapshots of  isocontours of $\lambda_2=-0.5$ in the range from $x=-3$ to $x=5$ (a,b,c)  
and snapshots of  streamlines for $x=-3$ to $x=7$ (d,e,f) 
}
\label{fig:streamlines_Re}
\end{figure}

The vortices are transported into the wake and are stretched due to inertia and the shear at the top wall. 
Thereby, their shape forms into a hairpins like structure \citep{Davidsonturbulence}.  After the hairpin is 
moved into the wake a new vortex is created by the Lorentz force. Hairpins are well known to appear in 
vortical shear layers at walls \citep{Jeong1997} and occur in  the transition to turbulence in the boundary layer.  
The spacial  development of the wake is determined with the help of the momentum balance 
(\ref{eq:integrals_bilanz}) in Fig.~\ref{fig:wake_bild}. This figure displays the terms  time averaged over one period $T$, i.\,e., it 
shows the mean of 3168 snapshots at intervals of 0.001 time units, for $Re=2000$ and $Ha=100$. 
The well pronounced area of distortion with high Lorentz forces in the interval $-5<x<5$ is followed by the 
region of vortex shedding $5<x<20$. In this region, the flow is periodic in time with the same frequency 
as in the area of distortion. Here, the time averaged momentum balance resembles the laminar flow as the 
pressure balances the wall stresses and the contribution of the Lorentz force as well as of the momentum flux  term  
are negligibly small. For $x > 25$ this changes to a transitional flow, where the momentum flux  term  dominates and 
the wall stresses  increase up to a higher constant value as known from turbulent flow. In the beginning of the 
transitional range, the velocity profiles are found to resemble a turbulent flow but possess a symmetry plane $y=0$. 
In this region, large scale 
vortices fill the whole cross section that break up further downstream. 
Our time averaging over one  period does not converge in this
range because the time-dependence  of the  structures is non-periodic.
Although the computational domain and the 
calculation time are too short to obtain statistically converged values for the wake, one may expect a fully turbulent flow 
after a certain transition length. The calculation for $Re=3000$ indicates that this length decreases for increasing 
Reynolds number.

\begin{figure}
\centering
  \hfill %
  \subfloat[]{ \includegraphics[width=0.48\textwidth]{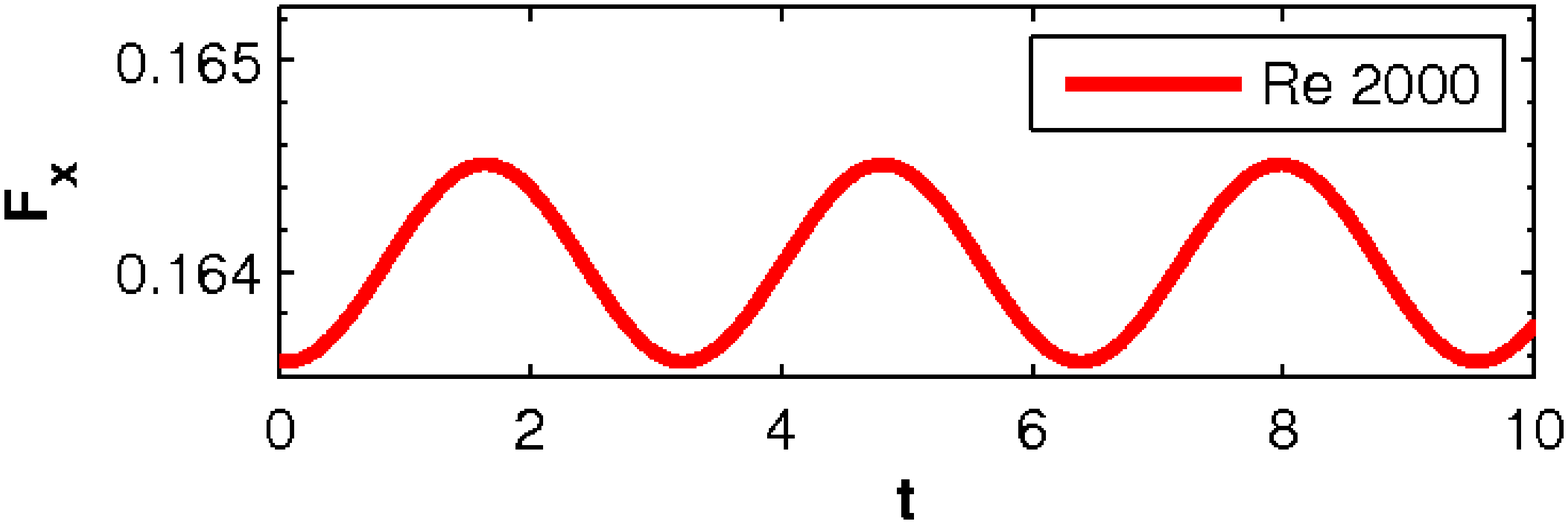}}
  \hfill %
  \subfloat[]{ \includegraphics[width=0.48\textwidth]{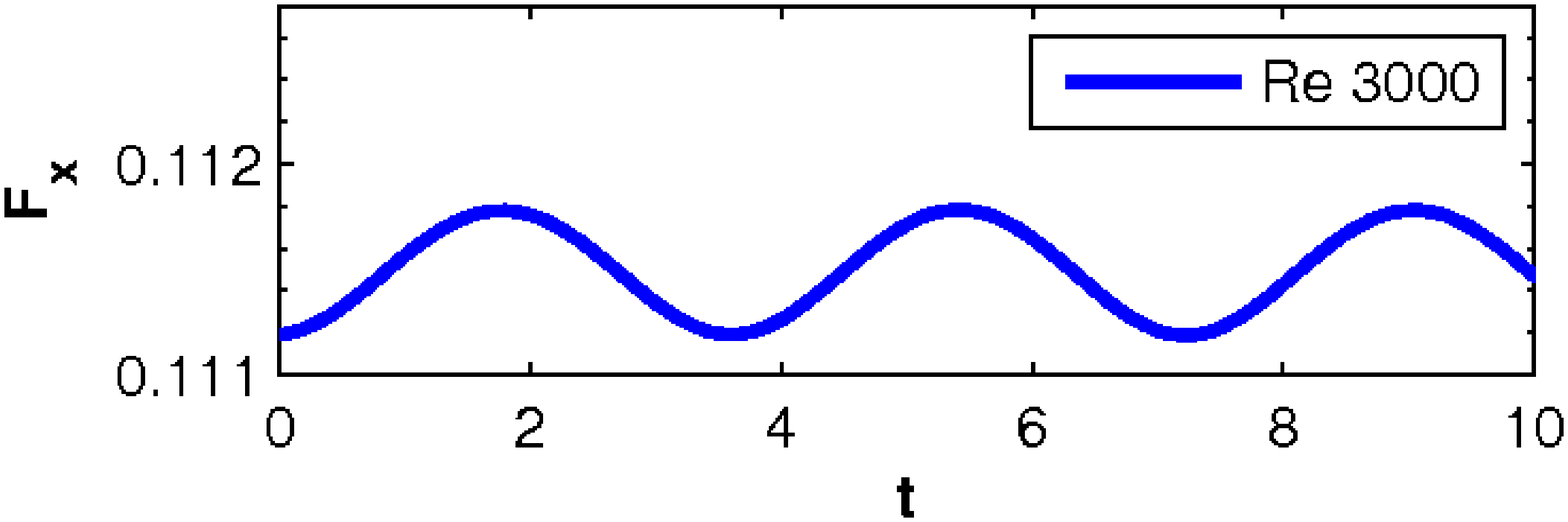}}
  \hfill %
\caption{(colour online)  Time-dependent behavior of the force: Sinusoidal time signal for (a) $Re=2000$ and (b) $Re=3000$ with  $Ha=100$ and spanwise dipole in a distance of $h=1.6$.
} \label{fig:force_per_t}
\end{figure}

In the following, we restrict the study of Reynolds number influence to three exemplary Reynolds numbers: $Re=1000, 2000$ and 3000 are examined for 
fixed Hartmann number of $Ha=100$. The streamlines for the three cases are shown in Fig.~\ref{fig:streamlines_Re}. 
The duct flow is stationary for $Re=1000$,  the wake remains laminar. As for $Re=2000$, 
the wake undergoes a transition to turbulence for $Re =3000$. The vortex structure below the dipole position remains 
similar for all three cases.  The $\lambda_2$-criterion reveals here the  differences (cf. top row of  
Fig.~\ref{fig:streamlines_Re}). For $Re=1000$ only one vortex is pronounced. In case of $Re\geq2000$,  additional 
vortices are produced in front of the dipole position, at $x<0$, and during the vortex shedding. The amount and the strength 
of these vortical structures is enhanced at $Re=3000$. 
The principal mechanism of the vortex shedding remains 
as explained above.  The higher the Reynolds number the shorter are the generated  hairpin structures in the 
wake. Besides the differences in the shape of the flow, the oscillation time is longer with increasing $Re$. For $Re =2000$ 
an oscillation time of 3.168 was calculated, while for $Re =3000$ it is 3.636 nondimensional time units. 
The time signal for both cases is depicted in Fig. \ref{fig:force_per_t}.

We conclude from this section that the vortex shedding is a complex mechanism initiated by the Lorentz force. 
The Reynolds number determines how likely it is that the flow forms vortical structures and transforms  into a turbulent 
flow in the wake. As a final step to this parameter study, it remains to investigate the influence of the Hartmann number 
on the created vortex. 

\subsection{Hartmann number dependence }\label{sec:turb_ha}

\begin{figure}
 \includegraphics[width = \textwidth]{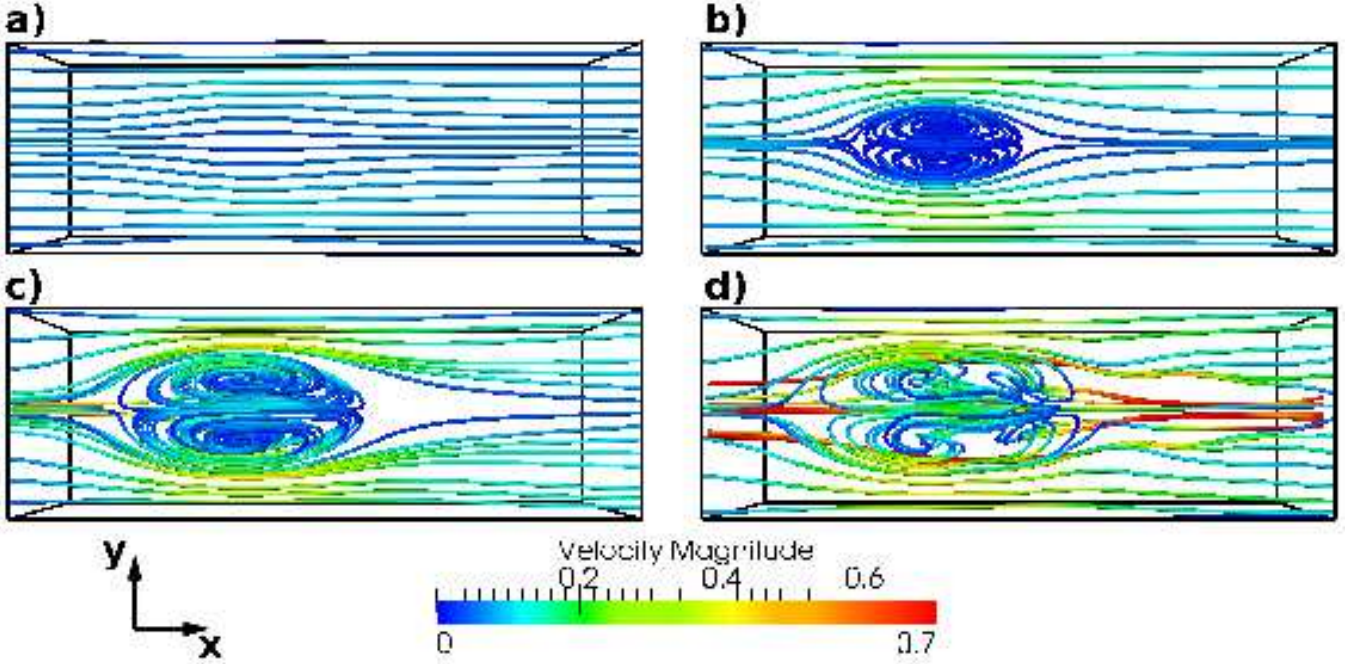}
\hfill
  \caption{(colour online)  Streamlines for different Hartmann numbers and a spanwise dipole orientation at $Re=2000$, $h=1.6$. (a) $Ha=25$, (b) $Ha=50$ (c) $Ha=70$ (d) $Ha=80$. 
Box indicates the duct from $x=-2$ to $ x=5$. Total length was $15\pi$. } 
\label{fig:streamlines_Ha}
\end{figure}

\begin{figure}
\centering
\includegraphics[width = 0.53 \textwidth]{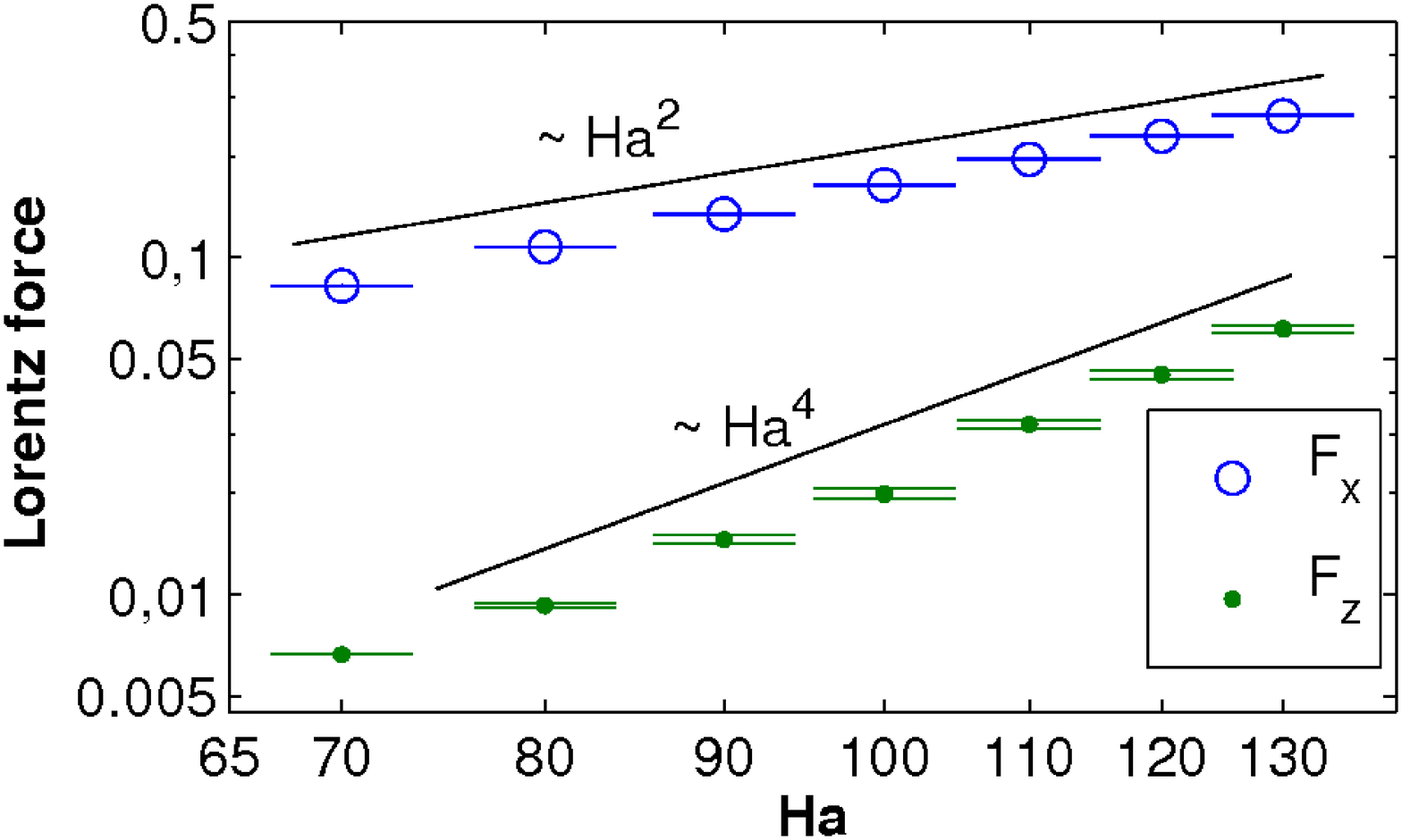}
\caption{(colour online)   Lift and drag forces follow approximately the same power laws as in low Reynolds case (cf. Fig.~\ref{fig:low_Re_FxFzTy_per_Ha}). 
Points are given at the mean value while the vertical bars indicate  the range of the variation of the  time signal. 
}\label{fig:power_high_Re} 
\end{figure}

To investigate  the influence of the Hartmann number, we consider  Reynolds number of 2000 and vary the Hartmann 
number between 25 and 130. For very small Hartmann number,  there is almost no deformation of the flow visible.
Fig.~\ref{fig:streamlines_Ha} shows examples of the streamlines for several Hartmann numbers. For $Ha =25$, one 
may already observe the local Hartmann layers, but the Lorentz force is not strong enough to create a flow reversal. 
Such reversed flow is indicated by the streamlines  from $Ha= 50$ and higher. A turbulent wake is observed for $Ha \geq 80$.

The dependence of the Lorentz force on the Hartmann number is summarized in Fig.~\ref{fig:power_high_Re}. 
Here, the mean values of the forces are displayed with points and the range of the variation due to the vortex shedding 
is marked with two bars. Similar to the investigations in Sec.~\ref{sec:steady_state_solutions}, one finds the same power 
laws for the total forces, i.\,e., $F_x\sim Ha^2$ and $F_z\sim Ha^4$. The  vortex shedding does not influence the power 
law as the oscillations of $F_x$ and $F_z$ are of the order of 1\% and 10\% of the forces, respectively.

As an additional characteristic the frequency of the vortex shedding can be obtained from the time signal of the force. 
Regarding the dipole as a magnetic obstacle allows one  to compare the frequency of vortex shedding with typical values 
for flows generated past a  solid cylinder (see e.\,g. Cuevas et al (2006)~\cite{Cuevas}). The nondimensional parameter for such comparison 
would be then the Strouhal number
\begin{equation}
 St = \frac{f D}{\bar u}. 
\end{equation}
Here, $f$ denotes the frequency, $\bar u$ the mean velocity and $D$ the characteristic length of the obstacle, e.\,g. the 
radius of the cylinder. Whereas the frequency and the velocity are known in our setting, it is not obvious how  to determine 
the characteristic length $D$ of the magnetic obstacle.  A good estimate for the size of the magnetic obstacle is the area  
in which the flow is detached from the wall, that was already discussed above (cf. Fig. \ref{fig:seperationline_Re2000_time}).  
This region is stretched along the streamwise direction and the shape may change in time due to the vortex shedding.
We define the characteristic length $D$ of the magnetic obstacle  as the maximal spanwise width of the area that is enclosed 
by $\del z u_x=0$ at $z=1$. 

\begin{figure}
\centering
  \hfill %
  \subfloat[]{ \includegraphics[width=0.48\textwidth]{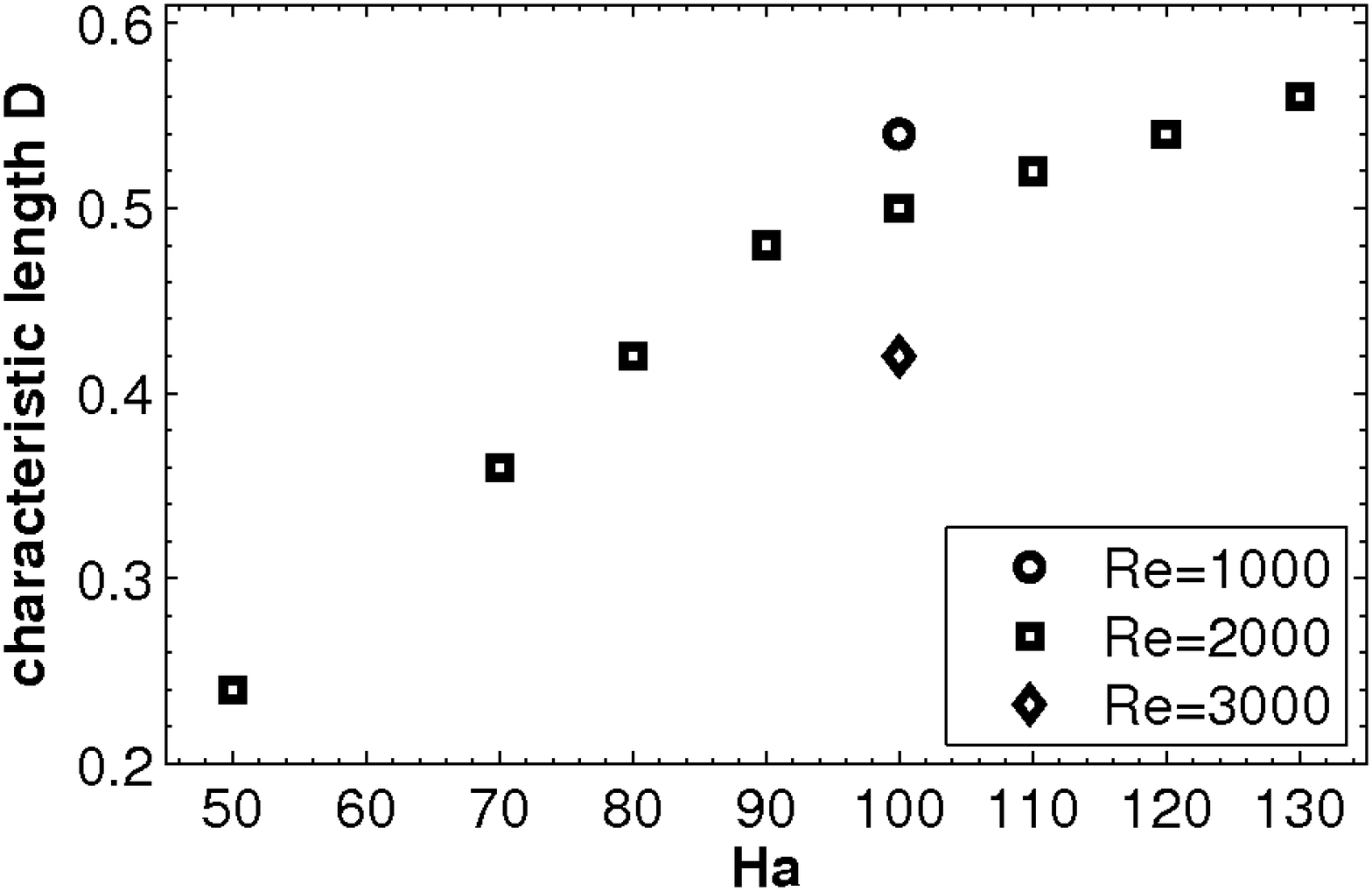}}
  \hfill %
  \subfloat[]{ \includegraphics[width=0.48\textwidth]{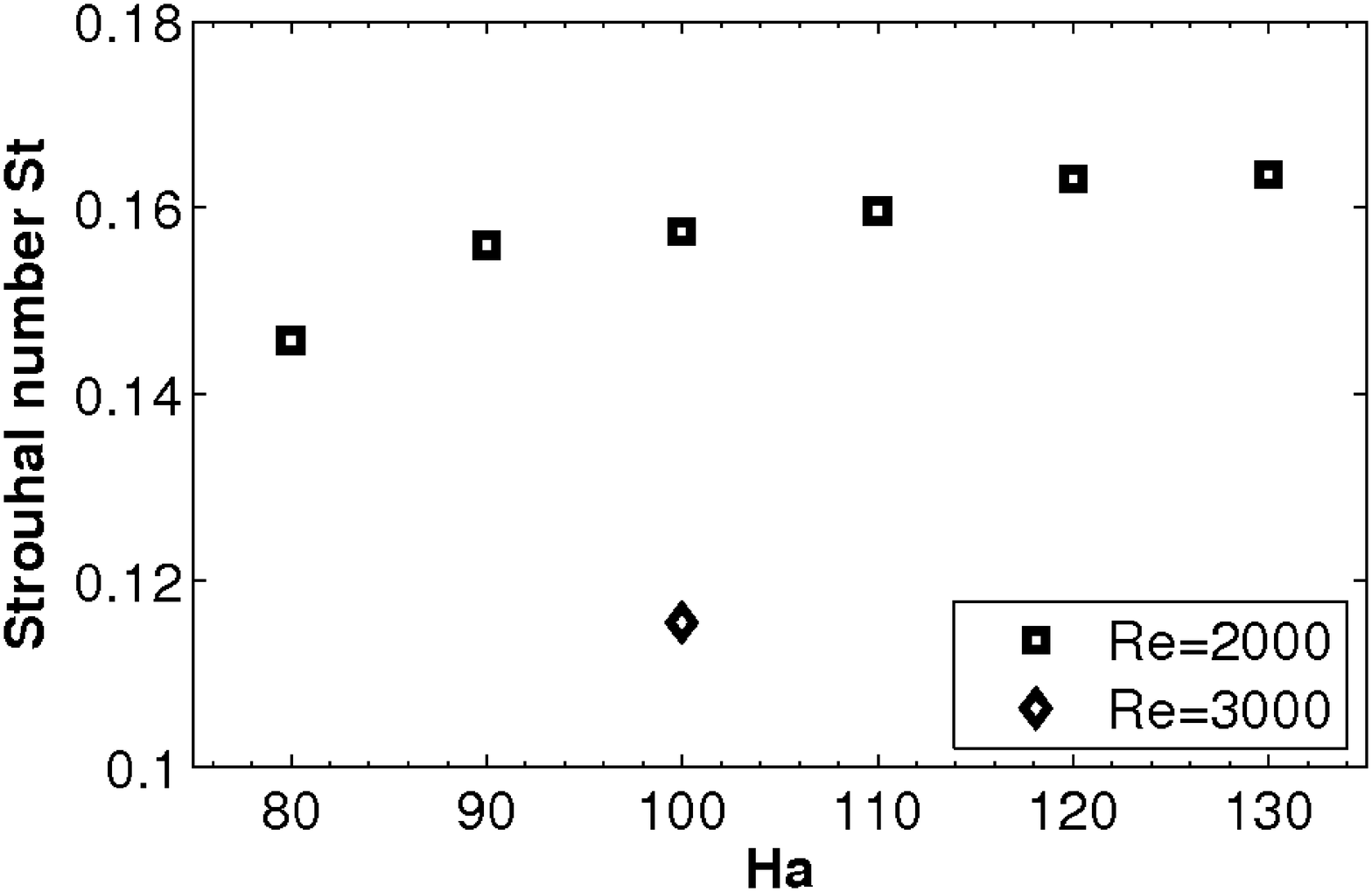}}
  \hfill %
\caption{ Characteristics of the vortex shedding. (a) 
The width of the magnetic obstacle increases linearly for small Hartmann numbers and comes to a saturation 
for higher $Ha$, where the vortex shedding appears. Increasing the Reynolds number leads to decreased width.  
(b) The Strouhal number increases slightly with increasing  Hartmann number and decreases with increasing Reynolds number.}
\label{fig:frequency_per_Ha}\label{fig:Strouhal_per_Ha}\label{fig:width_per_Ha}
\end{figure}

The  obtained widths increase with increasing Hartmann number as displayed in Fig. \ref{fig:width_per_Ha}~(a).
In addition, the width is decreased with increasing Reynolds number. It is worth to note the influence of the interaction parameter 
$N= Ha^2/Re$ in the following example. The width is approximately the same for the simulations at $Re=2000$ with $Ha=80$ 
and at $Re=3000$ with $Ha=100$. Both cases have similar interaction parameter of $N= 3.2$ and $3.\bar3$, respectively. This 
estimate does not hold for the regime without vortex shedding.  
 
Furthermore, the frequency is found to decrease with increasing Hartmann number. This behavior was also observed for the 
two-dimensional flow with a small magnetic obstacle \citep{Cuevas}. The Strouhal number in Cuevas et al. (2006)~\cite{Cuevas} was obtained 
with a characteristic length that was fixed by the size of the small magnet. Therefore, they describe a decrease of the Strouhal 
number for increasing Hartmann number with values around $St \sim 0.1$. In a similar parameter study, Kenjeres (2012)~\cite{Kenjeres2012} 
finds for the same interaction parameters as in our study $St = 0.282$ which is calculated from  power spectra at several positions 
close to the magnetic obstacle. In the present work, the width of the area of reversed flow is used as a characteristic length of 
the magnetic obstacle.  The width $D$ is therefore a dynamic parameter that depends on $Ha$. If the Hartmann number is 
enhanced, the resulting Strouhal number increases with a gentle slope and reaches a saturation around 
$St \sim 0.16$ (cf. Fig.~\ref{fig:Strouhal_per_Ha}~(b) for $Re=2000$).

The result can be compared to the flow around a solid cylinder. There, the Strouhal number is of the order $St\sim 0.2$ 
\cite{Williamson1996}. Dousset and Potherat (2008)~\cite{Dousset2008} investigated the influence of a homogeneous magnetic field on the flow around 
a cylinder.  In the regime of high Reynolds number, the Strouhal number was found to decrease with increasing Hartmann number. 
Thus, the homogeneous magnetic field damps the vortical structures. In our case, the inhomogeneous localized magnetic gives 
rise to the vortex formation and  subsequent shedding. 

\begin{figure}
\centering
 \includegraphics[width=.70\textwidth]{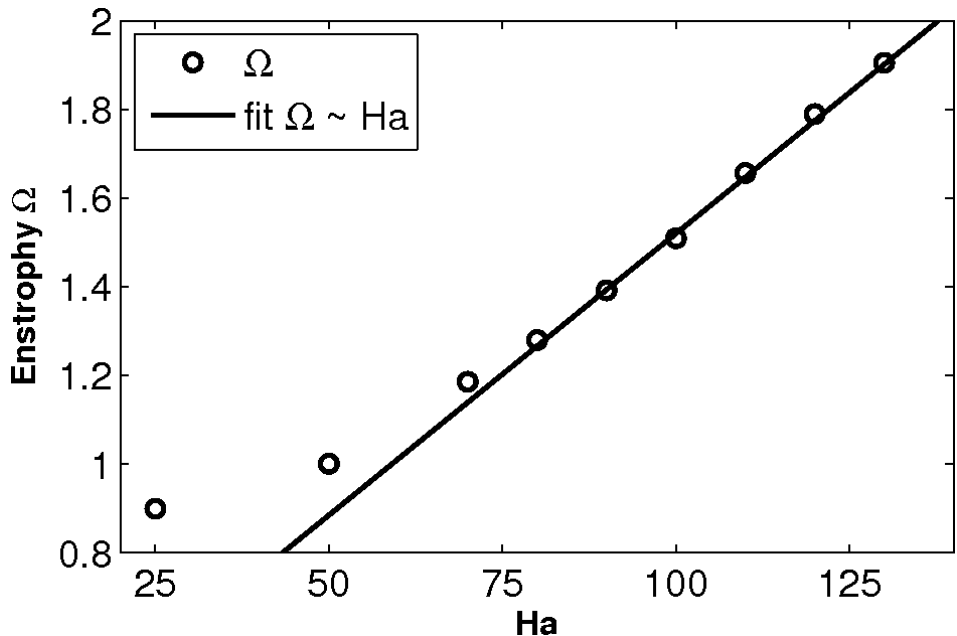}
\caption{ The total enstrophy $\Omega$ as calculated in (\ref{eq:enstrophy}) is linear dependent on the  Hartmann number. 
This behavior is originated in the  local Hartmann layers. Data is given for $Re=2000$, $h=1.6$ and spanwise oriented dipole.}
\label{fig:enstrophy_per_Ha}
\end{figure}

Apart from the  Strouhal number, it is desirable to have an additional quantity
for the deflection of the flow that is caused by the dipole. The aim is therefore to characterize  the structures in 
Fig.~\ref{fig:streamlines_Ha}.  A possible approach is to measure the vorticity, $\vec \omega = \nabla \times \vec u$, 
of the flow.  For this, we consider the enstrophy over a duct fraction right below the dipole which is defined as  
\begin{equation}\label{eq:enstrophy}
 \Omega = \int_{[-1,1]^3} \frac{1}{2} | \nabla \times \vec u |^2 d V. 
\end{equation}
As the wake contains turbulent vortices, the integral is not applied to the whole domain. Thus, we focus on the area  
that is directly influenced by the dipole. Fig.~\ref{fig:enstrophy_per_Ha} shows that the enstrophy increases linearly 
with the Hartmann number, which is caused by 
an  increase of the vorticity $\vec \omega= \nabla \times \vec u  $ 
in the  vortex as well as the Hartmann layers. The strongest contribution comes from the partial derivative $\del z u_x$, 
which reaches maximal magnitudes in the Hartmann layer, as already mentioned above (cf. Fig.~\ref{fig:seperationline_Re2000_time}). 
An integration shows that the integral $ \int_{[-1,1]^3} \frac{1}{2} | \del z u_x |^2 d V$ gives already about 80\% of the 
magnitude of the total enstrophy $\Omega$.  

To explain the linear dependence on the Hartmann number, we recall the properties of the Hartmann layers in homogeneous magnetic fields.
It is known from the analytic solution of the Hartmann channel flow~\citep{Mueller2001} that $u_x \sim 1-\exp(-Ha \, z)$, where 
$z$ measures the distance from the wall. We can therefore estimate  $(\del z u_x)^2 \sim \left(Ha\exp(-Ha \, z)\right)^2$ 
in the Hartmann layer. The width of the Hartmann layer is given by $\delta \sim 1/Ha$. Therefore, the integration 
of the enstrophy leads to $\Omega \sim \int_0^\delta  | (\del z u_x)^2 | d z\sim Ha $. 

It has to be remarked here, that this explanation is only valid for the total enstrophy of duct  flow in a uniform magnetic field when $Ha$ is large.
While the Hartmann layers  show  linear dependence of the total 
enstrophy on the Hartmann number, there are  additional contributions from the Shercliff layers  proportional to $Ha^{1/2}$. 
As a conclusion, the qualitative presentation of the flow structures in terms of the 
enstrophy (\ref{eq:enstrophy}) fails to provide details on the reversed flow and the vortex structures. Nevertheless, it 
gives a good criterion for the strength of the local Hartmann layers.


 
\section{Conclusion and Outlook}\label{sec:conclusion}

In this work, we have investigated the influence of a localized inhomogeneous magnetic field on liquid metal flow in 
the quasi-static approximation.  Our study focuses on the fundamental aspects of the structure formation. 
Therefore we use the simplest setting, a point dipole for the {\em localized} magnetic field and a simple shear flow in a square duct.
It should therefore be considered as a paradigm for many realistic liquid metal flows. The magnetic field gives rise 
to a Lorentz force in the liquid that deforms the streamlines. Depending on the orientation of the dipole, areas of 
reversed flow and localized or local Hartmann layers are formed at the top surface due to strong Lorentz forces. 
The notion of local Hartmann layer is considered as a generalization of the canonical Hartmann flow and layer which 
arises due to the finite spanwise extension in the square cross section of the duct and the locality of the magnetic field.
We have provided a systematic parameter dependence study which unraveled the local effects of the dipole field on the flow.

The parameter study is focused on the influence of the Reynolds and the Hartmann number.  As expected the flow 
remains laminar and stationary for low Reynolds numbers. When the Reynolds number is increased, we observe 
that the length of the wake is increased. We showed that the spatial downstream decay rate of the deflected flow in the 
wake is proportional to $1/Re$.  For $Re \geq 2000$, the spanwise oriented dipole in a distance of $h=1.6$ 
triggers vortex shedding in case of Hartmann number above 80. The flow becomes time-dependent and eventually turbulent.

The Hartmann number affects the total Lorentz force. The drag component is proportional to $Ha^2$, while the lift force 
behaves like $Ha^4$ for $Ha \leq 25$. The latter  is connected with the nonlinearity in the Navier-Stokes equation. 
In  case of time-dependent flow with a spanwise oriented dipole, the vortex shedding leads to a sinusoidal force 
signal. The frequency of the signal corresponds to a Strouhal number that increases with increasing Hartmann number. 
The Strouhal number saturates around 0.16 which is comparable with classical flow past a cylinder. Nevertheless, it is not 
only the primary vortex, that has strong influence of the flow structure, but also the local Hartmann layers. 
Due to the formation of these boundary layers, the enstrophy of the deformed flow increases linearly with the Hartmann 
number.

Our study opens some new perspectives for flow manipulation and thus also for flow control. The described setting 
can be used to trigger turbulence  in a laminar flow. This can be applied, e.\,g., for mixing injected 
substances or particles with the main flow. Such applications as well as experimental investigations of the problem 
have to be  done with magnetic fields that are localized and inhomogeneous but not necessarily of the shape of a point 
dipole. Future numerical studies should include a turbulent inflow to model the experimental setting  more realistically. 

To obtain a complete understanding of the mechanism of the vortex shedding,  the stability study can be extended. This 
could be provided by adding prescribed perturbations to a deformed but stationary flow. A more sophisticated way would 
be the direct optimal growth analysis as used by Barley et al. (2008)~\cite{Barkley2008} or a sensitivity study like Giannetti and Luchini (2007)~\cite{Giannetti2007}. 

Finally, the study should be extended to capture the influence of geometry parameters such as the distance of the dipole or 
its orientation. Here, the question arises whether the vortex shedding is still present if the dipole is oriented in an oblique 
orientation and what  the optimal local Hartmann layers are to trigger flow transition. In addition, the symmetry of the given 
setting could be broken if the dipole is not positioned in the centerline of the duct, but with a certain offset.  
The influences of walls could also be studied by using rectangular ducts with non-square cross-section or pipes.
These investigations open a huge parameter space that has to be considered  in future works.

\begin{acknowledgments}
The authors wish to thank Dmitry Krasnov for providing the core of the used numerical code and 
for many fruitful discussions. We have also benefited from discussions with Oleg Zikanov, Andr\'e Thess and Bernard Knaepen. 
The authors gratefully acknowledge the financial support from the Deutsche Forschungsgemeinschaft in the framework of the 
Research Training Group Lorentz Force Velocimetry and Lorentz Force Eddy Current Testing (grant DFG GK 1567/1). 
The numerical calculations have been performed on the cluster at  TU Ilmenau and on the Juropa cluster at NIC/JSC (J\"ulich).
\end{acknowledgments}


\end{document}